\documentclass[aps,pra,reprint,floatfix,superscriptaddress,nofootinbib]{revtex4-2}

\usepackage{graphicx}
\usepackage{dcolumn}
\usepackage{bm}
\usepackage[dvipsnames]{xcolor} 
\usepackage{amsthm,amsmath,amssymb,bbm}
\usepackage{physics}
\usepackage{tikz}
\usepackage[colorlinks=true, linkcolor=blue, citecolor=blue, urlcolor=blue]{hyperref}
\usepackage{soul}
\usepackage{tabularx}

\newtheorem{theorem}{Theorem}

\newtheorem{lemma}[theorem]{Lemma}
\newtheorem{prop}[theorem]{Proposition}

\begin{document}

\preprint{APS/123-QED}

\title{Recoverability of Gravitational Entanglement under Decoherence}

\author{Yi-Xuan Wang}
\affiliation{Institute of Theoretical Physics, Ulm University, Albert-Einstein-Allee 11, 89081 Ulm, Germany}
\affiliation{Center for Integrated Quantum Science and Technology (IQST), 89081 Ulm, Germany}

\author{Julen S. Pedernales}%
\affiliation{Institute of Theoretical Physics, Ulm University, Albert-Einstein-Allee 11, 89081 Ulm, Germany}
\affiliation{Center for Integrated Quantum Science and Technology (IQST), 89081 Ulm, Germany}
\affiliation{Department of Quantum Nanoscience, Kavli Institute of Nanoscience, Delft University of Technology, 2628 CJ Delft, The Netherlands.}

\author{Martin B. Plenio}
\affiliation{Institute of Theoretical Physics, Ulm University, Albert-Einstein-Allee 11, 89081 Ulm, Germany}
\affiliation{Center for Integrated Quantum Science and Technology (IQST), 89081 Ulm, Germany}

\begin{abstract}

Decoherence can make gravitationally generated entanglement arbitrarily small, or even vanishing, without necessarily rendering the effective matter dynamics separable. We ask whether this hidden non-separability can be revealed using strictly local quantum error correction. For Newtonian gravity supplemented by physically relevant localization noise, including environmental noise, and noise from collapse models such as GRW, CSL, and Di\'osi--Penrose, we analyze when the underlying entangling resource is locally recoverable. For GRW- and CSL-type noise, we analytically establish that strictly local quantum error correction can make the ratio of the effective gravitational
interaction strength to the residual decoherence rate arbitrarily large for any parametrization of the collapse model, even in regimes where the unencoded entanglement signal approaches zero. By contrast, the Di\'osi--Penrose model does not exhibit the same enhancement under bipartite protocol considered here, revealing a qualitative difference. These results show that the loss of detectable entanglement is not equivalent to the loss of a recoverable non-separable resource, and open a route toward practical recovery protocols.

\end{abstract}

\maketitle

\textit{Introduction.}---
Can gravity transmit quantum information? An empirical answer to this question remains far beyond current experimental reach. Nevertheless, recent advances in quantum optomechanics~\cite{Weig2022,doi:10.1126/science.abg3027} suggest that proposals to address it may no longer constitute pure Gedankenexperiments. Low-energy tests of quantum aspects of gravity, first envisioned by Feynman~\cite{dewitt2011role} and later extended and refined in various theoretical settings~\cite{lindner2005testing,Schmöle_2016,Kafri:2013wxa,PhysRevLett.119.120402,bose2017spin,marletto2017gravitationally,Tomasz2019Krisnanda, CoscoPP21, WeissRT+21}, offer one of the most promising routes for probing the interface between quantum mechanics and gravity~\cite{carney2019tabletop,RevModPhys.97.015003}.

From a quantum-information perspective, this question is naturally framed in terms of \emph{separable} versus \emph{non-separable} dynamical maps~\cite{PhysRevA.57.1619,Rains,plenio2005introduction}. Separable operations, quantum channels whose Kraus operators factorize across subsystems, cannot create entanglement from separable inputs.
The observation of any dynamical signature that requires a non-separable map for its description would provide an operational way to rule out gravitational models whose effective dynamics is separable. Proposed signatures include the generation of gravitationally induced entanglement~\cite{lindner2005testing,Kafri:2013wxa,Kafri_2014,Schmöle_2016,PhysRevLett.119.120402,bose2017spin,marletto2017gravitationally,Tomasz2019Krisnanda,CoscoPP21, WeissRT+21,Rovelli2019,Plenio2022Pedernales,Pikovski2022,Perche2023,Plenio2023Pedernales} and violations of inequalities satisfied by all separable dynamics~\cite{Kafri_2014,Wehner2016,Iyer2021,Weller-Davies2023,Plenio2024Lami,Bassi2025}.

It is often assumed that spontaneous-collapse noise~\cite{GRWPhysRevD.34.470,CSLPhysRevA.42.78,CSLPhysRevA.39.2277,bassi2013models,donadi2022seven} renders gravitational dynamics effectively separable, although recent work has suggested that non-separable features may persist even in strongly decohering regimes~\cite{PhysRevD.111.L121101,Aspelmeyer2026}. More generally, even when the dynamics remains non-separable, decoherence can suppress entanglement generation to arbitrarily low levels, making the underlying entangling resource exceedingly difficult to observe. This raises the natural question of whether the effect of noise can be reduced without simultaneously suppressing the gravitational signal.

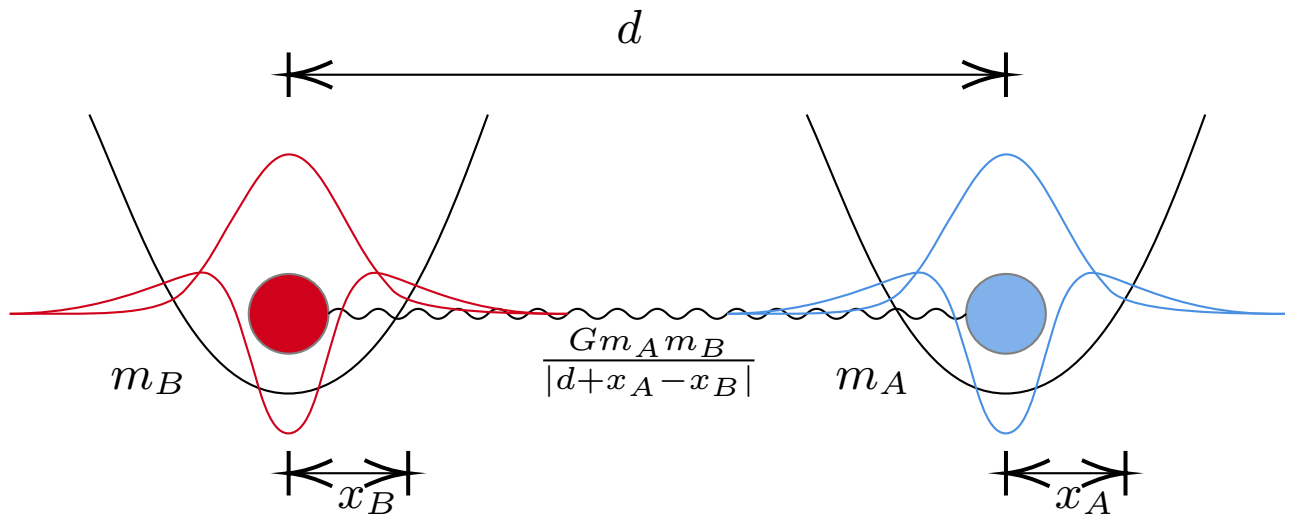
\begin{figure}
    \centering
      

\begin{tikzpicture}[x=0.75pt,y=0.75pt,yscale=-1,xscale=1]

\draw    (100,80) .. controls (101.67,78.33) and (103.33,78.33) .. (105,80) .. controls (106.67,81.67) and (108.33,81.67) .. (110,80) .. controls (111.67,78.33) and (113.33,78.33) .. (115,80) .. controls (116.67,81.67) and (118.33,81.67) .. (120,80) .. controls (121.67,78.33) and (123.33,78.33) .. (125,80) .. controls (126.67,81.67) and (128.33,81.67) .. (130,80) .. controls (131.67,78.33) and (133.33,78.33) .. (135,80) .. controls (136.67,81.67) and (138.33,81.67) .. (140,80) .. controls (141.67,78.33) and (143.33,78.33) .. (145,80) .. controls (146.67,81.67) and (148.33,81.67) .. (150,80) .. controls (151.67,78.33) and (153.33,78.33) .. (155,80) .. controls (156.67,81.67) and (158.33,81.67) .. (160,80) .. controls (161.67,78.33) and (163.33,78.33) .. (165,80) .. controls (166.67,81.67) and (168.33,81.67) .. (170,80) .. controls (171.67,78.33) and (173.33,78.33) .. (175,80) .. controls (176.67,81.67) and (178.33,81.67) .. (180,80) .. controls (181.67,78.33) and (183.33,78.33) .. (185,80) .. controls (186.67,81.67) and (188.33,81.67) .. (190,80) .. controls (191.67,78.33) and (193.33,78.33) .. (195,80) .. controls (196.67,81.67) and (198.33,81.67) .. (200,80) .. controls (201.67,78.33) and (203.33,78.33) .. (205,80) .. controls (206.67,81.67) and (208.33,81.67) .. (210,80) .. controls (211.67,78.33) and (213.33,78.33) .. (215,80) .. controls (216.67,81.67) and (218.33,81.67) .. (220,80) .. controls (221.67,78.33) and (223.33,78.33) .. (225,80) .. controls (226.67,81.67) and (228.33,81.67) .. (230,80) .. controls (231.67,78.33) and (233.33,78.33) .. (235,80) .. controls (236.67,81.67) and (238.33,81.67) .. (240,80) .. controls (241.67,78.33) and (243.33,78.33) .. (245,80) .. controls (246.67,81.67) and (248.33,81.67) .. (250,80) .. controls (251.67,78.33) and (253.33,78.33) .. (255,80) .. controls (256.67,81.67) and (258.33,81.67) .. (260,80) -- (260,80) ;
\draw  [color={rgb, 255:red, 0; green, 0; blue, 0 }  ,draw opacity=0.5 ][fill={rgb, 255:red, 208; green, 2; blue, 27 }  ,fill opacity=1 ] (80,80) .. controls (80,74.48) and (84.48,70) .. (90,70) .. controls (95.52,70) and (100,74.48) .. (100,80) .. controls (100,85.52) and (95.52,90) .. (90,90) .. controls (84.48,90) and (80,85.52) .. (80,80) -- cycle ;
\draw  [color={rgb, 255:red, 0; green, 0; blue, 0 }  ,draw opacity=0.5 ][fill={rgb, 255:red, 74; green, 144; blue, 226 }  ,fill opacity=0.7 ] (260,80) .. controls (260,74.48) and (264.48,70) .. (270,70) .. controls (275.52,70) and (280,74.48) .. (280,80) .. controls (280,85.52) and (275.52,90) .. (270,90) .. controls (264.48,90) and (260,85.52) .. (260,80) -- cycle ;
\draw    (90,20) -- (270,20) ;
\draw [shift={(270,20)}, rotate = 180] [color={rgb, 255:red, 0; green, 0; blue, 0 }  ][line width=0.75]    (0,5.59) -- (0,-5.59)(10.93,-3.29) .. controls (6.95,-1.4) and (3.31,-0.3) .. (0,0) .. controls (3.31,0.3) and (6.95,1.4) .. (10.93,3.29)   ;
\draw [shift={(90,20)}, rotate = 0] [color={rgb, 255:red, 0; green, 0; blue, 0 }  ][line width=0.75]    (0,5.59) -- (0,-5.59)(10.93,-3.29) .. controls (6.95,-1.4) and (3.31,-0.3) .. (0,0) .. controls (3.31,0.3) and (6.95,1.4) .. (10.93,3.29)   ;
\draw    (90,120) -- (120,120) ;
\draw [shift={(120,120)}, rotate = 180] [color={rgb, 255:red, 0; green, 0; blue, 0 }  ][line width=0.75]    (0,5.59) -- (0,-5.59)(10.93,-3.29) .. controls (6.95,-1.4) and (3.31,-0.3) .. (0,0) .. controls (3.31,0.3) and (6.95,1.4) .. (10.93,3.29)   ;
\draw [shift={(90,120)}, rotate = 0] [color={rgb, 255:red, 0; green, 0; blue, 0 }  ][line width=0.75]    (0,5.59) -- (0,-5.59)(10.93,-3.29) .. controls (6.95,-1.4) and (3.31,-0.3) .. (0,0) .. controls (3.31,0.3) and (6.95,1.4) .. (10.93,3.29)   ;
\draw    (40,30) .. controls (52.43,57.93) and (65.57,100.07) .. (90,100) .. controls (114.43,99.93) and (129.43,58.93) .. (140,30) ;
\draw [color={rgb, 255:red, 208; green, 2; blue, 27 }  ,draw opacity=1 ]   (20,80) .. controls (39.43,79.79) and (57.57,81.21) .. (64,74) .. controls (70.43,66.79) and (71.57,63.21) .. (76,56) .. controls (80.43,48.79) and (84.79,40.11) .. (90,40) .. controls (95.21,39.89) and (99.43,47.79) .. (104,56) .. controls (108.57,64.21) and (110.43,67.79) .. (116,74) .. controls (121.57,80.21) and (149.71,79.89) .. (160,80) ;
\draw [color={rgb, 255:red, 208; green, 2; blue, 27 }  ,draw opacity=1 ]   (20,80) .. controls (49.43,79.79) and (64.57,67.39) .. (70,70) .. controls (75.43,72.61) and (77.57,82.39) .. (80,90) .. controls (82.43,97.61) and (84.79,110.11) .. (90,110) .. controls (95.21,109.89) and (97.43,98.61) .. (100,90) .. controls (102.57,81.39) and (104.57,72.39) .. (110,70) .. controls (115.43,67.61) and (130.43,80.79) .. (160,80) ;
\draw    (220,30) .. controls (232.43,57.93) and (245.57,100.07) .. (270,100) .. controls (294.43,99.93) and (309.43,58.93) .. (320,30) ;
\draw    (270,120) -- (300,120) ;
\draw [shift={(300,120)}, rotate = 180] [color={rgb, 255:red, 0; green, 0; blue, 0 }  ][line width=0.75]    (0,5.59) -- (0,-5.59)(10.93,-3.29) .. controls (6.95,-1.4) and (3.31,-0.3) .. (0,0) .. controls (3.31,0.3) and (6.95,1.4) .. (10.93,3.29)   ;
\draw [shift={(270,120)}, rotate = 0] [color={rgb, 255:red, 0; green, 0; blue, 0 }  ][line width=0.75]    (0,5.59) -- (0,-5.59)(10.93,-3.29) .. controls (6.95,-1.4) and (3.31,-0.3) .. (0,0) .. controls (3.31,0.3) and (6.95,1.4) .. (10.93,3.29)   ;
\draw [color={rgb, 255:red, 74; green, 144; blue, 226 }  ,draw opacity=1 ]   (200,80) .. controls (219.43,79.79) and (237.57,81.21) .. (244,74) .. controls (250.43,66.79) and (251.57,63.21) .. (256,56) .. controls (260.43,48.79) and (264.79,40.11) .. (270,40) .. controls (275.21,39.89) and (279.43,47.79) .. (284,56) .. controls (288.57,64.21) and (290.43,67.79) .. (296,74) .. controls (301.57,80.21) and (329.71,79.89) .. (340,80) ;
\draw [color={rgb, 255:red, 74; green, 144; blue, 226 }  ,draw opacity=1 ]   (200,80) .. controls (229.43,79.79) and (244.57,67.39) .. (250,70) .. controls (255.43,72.61) and (257.57,82.39) .. (260,90) .. controls (262.43,97.61) and (264.79,110.11) .. (270,110) .. controls (275.21,109.89) and (277.43,98.61) .. (280,90) .. controls (282.57,81.39) and (284.57,72.39) .. (290,70) .. controls (295.43,67.61) and (310.43,80.79) .. (340,80) ;

\draw (171,2.4) node [anchor=north west][inner sep=0.75pt]    {$d$};
\draw (101,122.4) node [anchor=north west][inner sep=0.75pt]    {$x_{B}$};
\draw (44,92.4) node [anchor=north west][inner sep=0.75pt]    {$m_{B}$};
\draw (226,92.4) node [anchor=north west][inner sep=0.75pt]    {$m_{A}$};
\draw (281,122.4) node [anchor=north west][inner sep=0.75pt]    {$x_{A}$};
\draw (151,82.4) node [anchor=north west][inner sep=0.75pt]    {$\frac{Gm_{A} m_{B}}{|d+x_{A} -x_{B} |}$};

\end{tikzpicture}
    \caption{Two masses $m_A$ and 
    $m_B$ are trapped in harmonic potentials separated by a distance $d$. Each 
    mass is encoded into a local code subspace via a QEC protocol. The two 
    systems interact gravitationally, potentially generating entanglement.}
    \label{fig:setup}
\end{figure}

In this Letter, we show that several physically relevant collapse models, including GRW, CSL, and Di\'osi--Penrose, do not render the combined Newtonian-plus-noise dynamics separable, even when the collapse noise completely prevents entanglement generation in commonly considered initial state configurations~\cite{bose2017spin,marletto2017gravitationally}. In particular, appropriately chosen initial states can reveal an arbitrarily small but nonzero amount of
entanglement regardless of the collapse-noise strength. We then ask whether
this weak manifestation of the underlying entangling dynamics can be made
detectable using quantum-information techniques without introducing an
independent nonlocal channel between the masses. To this end, we construct a
strictly local quantum error-correction (QEC)~\cite{Terhal2015} protocol that suppresses relevant sources of decoherence while preserving the Newtonian interaction. This amplifies the absolute value of entanglement and sustains it for a longer time. Since all encoding, recovery, and processing operations are local, any entanglement revealed after recovery must originate from the underlying gravitational dynamics, and can be used to establish the non-separable character of the gravitational channel. We call this entangling resource \emph{locally recoverable}, relative to a given entanglement-detection protocol, when local error correction renders otherwise undetectable generated entanglement detectable. Applying this framework, we find that the entangling component of the GRW and CSL dynamics is locally recoverable, whereas the Di\'osi--Penrose model does not exhibit the same behavior under bipartite QEC protocols here.

\textit{The model.}---
We consider a bipartite quantum system consisting of two equal masses, $A$ and $B$, each of mass $m_A = m_B = m$, as depicted in Fig.~\ref{fig:setup}. The masses are confined in independent harmonic traps of frequency $\omega_A=\omega_B=\omega$, with equilibrium positions separated by a distance $d$. We restrict the dynamics to one dimension, along the line connecting the two traps, and denote the canonical operators by $(x_A,p_A)$ and $(x_B,p_B)$, where $x_A$ and $x_B$ represent the displacements of each mass from the equilibrium position of its respective trap center. The two masses interact via a Newtonian gravitational potential and are subject to additional noise sources, giving rise to the master equation

\begin{align}
\label{eq:lind}
    \frac{\dd }{\dd t}\rho = \mathcal{L} (\rho) 
    = -\frac{i}{\hbar}\left[\,H_0 
    - \frac{G m^2}{|d + x_A - x_B|},\,\rho \right] 
    + \mathcal{L}_{\mathrm{dec}}(\rho),
\end{align}
where $H_0 = \sum_{\alpha=A,B}\left({p_\alpha^2}/{2m} +  m \omega^2 x_\alpha^2/2\right)$
and $\mathcal{L}_{\mathrm{dec}}$ models Markovian decoherence via a Lindbladian superoperator. 
The specific form of $\mathcal{L}_{\mathrm{dec}}$ will depend on the considered decoherence 
model, but, in general it is expected to lead to a loss of coherence in the positional basis 
that hinders the generation of entanglement through the Newtonian interaction.

A wide variety of models relevant to low-energy tests of gravity lead to a dynamical equation of the form Eq.~\eqref{eq:lind}. If gravity possesses quantum degrees of freedom, the Newtonian potential follows from standard low-energy arguments, while $\mathcal{L}_{\mathrm{dec}}$ may describe experimental noise~\cite{rijavec2021decoherence}, gravitationally induced noise, as in the Anastopoulos--Hu model~\cite{anastopoulos2013master}, or hypothesized fundamental collapse mechanisms, such as those described in the GRW~\cite{GRWPhysRevD.34.470} and CSL~\cite{CSLPhysRevA.42.78,CSLPhysRevA.39.2277} models. If gravity is taken to be classical and coupled to quantum matter, tracing out the classical degrees of freedom often yields $\mathcal{L}_{\mathrm{dec}}$, as for example, in the  Di\'osi--Penrose (DP)~\cite{diosi1987universal,penrose1996gravity,PhysRevD.96.104045} and Kafri--Taylor--Milburn (KTM) models~\cite{Kafri_2014}.

\textit{Quantum-Error-Correction assisted tests}---
We introduce a strictly local quantum error-correction protocol into the
dynamics~\cite{Terhal2015} by defining a code space with tensor-product
structure $\mathcal{C}=\mathcal{C}_A\otimes\mathcal{C}_B$, where each local code
$\mathcal{C}_\alpha=\mathrm{span}\{\,|\bar k\rangle_\alpha\,\}_k \subset
\mathcal{H}_\alpha$ is a finite-dimensional subspace of the Hilbert space of
mass $\alpha\in\{A,B\}$, together with a local recovery map
\begin{equation}
\label{eq:qec channel}
    R(\rho_{AB})=(R_A\otimes R_B)(\rho_{AB}),\qquad 
    R_\alpha(\rho_\alpha)=\sum_i R_\alpha^i\,\rho_\alpha \,R_\alpha^{i\dagger}.
\end{equation}
Here $R_\alpha^i$ are correcting Kraus operators acting on subsystem $\alpha \in\{A,B\}$.

For a given set of error Kraus operators $\{E_k\}_k$, perfect correction on the
(code) subspace $\mathcal C$ is ensured by the Knill--Laflamme conditions~\cite{QECPhysRevA.55.900}
\begin{equation}
\label{eq:qec cond}
    \langle \bar i| E_k^\dagger E_l|\bar j\rangle = C_{kl}\,\delta_{ij},
\end{equation}
where $\{|\bar i\rangle\}_i$ denotes an orthonormal basis of the \emph{global}
code space $\mathcal C\subset\mathcal H_A\otimes\mathcal H_B$, and $C_{kl}$ is a
Hermitian matrix.

We assume that the local recovery operation $R$ is applied on a timescale much
shorter than the natural evolution generated by the Lindbladian $\mathcal{L}$ in
Eq.~\eqref{eq:lind}. If $R$ is idempotent ($R^2 = R$), then in the limit of
infinitely frequent recovery the stroboscopic dynamics converges to a Zeno-type
evolution (see App.~\ref{appendix:ZenoLimit} 
), $\lim_{N\to\infty}\left(R\circ e^{\mathcal{L}t/N}\right)^N (\rho)= e^{(R\circ\mathcal L )t} (\rho)$, where we have assumed that $\rho$ is a quantum state initially fully in the code space, i.e. $R(\rho)=\rho$. In this limit, the effective evolution within the
code space is thus generated by the superoperator
\begin{equation}
\mathcal L_{\mathrm{eff}} = R\circ\mathcal L.
\end{equation}
The recovery map $R$ restores the encoded state for correctable errors and maps any residual leakage, which originates from uncorrectable errors or the unitary evolution, back into the code-space as effective logical noise. Because $R$ is a strictly separable CPTP map, the non-separability of the logical dynamics $\mathcal{L}_{\rm eff}$ implies the non-separability of the underlying physical dynamics $\mathcal{L}$. While here we focus on the Zeno limit for analytical simplicity, the same implication holds at finite recovery rate.

Our task is to construct a local $R$ that suppresses noise while retaining the nonlocal coherent interaction in the resulting logical dynamics. This task is, however, non-
trivial and generally depends on the considered decoherence model. We illustrate the difficulty using the GRW model; constructions for other noise models appear in App.~\ref{app:Overview of the decoherence models} \&\ref{app:Effective error rates}
. For a single mass with position operator $x$, the GRW collapse mechanism is described by the Lindbladian
\begin{equation}
\label{eq:grwlind}
    \mathcal{L}_{\mathrm{GRW}}(\rho) 
    = \lambda_{\mathrm{GRW}}
      \left(
        \int \dd q\, \frac{1}{\sqrt{\pi r_{\mathrm C}^2}}
        e^{-\frac{(x - q)^2}{2 r_{\mathrm C}^2}} 
        \rho\,
        e^{-\frac{(x - q)^2}{2 r_{\mathrm C}^2}}
        - \rho
      \right),
\end{equation}
where $\lambda_{\mathrm{GRW}}$ is the collapse rate and $r_{\mathrm C}$ the localization length. The effect of this term is to suppress spatial superpositions: an off-diagonal positional matrix element $\operatorname{Tr}[\rho |x_1\rangle\langle x_2|]$ decays exponentially at a rate $\lambda_{\mathrm{GRW}} \left(1 - \exp[-(x_1 - x_2)^2/4 r_{\mathrm C}^2]\right)$. We further consider the commonly studied regime in which both the interaction and the noise are linearized in the position operators, so that the coherent interaction takes the form $x_A\otimes x_B$, while the dominant local noise acts through jump operators $x_A\otimes\mathbb{I}_B$ and $\mathbb{I}_A\otimes x_B$. In the Zeno limit of continuous recovery, perfect correction of the single-body errors generated by $x_A$ and $x_B$ eliminates the bipartite interaction $x_A \otimes x_B$, independent of whether the recovery is local or nonlocal; see App.~\ref{appendix:LimitationsZeno}
. This obstruction is specific to noise generated by the \emph{same operators} that enter the interaction Hamiltonian; if the dominant local noise acted through operators orthogonal to $x_A$ and $x_B$, error correction (including local error correction) could in principle suppress decoherence without affecting the interaction, e.g. $p^2$ errors as in Ref.~\cite{anastopoulos2013master}, see also discussion section and App.~\ref{app:Asymptotic entangled state in the AH model}
.

To circumvent this limitation, we instead exploit the full functional form of both the gravitational interaction and the decoherence mechanisms beyond their leading-order contributions. Rather than attempting to preserve a first-order interaction, we design the encoding such that an effective two-qubit dynamics is retained, and logical $\sigma^x$-like interaction and errors arise from the higher-order contributions. Crucially, the interaction and the noise generally scale differently at these higher orders, which can lead to an effective enhancement of the signal-to-noise ratio in the logical evolution.

We encode each mass into a two-dimensional logical subspace spanned by Fock 
states of the motional degree of freedom,
\begin{equation}
\mathcal{C}_\alpha= \operatorname{span}\{|\bar 0\rangle_\alpha=|0\rangle_\alpha,
|\bar 1\rangle_\alpha=|2N+1\rangle_\alpha\},
\end{equation}
with $\alpha \in\{A,B\}$, where the non-negative integer $N$ plays a role analogous to that of the code distance. This code is inspired by the Gottesman–Kitaev–Preskill (GKP) code and binomial quantum code \cite{GKPPhysRevA.64.012310, PhysRevX.6.031006} but employs simple superpositions in the Fock basis. An ideal GKP encoding would eliminate the logical interaction to all orders, contrary to our objective.

The key algebraic properties of this encoding, derived in detail in App.~\ref{app:Local imperfect QEC code}
, are as follows. For $i=1,\ldots,2N$, the off-diagonal matrix elements vanish, $\langle\bar{0}|x^i|\bar{1}\rangle=0$, while the diagonal elements satisfy $\langle\bar{0}|x^i|\bar{0}\rangle=\langle\bar{1}|x^i|\bar{1}\rangle=0$ for odd $i$ and are nonzero for even $i$. For odd $i>2N$, the matrix elements $\langle\bar{0}|x^i|\bar{1}\rangle$ become nonzero and generate a finite nonlocal coupling through the Newtonian interaction,
$$
{}_A\langle\bar{0}|{}_B\langle\bar{0}|
\frac{Gm_A m_B}{|d+x_A-x_B|}
|\bar{1}\rangle_A|\bar{1}\rangle_B\neq0,
$$
which, upon projection onto the code space, yields an effective
$\sigma_A^x\otimes\sigma_B^x$ interaction.

These properties also determine the error-correction behavior. Since $\langle\bar{0}|x^i|\bar{1}\rangle=0$ for $i\leq2N$, the error states
$x^i|\bar{0}\rangle$ and $x^j|\bar{1}\rangle$ with $i,j\leq N$ are
orthogonal and can be unambiguously distinguished; hence all logical-$\sigma_x$ errors from monomial
errors $x^k$ with $k\leq N$ are correctable, while logical-$\sigma^z$ errors remain, defining the effective $\sigma_x$-direction code
distance. The unequal diagonal elements
$\langle\bar{0}|x^i|\bar{0}\rangle\neq\langle\bar{1}|x^i|\bar{1}\rangle$
for even $i$ induce logical $\sigma^z$ errors. Such $\sigma^z$ errors can be suppressed with a party-local concatenated construction while preserving the relevant $\sigma_A^x\otimes\sigma_B^x$ sector, as shown in App.~\ref{app:Alternative QEC toolkit}.
.

The corresponding recovery channel is constructed by introducing auxiliary 
states $|\bar 0_i\rangle=|i\rangle$ and 
$|\bar 1_i\rangle=c_i|2N+1-i\rangle+d_i|2N+1+i\rangle$ for $i=1,\ldots,N$, with 
coefficients $c_i,d_i$ determined in 
App.~\ref{app:Effective error rates}. Defining the Kraus operators
$$
R^0=\sum_{k=0,1}|\bar k\rangle\langle \bar k|,\qquad
R^i=\sum_{k=0,1}|\bar k\rangle\langle \bar k_i| \;\;(1\le i\le N),
$$
and $R^i=|\bar 1\rangle\langle \bar 1_i|$ for $i>N$, where $|\bar 1_i\rangle$ are complementary bases for the rest of the whole Fock space, to ensure this channel is trace-preserving, see also App.~\ref{app:Effective error rates}
. The local recovery channel 
for each mass reads
\begin{equation}
R_\alpha(\rho)=\sum_{i=0}^{\infty} R_\alpha^i \,\rho\,R_\alpha^{i \dagger} ,
\end{equation}
with identical constructions for each subsystem $\alpha \in\{A,B\}$. 

With this encoding and recovery map, the effective Lindbladian in the logical code space is of the following form
\begin{align}
\label{eq:eff lind full}
\mathcal{L}_{\mathrm{eff}}(\rho)
=&-\frac{i}{\hbar}[H_{\mathrm{eff}},\rho] + \sum_{\alpha=A,B}\sum_{\mu\in\{x,z,+,-\}} \Gamma_{\mathrm{eff}}^{\mu}\mathcal{D}[\sigma_\alpha^\mu](\rho) \nonumber\\
&+\sum_{\mu,\nu\in\{x,z\}}\Gamma_{\mathrm{eff}}^{\mu\nu} \left( \mathcal{D}[\sigma_A^\mu-\sigma_B^\nu]- \mathcal{D}[\sigma_A^\mu]-\mathcal{D}[\sigma_B^\nu] \right)(\rho),\nonumber\\
H_{\mathrm{eff}}=&H_{\mathrm{loc}}+ \hbar\sum_{\mu,\nu\in\{x,z\}}\lambda_{\mathrm{eff}}^{\mu\nu} \sigma_A^\mu\sigma_B^\nu.
\end{align}
Here $\mathcal{D}[L](\rho)=L\rho L^\dagger- \{L^\dagger L,\rho\}/2$, and $\sigma_\alpha^\mu$ are logical operators, with $\sigma^+=|\bar1\rangle\langle\bar0|$ and $\sigma^-=(\sigma^+)^\dagger$, and $H_\mathrm{loc}$ accounts for the local Hamiltonians, which does not affect separability classification. The symmetry of our encoding gives $\lambda_{\mathrm{eff}}^{zx}=-\lambda_{\mathrm{eff}}^{xz}$ and   $\Gamma_{\mathrm{eff}}^{zx}=-\Gamma_{\mathrm{eff}}^{xz}$. These effective interaction and error rates with respect to each specific model and recovery map are calculated numerically, given in App.~\ref{app:Logical error rates} and App.~\ref{app:Effective error rates}. For the GRW model considered here, all $\Gamma^{\mu\nu}_\mathrm{eff} = 0$.

Then let us define \begin{align}
\label{eq:eta main}
\eta=\left\|
\begin{pmatrix}
\frac{\Gamma_{\mathrm{eff}}^{xx}-i\lambda_{\mathrm{eff}}^{xx}}{\bar{\Gamma}_{\mathrm{eff}}^x} &-\frac{\Gamma_{\mathrm{eff}}^{xz}-i\lambda_{\mathrm{eff}}^{xz}}     {\sqrt{\bar{\Gamma}_{\mathrm{eff}}^x\Gamma_{\mathrm{eff}}^z}}\\
\frac{\Gamma_{\mathrm{eff}}^{xz}-i\lambda_{\mathrm{eff}}^{xz}}{\sqrt{\bar\Gamma_{\mathrm{eff}}^x\Gamma_{\mathrm{eff}}^z}}
&\frac{\Gamma_{\mathrm{eff}}^{zz}-i\lambda_{\mathrm{eff}}^{zz}}
       {\Gamma_{\mathrm{eff}}^z}
\end{pmatrix}
\right\|_2,\\
\text{where } \bar{\Gamma}_{\mathrm{eff}}^x=\Gamma_{\mathrm{eff}}^x+ \frac{\Gamma_{\mathrm{eff}}^+\Gamma_{\mathrm{eff}}^-}{\Gamma_{\mathrm{eff}}^++\Gamma_{\mathrm{eff}}^-}.\nonumber
\end{align}
and $\|\cdot\|_2$ is the largest singular value.

\begin{theorem}[Separability of the effective dynamics]
\label{thm:classification}
For the effective Lindbladian $\mathcal{L}_{\mathrm{eff}}$ defined in 
Eq.~\eqref{eq:eff lind full}:
\begin{itemize}
\item If $\eta> 1$, the induced dynamical 
map is \emph{non-separable}.
\item If $\eta \le 1$, the induced dynamical map is 
\emph{separable}.
\end{itemize}
\end{theorem}

The proof is given in App.~\ref{app:Locality of the effective Lindbladian}. For the GRW model considered here, note that all $\Gamma^{\mu\nu}_\mathrm{eff} = 0$, and $\eta^x\equiv |\Gamma_{\mathrm{eff}}^{xx}-i\lambda_{\mathrm{eff}}^{xx}|/\bar{\Gamma}^x_\mathrm{eff } = |\lambda_{\mathrm{eff}}^{xx}|/\bar{\Gamma}^x_\mathrm{eff } \le \eta $, such that $\eta^x>1$ already verifies non-separability. Moreover, in this case, a sufficient condition for entanglement is:
\begin{equation}
\label{eq:entagling cond main}
\sqrt{(\lambda_{\mathrm{eff}}^{xx})^2+(\Gamma_{\mathrm{eff}}^{xx})^2} > \min(\Gamma_{\mathrm{eff}}^x+\Gamma_{\mathrm{eff}}^+, \Gamma_{\mathrm{eff}}^x+\Gamma_{\mathrm{eff}}^-),
\end{equation}
see Prop.~\ref{prop:entangling}.

\textit{Recoverability of entanglement under collapse noise.}---
We now analyze the scaling of the effective $\sigma^x$-sector ratio $|\lambda_{\mathrm{eff}}^{xx}|/\bar{\Gamma}^x_\mathrm{eff }$. For the QEC construction considered here, recoverability is quantified by the extent to which error correction enhances the entanglement generated by the effective logical dynamics. In particular, the separability parameter $\eta$ provides a convenient diagnostic of this enhancement. 

For a fixed code distance $N$,
the leading non-vanishing $\sigma^x_A\sigma^x_B$ component contribution to the gravitational coupling  arises at
order $4N+2$ in the Taylor expansion of the Newtonian potential,
\begin{align*}
    &-\frac{G m_A m_B}{|d+x_A-x_B|}\\
=&\cdots+
\frac{G m_A m_B (4N+2)!}{d^{4N+3}(2N+1)!(2N+1)!}\,
x_A^{2N+1}x_B^{2N+1}+\cdots .
\end{align*}
By contrast, the leading contribution of the GRW dynamics that induces logical
$\sigma^x$ errors appears at order $2N+2$,
\[
\mathcal{L}_{\mathrm{GRW}}(\rho)
=\cdots+
\frac{\lambda_{\mathrm{GRW}}(2N+2)!}{((N+1)!)^3 4^{N+1}r_C^{2N+2}}\,
x^{N+1}\rho x^{N+1}
+\cdots .
\]
Using Stirling's formula, this yields a rough estimate of the scaling
\begin{equation}
\label{eq:scaling}
\eta^x \propto
\frac{
\frac{G m_A m_B (4N+2)! x_0^{4N+2}}{d^{4N+3}(2N+1)!}
}{
\frac{\lambda_{\mathrm{GRW}}(2N+2)! x_0^{2N+2}(2N+1)!}{[(N+1)!]^3 4^{N+1}r_C^{2N+2} N!}
}
\propto N^2\left(\frac{16 N^2}{e^2}\frac{r_C^2 x_0^2}{d^4}\right)^N,
\end{equation}
where $x_0=\sqrt{\hbar/2m\omega}$ and we have used $\langle 0|x^{2N+1}|2N+1\rangle = x_0^{2N+1}\sqrt{(2N+1)!}$, and the largest error rate $|\langle N|x^{N+1}|2N+1\rangle|^2=x_0^{2N+2} (2N+1)!/N!$. The ratio $|\lambda_{\mathrm{eff}}^{xx}|/\bar{\Gamma}^x_\mathrm{eff }$ therefore increases super-exponentially with $N$ and can be made arbitrarily large for any choice of $\lambda_{\mathrm{GRW}}$ and $r_C$, so long as $16 r_C^2 x_0^2 N^2>e^2 d^4$. Physically, this reflects the fact that GRW localization saturates at large length scales, whereas the Newtonian interaction decreases only polynomially. Consequently, for sufficiently large code distance the condition
$\eta > 1$ is always satisfied, and the effective
logical dynamics becomes entangling. Moreover, the unbounded growth of
$\eta^x$ shows that local QEC can progressively
enhance the accessibility of this entanglement, providing the basis for its
recoverability even when the bare entanglement is arbitrarily suppressed. In Theorem~\ref{thm:recoverability}
in the App.~\ref{app:Rigorous proof of the entanglement recoverability}, for GRW/CSL we further show that this asymptotic advantage can be retained when $x_0$ and $d$ are scaled with $N$ to keep the contact probability between the particles below a prescribed threshold, thereby establishing a controlled form of unbounded local recoverability. At the same time, the DP model exhibits an upper bound $\eta\le\sqrt{2}$ for any encoding in a bipartite setting.

\begin{figure}
    \centering
    \includegraphics[width=\linewidth]{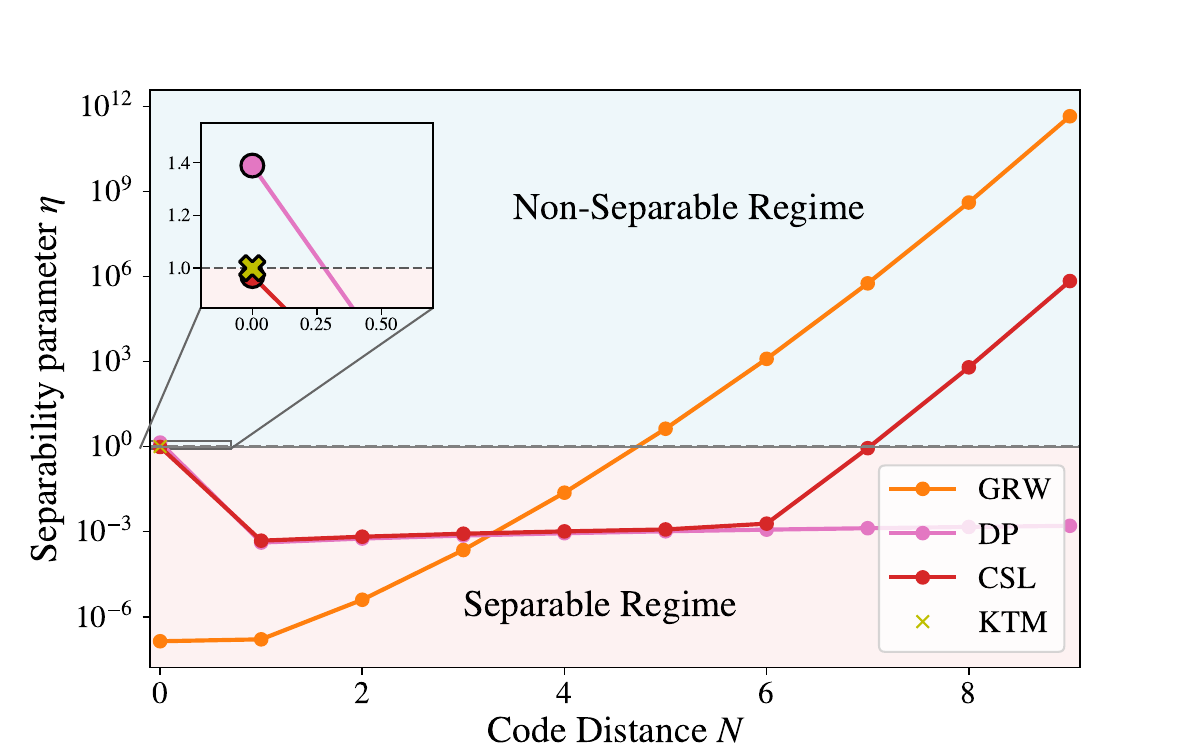}
    \caption{Separability parameter $\eta$, shown as a function of the code distance $N$ for Newtonian gravity combined with GRW, CSL, DP, and KTM noise models. Increasing the code distance progressively suppresses the logical noise relative to the logical interaction for the GRW and CSL models, demonstrating that their underlying entangling resource remains locally recoverable. In contrast, the DP model does not exhibit a comparable enhancement under the bipartite setting here, despite remaining non-separable at $N=0$. For reference, we depict at $N=0$ the KTM model, which is separable by construction. For all models considered, $\eta>1$ implies non-separability, and $\eta\leq1$ implies separability.}
    \label{fig:classification}
\end{figure}

In Fig.~\ref{fig:classification} we plot the separability parameter $\eta$ for the effective dynamics
resulting from Newtonian gravitational interaction in the presence of GRW/CSL/KTM
and Di\'osi--Penrose (DP) noise. It is worth noting that for the DP and CSL models the noise is intrinsically
nonlocal and, consequently, $\Gamma^{\mu\nu}_\mathrm{eff}\neq 0$ should also be considered. For more details, see App.~\ref{app:Locality of the effective Lindbladian}. For each case, we apply the local QEC protocol
defined in Eq.~\eqref{eq:qec channel}, construct the corresponding effective
Lindbladian of the form in Eq.~\eqref{eq:eff lind full}, and extract numerically the effective
interaction rate $\lambda_{\mathrm{eff}}^{\mu\nu}$ and the effective decoherence rates
$\Gamma_{\mathrm{eff}}^\mu$ and $\Gamma_{\mathrm{eff}}^{\mu\nu}$. The separability or non-separability of the resulting
effective dynamics is then determined using the separability parameter $\eta$  and Theorem~\ref{thm:classification}. 

For the numerical analysis in Fig.~\ref{fig:classification} we choose the following parameters:
$m_A=m_B=10^{-21}\,\mathrm{kg}$ and
$x_0=\sqrt{\hbar/2m\omega}=2\times 10^{-8}\,\mathrm{m}$. Assuming the masses are composed
of silica with density $\rho\simeq2.2\times10^3\,\mathrm{kg/m^3}$, 
this corresponds to a physical radius
$r_0=(3m_A/4\pi\rho)^{1/3}\simeq4.8\times10^{-9}\,\mathrm{m}$. We further take the
trap separation to be $d=4\times10^{-7}\,\mathrm{m}$. For the GRW/CSL models we
consider a large localization length
$r_{\mathrm C}=2\times10^{-6}\,\mathrm{m}$, five times the inter-mass
separation. For the DP model we analogously
consider a large smearing parameter $\sigma=2\times 10^{-6}\,\mathrm{m}$. Larger localization lengths or smearing parameters correspond
physically to the survival of spatial coherences below these length scales, and
thus to a reduced ability to suppress entanglement between nearby particles. 

For each set of parameters and each code distance $N$, the separability parameter $\eta$ is computed numerically
from the full error-corrected generator, rather than inferred solely from
leading-order scaling arguments; details of the numerical procedure are given in
App.~\ref{app:Calculation of the effective Lindbladian}. The non-separable data points also satisfy the entangling criterion in Prop.~\ref{prop:entangling}
, hence the effective dynamics is not only non-separable but also generates entanglement. 

Notably, for the GRW/CSL and DP models the onset of non-separability does not
require macroscopic masses. This contrasts with the case without QEC, where observing entanglement under increasingly strong collapse noise requires correspondingly larger particle sizes, see App.~\ref{app:Gravity-induced entanglement with spatially-superposed states}. Our results highlight an important distinction between non-separability and
the practical observability of entanglement. Collapse noise can completely
suppress entanglement generation for commonly considered spatial-superposition
states without rendering the underlying dynamics separable. For suitably
chosen code states, entanglement remains nonzero for arbitrary collapse
strength, but in the absence of error correction both its magnitude and
lifetime can become arbitrarily small, making it effectively inaccessible.

The role of local QEC is therefore not to make an otherwise separable dynamics non-separable, but to render this latent entangling resource observable. For GRW and CSL noise, error correction suppresses decoherence relative to the gravitational interaction, allowing the generated entanglement to reach larger values and persist for longer times within the code subspace. In this sense, QEC makes otherwise inaccessible gravitational entanglement locally recoverable. By contrast, the DP model shows no comparable improvement under the bipartite QEC protocol. Nevertheless, the DP model remains non-separable already at zeroth order, in accordance with Ref.~\cite{PhysRevD.111.L121101}.

\textit{Discussion and Conclusions.}---
Our results show that collapse-induced decoherence need not render
Newtonian-plus-noise dynamics separable even when observable entanglement is
strongly suppressed.  Local QEC reveals a further distinction: for GRW and
CSL, the effective entangling scale is suppressed more slowly than the
relevant logical noise, rendering the entangling resource locally recoverable.
By contrast, the DP model shows no comparable enhancement under bipartite QEC protocol because its interaction and dominant noise arise from the same
kernel and scale in parallel. Recoverability therefore refines the
separable/non-separable classification of noisy gravitational dynamics.
Related QEC ideas have also been explored in quantum
metrology~\cite{PhysRevLett.112.080801,PhysRevLett.112.150802,PhysRevLett.112.150801,zhou2018achieving}.

The specific structure of the decoherence channel plays a central role in determining recoverability. In the collapse models studied in the main text, both the interaction and the dominant noise act through position operators, making local discrimination between signal and noise particularly challenging. In contrast, other physically relevant scenarios, including the Anastopoulos--Hu model and several forms of environmental decoherence, involve momentum-type noise generated by operators distinct from those responsible for the gravitational interaction. As shown in App.~\ref{app:Asymptotic entangled state in the AH model}, such noise can be suppressed by local QEC without simultaneously reducing the coherent gravitational coupling. This demonstrates that the limitations encountered for position-type noise are not generic, but rather arise from the particular operator structure shared by the interaction and decoherence. Nevertheless, position-type noise is a generic feature of universal collapse models~\cite{PhysRevLett.130.230202}. It is also noteworthy that, above, we have exploited local imperfect error-correction; if one allows non-local imperfect error-correction, the interaction-to-noise ratio can be made arbitrarily large even at leading order, see App.~\ref{app:Beyond local QEC code}.

The present work is not intended as a near-term experimental proposal. Increasing the code distance generally suppresses not only the effective noise but also the effective interaction strength, and the resulting recovery procedures may therefore be impractical for realistic implementations even when the required motional states can be prepared. Nevertheless, our results establish a more fundamental point: collapse-induced decoherence does not necessarily destroy the underlying entangling resource. For physically relevant models such as GRW and CSL, this resource remains locally recoverable in principle. More broadly, our results suggest that recoverability may provide a useful organizing principle for noisy gravitational dynamics, complementing the traditional separable/non-separable classification and motivating the search for more efficient local protocols capable of exploiting this recoverability in future experiments. A possible route would be to relax the requirement that the QEC protocol recover the ideal, error-free dynamics, and instead consider error-filtering protocols (possibly probabilistic) designed to produce noisy yet entangling dynamics within the code space.

\textit{Acknowledgments.} --- We thank Kirill Streltsov for useful discussions. OpenAI's GPT-5.6 Sol was used to assist in developing and checking the proof of Theorem~\ref{thm:recoverability} in Appendix~\ref{app:Rigorous proof of the entanglement recoverability}. The resulting proof was formulated and independently verified by the authors, who take the full responsibility for its correctness. This work was supported by the ERC Synergy Grant HyperQ (Grant No. 856432), the BMBF via project PhoQuant (Grant No. 13N16110) and the EU project SPINUS (Grant No. 101135699).


\bibliography{ref}

\clearpage
\appendix
\onecolumngrid
\begin{center}
{\large\bfseries Supplemental Material for}\\[0.5em]
{\large\bfseries Recoverability of Gravitational Entanglement under Decoherence}
\end{center}

\makeatletter
\let\l@subsubsection\@gobbletwo
\makeatother
\tableofcontents
\addtocounter{theorem}{-1}

\section{Overview of the decoherence models}\label{app:Overview of the decoherence models}

This appendix summarizes the collapse and decoherence models considered throughout the paper and collects the corresponding master equations for reference. Since the analysis presented in the main text and subsequent appendices applies the same quantum error-correction framework to several different physical mechanisms, we provide here a concise overview of their defining features and effective dynamics.

\subsection{Fundamental decoherence models}\label{app:Collapse models}

We begin by reviewing the fundamental collapse and decoherence models that will be analyzed throughout the paper. For each model, we summarize the underlying physical motivation and present the corresponding master equation used in our calculations.

\subsubsection{Ghirardi–Rimini–Weber (GRW) Model}

The GRW model provides the simplest phenomenological realization of spontaneous wavefunction collapse through random localization events acting independently on each constituent particle~\cite{GRWPhysRevD.34.470}. In this model, the decoherence rate is negligible for small spatial superpositions and approaches a constant for sufficiently large superposition sizes. A key feature of the model is the amplification mechanism, whereby the effective decoherence rate increases with the mass (or number of constituents) of the system. As a consequence, standard Schr\"odinger dynamics is recovered for microscopic systems and small superpositions, while macroscopic superpositions at large masses and separations are strongly suppressed. At the level of pure-state dynamics, the GRW modification introduces stochastic and nonlinear corrections to the Schr\"odinger equation in the form of random, instantaneous localization events that interrupt the otherwise unitary evolution; however, the resulting ensemble-averaged evolution of the density matrix remains linear and can be expressed in Lindblad form. For a single particle, the corresponding one-dimensional GRW decoherence is given by the following Lindbladian, which we simplify in the position representation as follows:

\begin{align}
    \mathcal{L}_{\mathrm{GRW}}(\rho) =&-\lambda_{\mathrm{GRW}} \left( \rho- \int dq \left(\frac{1}{\pi r_\mathrm{C}^2}\right)^{\frac{1}{2}} e^{-\frac{(x-q)^2}{2r_\mathrm{C}^2}}\rho e^{-\frac{(x-q)^2}{2r_\mathrm{C}^2}}\right)\nonumber\\
    =&-\lambda_{\mathrm{GRW}} \left( \rho-\int dq dx_1 dx_2 \left(\frac{1}{\pi r_\mathrm{C}^2}\right)^{\frac{1}{2}} e^{-\frac{(x_1-q)^2}{2r_\mathrm{C}^2}}|x_1\rangle \langle x_1|\rho |x_2\rangle \langle x_2| e^{-\frac{(x_2-q)^2}{2r_\mathrm{C}^2}}\right)\nonumber\\
    =&-\lambda_{\mathrm{GRW}} \left( \rho - \int dx_1 dx_2  e^{-\frac{(x_1-x_2)^2}{4r_\mathrm{C}^2}}|x_1\rangle \langle x_1|\rho |x_2\rangle \langle x_2| \right).
\end{align}
The model depends on two phenomenological parameters, $\lambda_{\mathrm{GRW}}$ and $r_{\mathrm{C}}$, referred to as the collapse rate and the localization length, respectively. In the position representation, the GRW dynamics induces an exponential suppression of the off-diagonal elements of the density matrix $\Tr\left(\rho\,\ketbra{x_1}{x_2}\right)$ at a rate $\lambda_{\mathrm{GRW}}\left[1-\exp \left(-(x_1-x_2)^2/4r_{\mathrm{C}}^2\right)\right]$ that increases with spatial separation.

For a body composed of many constituent particles, we focus on the center-of-mass (COM) position $\hat{\boldsymbol{X}}_{\mathrm{CM}}$. The position of each constituent particle can be written as $\hat{\boldsymbol{x}}_i=\hat{\boldsymbol{X}}_{\mathrm{CM}}+\hat{\boldsymbol{r}}_i$ in terms of the relative coordinates $\hat{\boldsymbol{r}}_i$.  We assume a rigid-body regime in which the internal (relative) degrees of freedom remain sharply peaked around their equilibrium values (equivalently, their characteristic frequencies are much larger than the GRW-induced rates), such that $\hat{\boldsymbol{r}}_i \approx \boldsymbol{r}_i$ can be treated as fixed parameters. In the GRW model, localization events act independently on each constituent particle, with each degree of freedom undergoing stochastic collapse events at the single-particle rate. In the rigid-body limit, the action of a collapse on any individual constituent induces an effective collapse on the COM coordinate $\hat{\boldsymbol{X}}_{\mathrm{CM}}$, and the $n$ independent collapse generators contribute additively, giving rise to a collapse rate that scales linearly with the number of constituent particles. In the more general three-dimensional case, this leads to the generator

\begin{equation}
    \mathcal{L}_{\mathrm{GRW}}(\rho)= n \times \lambda_{\mathrm{GRW}} \left( \int d^3 \boldsymbol{x}_1 d^3 \boldsymbol{x}_2  e^{-\frac{|\boldsymbol{x}_1-\boldsymbol{x}_2|^2}{4r_\mathrm{C}^2}}|\boldsymbol{x}_1\rangle \langle \boldsymbol{x}_1|\rho |\boldsymbol{x}_2\rangle \langle \boldsymbol{x}_2| -\rho\right).
\end{equation}
Generally, we take $n$ to be the number of nucleons.

\subsubsection{Continuous Spontaneous Localization (CSL) Model}

The Continuous Spontaneous Localization (CSL) model provides a continuous-in-time realization of the GRW collapse mechanism and constitutes a widely used phenomenological framework for spontaneous wavefunction collapse~\cite{CSLPhysRevA.42.78,CSLPhysRevA.39.2277}. As in the GRW model, CSL introduces stochastic and non-linear modifications of the Schr\"odinger equation at the level of pure states, while the corresponding ensemble-averaged evolution of the density matrix remains linear and completely positive. The key distinction with respect to GRW is that, in CSL, localization occurs continuously in time rather than through discrete and instantaneous collapse events.

This continuous dynamics is commonly modeled by coupling the system to a postulated classical stochastic field, which acts as a fundamental source of noise and drives the collapse process even in the absence of environmental interactions. In contrast to the GRW model, where localization events act independently on individual particles, in CSL all constituents couple to the same stochastic field. As a result, the collapse noise can be spatially correlated over length scales set by the characteristic localization length $r_{\mathrm{C}}$. This feature will become relevant in App.~\ref{app:Technical details of the quantum error-correction protocol}.

The CSL dynamics is most naturally formulated in terms of the mass density operator,
\begin{equation}
M(\boldsymbol{x}) = \sum_{j,s} m_j\, a_j^\dagger(\boldsymbol{x},s)\, a_j(\boldsymbol{x},s),
\end{equation}
where $a_j^\dagger(\boldsymbol{x},s)$ creates a particle of mass $m_j$ and spin $s$ at position $\boldsymbol{x}$. Upon averaging over the stochastic field, the resulting collapse dynamics is described by the Lindblad generator
\begin{equation}
\mathcal{L}_{\mathrm{CSL}}(\rho)
=
-\frac{\lambda_{\mathrm{CSL}}}{2 m_0^2}
\int d^3\boldsymbol{x}\, d^3\boldsymbol{y}\,
g(\boldsymbol{x}-\boldsymbol{y})
\bigl[M(\boldsymbol{x}),[M(\boldsymbol{y}),\rho]\bigr],
\end{equation}
where $\lambda_{\mathrm{CSL}}$ denotes the collapse rate, $m_0$ is a reference mass conventionally chosen as the nucleon mass, and
\begin{equation}
g(\boldsymbol{x}-\boldsymbol{y})
=
\exp\!\left(-\frac{|\boldsymbol{x}-\boldsymbol{y}|^2}{4 r_{\mathrm{C}}^2}\right)
\end{equation}
is a spatial correlation function with characteristic localization length $r_{\mathrm{C}}$.

As in the GRW model, it is often useful to describe the collapse dynamics in terms of the center-of-mass (COM) degree of freedom of a rigid object. Writing the position of each constituent particle as
\begin{equation}
\hat{\boldsymbol{x}}_i=\hat{\boldsymbol{X}}_{\mathrm{CM}}+\hat{\boldsymbol{r}}_i,
\end{equation}
we assume that the internal coordinates remain sharply localized around their equilibrium values, such that $\hat{\boldsymbol r}_i\approx \boldsymbol r_i$ can be treated as fixed parameters. Under this rigid-body approximation, the mass density operator depends on the COM coordinate through
\begin{equation}
M(\boldsymbol{x})=\sum_i m_i\,\delta(\boldsymbol{x}-\hat{\boldsymbol X}_{\mathrm{CM}}-\boldsymbol r_i),
\end{equation}
or, equivalently, $M(\boldsymbol{x})=\mu(\boldsymbol{x}-\hat{\boldsymbol X}_{\mathrm{CM}})$ for a continuous mass distribution $\mu$.

Unlike the GRW model, where localization events act independently on each constituent particle, all constituents in CSL couple to the same stochastic field. As a consequence, the collapse noise can be spatially correlated over distances of order $r_C$, leading to a qualitatively different amplification mechanism. For a rigid body, the COM decoherence rate associated with a spatial delocalization $\Delta\boldsymbol X$ can be written as
\begin{equation}
\label{eq:CSL decoherence rate}
\Gamma_{\mathrm{CSL}}(\Delta\boldsymbol X) 
=
\frac{\lambda_{\mathrm{CSL}}}{m_0^2} (4\pi r_C^2)^{3/2}
\int \frac{d^3\boldsymbol{k}}{(2\pi)^3}\,
e^{-r_C^2 k^2}
\,|\tilde{\mu}(\boldsymbol{k})|^2
\left[1-\cos(\boldsymbol{k}\cdot\Delta\boldsymbol X)\right],
\end{equation}
where $\tilde{\mu}(\boldsymbol{k})$ denotes the Fourier transform of the mass density distribution. This expression shows that the amplification mechanism depends not only on the total mass but also on the geometry of the object through the form factor $|\tilde{\mu}(\boldsymbol{k})|^2$. In the limit where the size of the object is much smaller than $r_C$, one has $\tilde{\mu}(\boldsymbol{k})\simeq M$ over the relevant range of wavevectors, leading to the coherent scaling
\begin{equation}
\Gamma_{\mathrm{CSL}}\propto
\lambda_{\mathrm{CSL}}
\frac{M^2}{m_0^2},
\end{equation}
which is quadratic in the total mass. For larger objects, the amplification is reduced because different regions of the body no longer experience the same realization of the collapse noise.

The same spatial correlations imply that two distinct rigid bodies can experience correlated noise whenever their separation is comparable to or smaller than $r_C$. As discussed in App.~\ref{app:Effective error rates}, these correlated noise terms are important when analyzing the error-corrected dynamics.

\subsubsection{Di\'osi--Penrose (DP) Model}
The DP model attributes decoherence to fluctuations associated with the gravitational field itself, leading simultaneously to gravitational interactions and gravity-induced decoherence~\cite{penrose1996gravity,diosi1987universal}. In this framework, quantum matter probes the gravitational field, and the uncertainty principle motivates an intrinsic uncertainty in the spacetime metric experienced by a delocalized mass distribution. This can be modeled as a stochastic perturbation of the metric, which in turn induces effective decoherence of the matter degrees of freedom.

As a representative setting, consider two masses $m_A$ and $m_B$ trapped in harmonic potentials. The Di\'osi--Penrose model of gravitational interaction and decoherence is described by the following generator:
\begin{align}
\label{eq:dp}
    \mathcal{L}_{\mathrm{DP}}(\rho)=&\frac{iG}{\hbar}\Big[\int \dd^3 \boldsymbol{x} \dd^3 \boldsymbol{y}\frac{M(\boldsymbol{x})M(\boldsymbol{y})}{2|\boldsymbol{x}-\boldsymbol{y}|},\rho\Big]+ \frac{G}{\hbar} \int \frac{\dd^3 \boldsymbol{x} \dd^3 \boldsymbol{y}}{|\boldsymbol{x}-\boldsymbol{y}|}\left(M(\boldsymbol{x})\rho M(\boldsymbol{y}) - \frac{1}{2}\{M(\boldsymbol{x}) M(\boldsymbol{y}),\rho\}\right).
\end{align}
Here $M(\boldsymbol{x})=\sum_{\alpha=A,B} m_\alpha \mu_{\sigma}(\boldsymbol{x}-\boldsymbol{x}_\alpha)$ is the total mass density operator, where $\boldsymbol{x}_\alpha$ is the center-of-mass position operator for $m_\alpha$. The mass density distribution is assumed to be Gaussian,
\begin{equation}
    \mu_{\sigma}(\boldsymbol{x}-{\boldsymbol{x}_\alpha})=\frac{1}{(2\pi \sigma^2)^{3/2}}\exp\left(-\frac{|\boldsymbol{x}-{\boldsymbol{x}_\alpha}|^2}{2\sigma^2}\right),
\end{equation}
where $\sigma$ is a smearing radius for the mass density function.

Note that the DP model can also be obtained as an effective dynamics describing interactions between quantum systems mediated by a classical gravitational field~\cite{diosi2011gravity}.

As in the CSL model, the DP generator is nonlocal in space, as it couples mass-density operators evaluated at different positions through a two-point kernel. In fact, the mathematical structure of the DP model is closely related to that of CSL. For example, after introducing the smeared mass density, the DP decoherence rate associated with a rigid body can be expressed in Fourier space as
\begin{equation}
\Gamma_{\mathrm{DP}}(\Delta\mathbf X)
\propto
G
\int \frac{d^3\mathbf k}{(2\pi)^3}
\frac{e^{-\sigma^2 k^2}}{k^2}
|\tilde{\mu}(\mathbf k)|^2
\left[1-\cos(\mathbf k\cdot\Delta\mathbf X)\right],
\end{equation}
where $\tilde{\mu}(\mathbf k)$ denotes the Fourier transform of the mass distribution. This should be compared with the corresponding CSL expression in Eq.~\eqref{eq:CSL decoherence rate}. In both cases, the decoherence rate is governed by the form factor $|\tilde{\mu}(\mathbf k)|^2$, implying a coherent amplification mechanism. For objects whose size is small compared to the relevant correlation length ($r_C$ in CSL and $\sigma$ in DP), one has $\tilde{\mu}(\mathbf k)\simeq M$ over the dominant range of wavevectors, leading to an effective decoherence rate that scales approximately as $M^2$. For larger objects, the factor $e^{-\sigma^2 k^2}$ no longer allows one to approximate $\tilde{\mu}(\mathbf{k})$ by the total mass $M$ over the relevant range of wavevectors. The resulting decoherence rate is therefore reduced by the form factor $|\tilde{\mu}(\mathbf{k})|^2$ and depends explicitly on the geometry of the object.

Despite these similarities, there are important differences between the DP and the CSL model. In CSL, the Gaussian kernel itself determines the spatial correlations of the collapse noise, which become exponentially suppressed beyond the length scale $r_C$. In contrast, the fundamental DP kernel remains the Newtonian kernel $1/|\mathbf x-\mathbf y|$, and is therefore intrinsically long-ranged. The Gaussian factor $e^{-\sigma^2 k^2}$ does not replace this kernel; rather, it arises from the smearing of the mass density and acts as a short-distance regularization. Equivalently, the Newtonian interaction is modified from $1/r$ to $\frac{\operatorname{erf}(r/2\sigma)}{r}$, which remains finite at short distances while recovering the standard Newtonian behavior for $ r\gg \sigma $.

The introduction of the smearing length $\sigma$ is motivated by the fact that the unsmeared DP model leads to divergent gravitational self-energies. Historically, $\sigma$ was taken to be of the order of the nucleon radius, reflecting the idea that gravity should couple to a coarse-grained rather than perfectly localized mass density. More recent treatments often regard $\sigma$ as an independent phenomenological parameter. From this perspective, $\sigma$ may be interpreted as the spatial resolution with which the gravitational field probes the mass distribution, and need not coincide with either the size of the gravitating object or that of its microscopic constituents. Consequently, in general, one does not require the separation between two rigid bodies to exceed $\sigma$; rather, $\sigma$ determines the spatial scales below which the DP dynamics ceases to resolve details of the mass distribution.

A final distinction from CSL, which will play a central role in the QEC analysis of App.~\ref{app:Effective error rates}, is that the same Newtonian kernel generates both the decoherence and the coherent gravitational interaction. As a consequence, the interaction and the dominant error channels remain intrinsically linked, in contrast to CSL where the collapse noise and gravitational interaction originate from distinct kernels.

\subsubsection{Kafri--Taylor--Milburn (KTM) Model}

The KTM model provides a representative example of a classical mediator theory in which Newtonian interactions arise from local measurements and classical feedback and are therefore separable by construction~\cite{Kafri_2014}. In this framework, the overall dynamics is LOCC (and hence separable) by construction.

To state the general structure, consider a bipartite system with subsystems $A$ and $B$. Suppose a classical field measures subsystem $A$ with operator $M_A$ and feeds back the measurement record onto subsystem $B$ through operator $F_B$. The resulting LOCC Lindbladian can be written as~\cite{streltsov2022significance}
\begin{align}
    \frac{\dd \rho}{\dd t}=& -\frac{i}{2} [ M^\dagger_A F_B +M_A F_B, \rho ] +\mathcal{D}[M_A -i F_B](\rho),
\end{align}
where $\mathcal{D}[A](\rho)=A\rho A^\dagger - \{A^\dagger A/2,\rho\}$. More specifically, let $O_A$ and $O_B$ be Hermitian operators acting on $\mathcal{H}_A$ and $\mathcal{H}_B$, respectively. Applying these as measurement and feedback operators in both directions, with (potentially) different rates, yields
\begin{align}
    \frac{\dd \rho}{\dd t}=&-i[a_A O_A \otimes a_B O_B,\rho]+\mathcal{D}[a_A O_A - i a_B O_B](\rho)-i[b_A O_A \otimes b_B O_B,\rho]+\mathcal{D}[b_B O_B - i b_A O_A](\rho)\nonumber\\
    =&-i(a_A a_B+b_A b_B)[O_A \otimes O_B,\rho]+(a_A^2+b_A^2)\mathcal{D}[O_A](\rho)+(a_B^2+b_B^2)\mathcal{D}[O_B](\rho)+i (a_A a_B-b_A b_B)(O_A\rho O_B -O_B\rho O_A ).
\end{align}
When the products of measurement and feedback rates are symmetric, i.e. $a_A a_B =b_A b_B$, the nonlocal error terms cancel, leaving an effective nonlocal interaction accompanied by two local decoherence channels. Moreover, the product of the local error rates is bounded from below by the square of the interaction rate. In particular, by the Cauchy--Schwarz inequality,
\begin{equation}
    \sqrt{(a_A^2+b_A^2)(a_B^2+b_B^2)}\ge (\sqrt{a_A^2 a_B^2} +\sqrt{b_A^2 b_B^2})=a_Aa_B + b_A b_B.
\end{equation}

In this way, a Newtonian interaction can be synthesized in an LOCC fashion. By setting $O_A=x_A$, $O_B=x_B$, $a_A=b_B=\sqrt{\Gamma/\hbar}$, and $a_B=b_A=Gm_Am_B/(d^3\sqrt{\Gamma\hbar})$, the KTM model reads
\begin{equation}
\label{eq:KTM}
    \mathcal{L}_{\mathrm{KTM}}(\rho) = -\frac{2iGm_A m_B}{\hbar d^3}[x_A x_B,\rho]+\left(\frac{\Gamma}{\hbar}+\frac{G^2 m_A^2 m_B^2}{\Gamma\hbar d^6}\right)(\mathcal{D}[x_A]+\mathcal{D}[x_B])(\rho),
\end{equation}
which is LOCC and hence separable by construction. The local decoherence rate is minimized when $\Gamma/\hbar=G^2 m_A^2 m_B^2/\Gamma\hbar d^6$, in which case the minimum equals the interaction rate, $2Gm_Am_B/\hbar d^3$. The unitary part reproduces the leading-order Newtonian interaction, as seen from the expansion
$$
-\frac{G m_A m_B}{|d+x_A-x_B|}=-\frac{G m_A m_B}{d}+\frac{Gm_Am_B(x_A-x_B)}{d^2}-\frac{Gm_Am_B(x_A^2+x_B^2-2x_A x_B)}{d^3}+\cdots.
$$
At the same time, the accompanying local $x_A,x_B$ errors prevent the generation of nonlocal quantum correlations, consistent with the LOCC character of the model.

\subsubsection{Anastopoulos--Hu Model}

The Anastopoulos–Hu model describes decoherence induced by fluctuations of the quantized gravitational field and predicts a qualitatively different momentum-basis decoherence mechanism~\cite{anastopoulos2013master}. In this framework, quantum matter is coupled to linearized general relativity around Minkowski spacetime (derived from the Einstein--Hilbert action in the weak-field regime). Assuming the gravitons are in a Gaussian thermal state characterized by an effective ``temperature'' $\Theta$ (with units of frequency, i.e. $\mathrm{s^{-1}}$), tracing out the graviton environment yields a master equation for the matter degrees of freedom.

For a single particle in the non-relativistic limit in one dimension, $|p|\ll m c$, the resulting gravitational decoherence takes the form
\begin{equation}
    \mathcal{L}_{\mathrm{AH}}(\rho)= -\frac{4 \pi G \Theta}{9 \hbar m^2 c^5} [p^2,[p^2,\rho]].
\end{equation}
This term acts in the momentum basis, in contrast to the position-basis decoherence mechanisms discussed above and below.

\subsection{Environmental decoherence}

In addition to the fundamental mechanisms discussed above, realistic experiments are subject to a variety of environmental decoherence processes. Many of these mechanisms lead to effective master equations of a form closely related to the GRW model, which motivates the focus on GRW-type dynamics throughout the paper. For instance, a levitated mass in an optical cavity is never in a perfect vacuum: residual gas molecules can collide with the object and impart momentum kicks. Moreover, the cavity contains thermal photons (e.g., from black-body radiation), and the levitated particle may scatter, emit, or absorb such photons~\cite{rijavec2021decoherence}.

For a particle prepared in a spatial superposition, let $\rho$ denote its density matrix. Many environmental decoherence processes lead to an exponential decay of the off-diagonal elements of the form
\begin{equation}
    \frac{\dd}{\dd t} \langle x|\rho|y\rangle=-\Gamma\left(1-\exp\left(-\frac{(x-y)^2}{4 a^2}\right)\right)\langle x|\rho|y\rangle,
\end{equation}
where the decoherence rate $\Gamma$ and localization length scale $a$ depend on the specific microscopic mechanism.

Finally, note that this functional form is closely analogous to the GRW model discussed above. For this reason, in the following discussions we focus on the GRW model; similar arguments apply to the environmental case.

\subsection{Gravity-induced entanglement with spatially-superposed states}\label{app:Gravity-induced entanglement with spatially-superposed states}

Before proceeding, let us revisit the gravity-induced entanglement tests in the presence of collapse noises. We will consider the BMV protocol~\cite{bose2017spin,marletto2017gravitationally} in the presence of collapse noise, and show that the attainable entanglement and the lifetime of entanglement are restricted by collapse noises. In the case of GRW noise, one may need to increase the mass macroscopically to observe entanglement.

Without loss of generality, we consider two masses $A$ and $B$, where $m_A=m_B=m$  each prepared in a spatial cat state encoded in the two-dimensional subspace spanned by two position eigenstates, $|x_1\rangle$ and $|x_2\rangle$. The two cat states are aligned along the same axis, and the distance between their midpoints is $d$. We define the cat-state separation as $\Delta x=|x_2-x_1|$ and choose coordinates
\begin{align}
x_{A,1}=-\frac{\Delta x}{2},\qquad x_{A,2}=\frac{\Delta x}{2},\qquad x_{B,1}=d-\frac{\Delta x}{2},\qquad x_{B,2}=d+\frac{\Delta x}{2}.
\end{align}
We assume $d>\Delta x$. The qubit basis is defined by $\sigma_A^z|x_1\rangle_A=|x_1\rangle_A, \sigma_A^z|x_2\rangle_A=-|x_2\rangle_A$,
and similarly for mass $B$.

By directly evaluating Newtonian potential energy on these bases, we can rewrite the Newtonian interaction Hamiltonian as follows:
\begin{align}
H_{\mathrm{NG}}/\hbar=E_0+E_A\sigma_A^z+E_B\sigma_B^z+\lambda_{\mathrm{NG}}\sigma_A^z\sigma_B^z,
\end{align}
where
\begin{align}
\lambda_{\mathrm{NG}}=\frac{Gm_A m_B}{4\hbar}\left(\frac{1}{d+\Delta x}+\frac{1}{d-\Delta x}-\frac{2}{d}\right)
=\frac{Gm_A m_B\Delta x^2}{2\hbar d(d^2-\Delta x^2)}.
\end{align}

The GRW collapse induces pure dephasing in the position basis. For mass $\alpha=A,B$, the decay rate of the spatial coherence between $|x_1\rangle_\alpha$ and $|x_2\rangle_\alpha$ is
\begin{align}
\Gamma_\alpha^{\rm dep}=\lambda_\mathrm{GRW} n\left[1-\exp\left(-\frac{\Delta x^2}{4r_C^2}\right)\right].
\end{align}
Dropping the irrelevant global energy and local Hamiltonian terms, the effective two-qubit Lindbladian is then
\begin{align}
\frac{\dd}{\dd t}\rho
=-i\lambda_{\mathrm{NG}}
[\sigma_A^z\sigma_B^z,\rho]
+\Gamma\mathcal{D}[\sigma_A^z](\rho)
+\Gamma\mathcal{D}[\sigma_B^z](\rho),
\end{align}
where $\mathcal D[O](\rho)=O\rho O^\dagger-\frac{1}{2}\{O^\dagger O,\rho\}$, and the corresponding error rate is
$\Gamma=\Gamma_A^{\rm dep}/2=\Gamma_B^{\rm dep}/2$.

\begin{figure}
    \centering
    \includegraphics[width=0.5\linewidth]{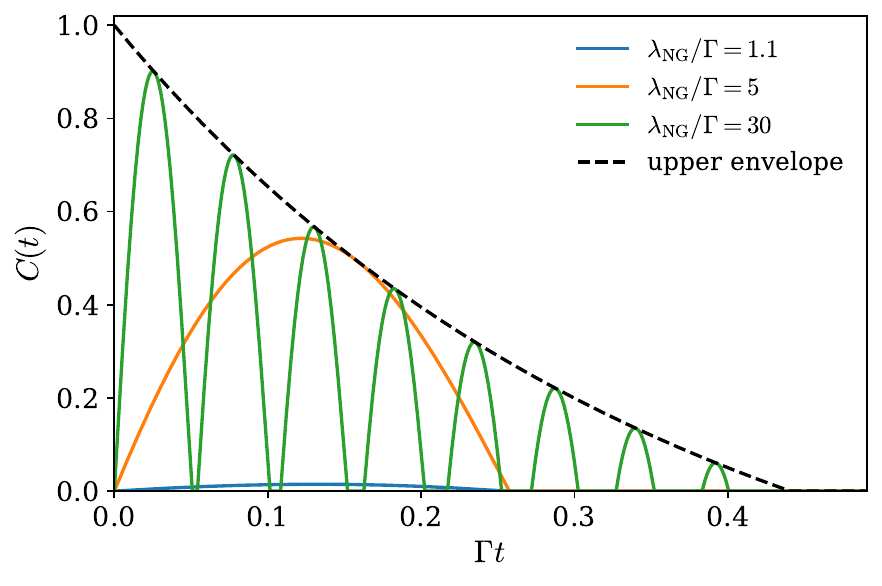}
    \caption{The concurrence $C(t)$ is plotted as a function of the dimensionless time $\Gamma t$ for different ratios $\lambda_{\mathrm{NG}}/\Gamma$. For $\lambda_{\mathrm{NG}}/\Gamma\leq 1$, no entanglement is generated from the initial product state $|+\rangle_A|+\rangle_B$, while larger $\lambda_{\mathrm{NG}}/\Gamma$ produces oscillatory entanglement windows. The black dashed curve shows the upper envelope $e^{-2\Gamma t}[1-\sinh(2\Gamma t)]$. }
    \label{fig:concurrence_vs_time}
\end{figure}

Let us then analyze entanglement generation in this setting. We take the initial state to be $|\psi(0)\rangle = |+\rangle_A|+\rangle_B$. By direct evaluation, the concurrence at time $t$ is
\begin{align}
C(t) = e^{-2\Gamma t}
\max\{0,|\sin(2\lambda_{\mathrm{NG}}t)|-\sinh(2\Gamma t)\}.
\end{align}
Thus entanglement exists only when $|\sin(2\lambda_{\mathrm{NG}}t)|>\sinh(2\Gamma t).$ In the regime $|\lambda_{\mathrm{NG}}|>\Gamma$, the final entanglement sudden-death time $t_{\rm ESD}$ is determined by the largest positive root
\begin{align}
|\sin(2\lambda_{\mathrm{NG}}t_{\rm ESD})|
=
\sinh(2\Gamma t_{\rm ESD}).
\end{align}
Since $|\sin(2\lambda_{\mathrm{NG}}t)|\leq 1$, no entanglement is possible once $\sinh(2\Gamma t)\geq 1$, therefore
\begin{align}
t_{\rm ESD}\le\frac{\operatorname{arsinh}(1)}{2\Gamma}\approx \frac{0.44}{\Gamma}.
\end{align}
This upper bound can be achieved in the strong-coupling limit $|\lambda_{\mathrm{NG}}|/\Gamma\to\infty$. To illustrate the behaviors of $C(t)$, we show the curves of $C(t)$ for some representative values of $\lambda_{\mathrm{NG}}/\Gamma$ in Fig.~\ref{fig:concurrence_vs_time}.

On the other hand, for fixed $\lambda_{\mathrm{NG}}$ and $\Gamma$, the maximum concurrence $C_{\max}=\max_{t\geq 0}C(t)$ in general does not have a closed form. It satisfies $C_{\max}\to 0$ in the weak-interaction limit $|\lambda_{\mathrm{NG}}|/\Gamma\to 1^+$, and in this limit 
\begin{align}
    C_{\max} \simeq \frac{2}{3} \left(\frac{|\lambda_{\mathrm{NG}}|}{\Gamma}-1\right)^{3/2}.
\end{align} 
Then $C_{\max}$ monotonically increases as $|\lambda_{\mathrm{NG}}|/\Gamma$ increases, with an upper bound 1.
Therefore the ratio $|\lambda_{\mathrm{NG}}|/\Gamma$ controls both the largest attainable entanglement and the final entanglement lifetime. 

Let us then study the limit of this scheme by imposing an upper bound on $|\lambda_{\mathrm{NG}}|/\Gamma$. As for the scheme based on cat states considered here, we note that 
\begin{align}
    \frac{|\lambda_{\mathrm{NG}}|}{\Gamma}= \frac{Gm_A m_B\Delta x^2}{\hbar \lambda_{\mathrm{GRW}} n d(d^2-\Delta x^2)\left[1-\exp\left(-\frac{\Delta x^2}{4r_C^2}\right)\right]}
\end{align}
Since this model should only apply if the two spheres don't touch each other, let us say there is a minimum $d-\Delta x = d_{\min}$. Then we note that
\begin{align}
    \frac{|\lambda_{\mathrm{NG}}|}{\Gamma} \le \frac{Gm_A m_B}{\hbar \lambda_{\mathrm{GRW}}n d_{\min} } \max\left(\frac{4 r_C^2}{d_{\min}^2},\frac{1}{2}\right)
\end{align}
This maximum serves as an upper bound on $|\lambda_{\mathrm{NG}}|/\Gamma$. Therefore, within this framework of gravity-induced entanglement tests, the maximal entanglement and its lifetime would be quite limited. And one needs to choose sufficiently large mass to observe entanglement.

As a comparison, let us also study the CSL model and the Di\'osi--Penrose model. In this case there will be an additional correlated noise in the master equation as follows:
\begin{align}
\frac{\dd}{\dd t}\rho
=-i\lambda_{\mathrm{NG}}
[\sigma_A^z\sigma_B^z,\rho]
+\Gamma\mathcal{D}[\sigma_A^z](\rho)
+\Gamma\mathcal{D}[\sigma_B^z](\rho)+\Gamma_{AB}(\sigma^z_A\rho \sigma^z_B +\sigma^z_B \rho \sigma^z_A - \{\sigma^z_A \sigma^z_B,\rho\}),
\end{align}
where $\Gamma_{AB}$ contributes the nonlocal error. By direct evaluation, we take the initial state to be $|+\rangle_A|+\rangle_B$, the concurrence at time $t$ reads as follows:
\begin{align}
    C(t) = \frac{1}{2} \max \{&0, e^{-2\Gamma t}\sqrt{4 \sin^2 (2\lambda_{\mathrm{NG}} t) + e^{-4\Gamma t} \sinh^2 (4\Gamma_{AB}t)}-|1-e^{-4\Gamma t} \cosh 4\Gamma_{AB}t|\}.
\end{align}
Then one can show that in the weak interaction limit $\sqrt{\lambda_{\mathrm{NG}}^2+\Gamma_{AB}^2}/\Gamma \to 1^+$, assuming $|\Gamma_{AB}|\neq \Gamma$ (complete positivity requires $|\Gamma_{AB}|\le \Gamma$), the maximal concurrence is roughly
\begin{align}
    C_{\max} \simeq \frac{2}{3\sqrt{1-\frac{\Gamma_{AB}^2}{\Gamma^2}}} \left(\frac{\sqrt{\lambda_{\mathrm{NG}}^2+\Gamma_{AB}^2}}{\Gamma}-1\right)^{3/2}.
\end{align}
Therefore, similarly the ratio $\sqrt{\lambda_{\mathrm{NG}}^2+\Gamma_{AB}^2}/\Gamma $ is crucial to the attainable entanglement.

As for the CSL model, for simplicity we consider the radius of the two masses, $r_0\ll r_C$, such that decoherence is not too large, then one has 
\begin{align}
    \Gamma = \frac{\lambda_\mathrm{CSL}}{2} \left(\frac{m}{m_0}\right)^2\left[1-\exp\left(-\frac{\Delta x^2}{4r_C^2}\right)\right], \qquad \Gamma_{AB}= \frac{\lambda_\mathrm{CSL}}{4} \left(\frac{m}{m_0}\right)^2 \left[2e^{-\frac{d^2}{4 r_C^2}}-e^{-\frac{(d+\Delta x)^2}{4 r_C^2}}-e^{-\frac{(d-\Delta x)^2}{4 r_C^2}}\right]
\end{align}
Similar analysis as GRW model would also indicate that $\sqrt{\lambda_{\mathrm{NG}}^2+\Gamma_{AB}^2}/\Gamma $ is upper bounded.

As for the Di\'osi--Penrose model, let us also consider the limit $r_0\ll \sigma$. By direct calculation one can derive
\begin{align}
\Gamma = 
\frac{Gm^2}{2\hbar}
\left[
\frac{1}{\sqrt{\pi}\sigma}
-\frac{\mathrm{Erf}\left(\frac{\Delta x}{2\sigma}\right)}{\Delta x}
\right], \qquad -\lambda_{\mathrm{NG}}=\Gamma_{AB}=\frac{Gm^2}{4\hbar}
\left[
2\frac{\mathrm{Erf}\left(\frac{d}{2\sigma}\right)}{d}-\frac{\mathrm{Erf}\left(\frac{d+\Delta x}{2\sigma}\right)}{d+\Delta x}-\frac{\mathrm{Erf}\left(\frac{d-\Delta x}{2\sigma}\right)}{d-\Delta x}
\right].
\end{align}
In this situation, note that the critical ratio $\sqrt{\lambda_{\mathrm{NG}}^2+\Gamma_{AB}^2}/\Gamma$ can be larger than 1 only if $d_{\min}/\sigma$ is sufficiently small, see also~\cite{PhysRevD.111.L121101}. In the limit that $d_{\min}/\sigma, \Delta x/ \sigma \to 0^+$, $\sqrt{\lambda_{\mathrm{NG}}^2+\Gamma_{AB}^2}/\Gamma = \sqrt{2}$ takes the maximum.

In summary, we show that the current protocols of gravity-induced entanglement tests may be quite limited by fundamental collapse models.

\newpage
\section{Technical details of the quantum error-correction protocol}\label{app:Technical details of the quantum error-correction protocol}

This appendix provides additional technical details and supporting results for the quantum error-correction framework introduced in the main text. We begin by clarifying the meaning of error correction in the continuous-recovery (Zeno) limit and showing that perfect error correction in this regime necessarily removes the same interactions one aims to preserve. We then introduce an imperfect local QEC strategy that, although it does not completely remove the noise, can improve the signal-to-noise ratio by retaining the nonlocal interaction while suppressing part of the decoherence. We also derive the corresponding logical error channels, and establish the separability criterion for the resulting effective dynamics. Finally, we discuss possible extensions beyond strictly local codes and alternative recovery strategies.

\subsection{Error correction in the Zeno limit}
\label{appendix:ZenoLimit}
In the main text we make use of the continuous-recovery (Zeno) limit as a convenient theoretical description of error correction. This subsection briefly reviews this limit and clarifies its relation to more realistic finite-rate recovery protocols.

Consider a total evolution time $t$, during which a recovery map $R$ is applied $N$ times uniformly in time. The resulting channel is given by
\begin{equation}
(R\circ e^{\mathcal{L}t/N})^N .
\end{equation}
Restricting to initial states within the code space, for which $R(\rho)=\rho$, and using idempotency of $R$, i.e. $R^2=R$, this
channel can be expanded as
\begin{align}
(R\circ e^{\mathcal{L}t/N})^N(\rho)
=&\left(\mathbbm{1}
+ R\circ\mathcal{L}\,\frac{t}{N}
+ \frac{t^2}{2N^2} R\circ\mathcal{L}^2  
+ \cdots \right)^N(\rho)
\nonumber\\
=&\exp\!\left(
R\circ\mathcal{L}\,t
+ \frac{t^2}{2N}
\big[R\circ\mathcal{L}^2-(R\circ\mathcal{L})^2\big]
+ O\!\left(\frac{1}{N^2}\right)
\right)(\rho),
\end{align}
where higher-order terms arise from the noncommutativity of the recovery and dynamical maps. Note that we use the identity operator as just defined in the code subspace.

In the limit $N\to\infty$, all subleading corrections vanish and the evolution converges to a Zeno-type dynamics generated by the effective Lindbladian
\begin{equation}
\mathcal{L}_{\mathrm{eff}} = R\circ\mathcal{L}.
\end{equation}
In this regime, recovery acts after every infinitesimal time step, continuously projecting the dynamics onto the code space. As a result, any component of the generator that is perfectly corrected by $R$ is removed at the level of the effective generator itself. This is the sense in which the Zeno limit implements \emph{perfect dynamical error correction}.

At finite $N$, by contrast, error correction is applied at a finite rate $N/t$, and the effective evolution over a time $t$ is described by the channel
\begin{equation}
R\circ e^{\mathcal{L}\frac{t}{N}}\circ R\circ e^{\mathcal{L}\frac{t}{N}}\circ\cdots R\circ e^{\mathcal{L}\frac{t}{N}},
\end{equation}
rather than by a Lindbladian generator of the form $R\circ\mathcal{L}$. In this case, recovery is applied only after a finite evolution time, allowing coherent interactions and errors to act between recovery steps. As a result, interactions that are projected out in the Zeno limit may survive at finite recovery rates, albeit at the cost of residual logical noise.

While the Zeno limit represents an idealized form of error correction, it serves here as a useful theoretical tool for classifying the separability properties of the underlying generator $\mathcal{L}$. In particular, since the recovery map $R$ is local, the non-separability of the Zeno-projected dynamics already implies that of $\mathcal{L}$ itself. Of course, observing entanglement in the finite-rate channel $R\circ e^{\mathcal{L}t}$ likewise implies the non-separability of $\mathcal{L}$, however, the detailed dynamics in this case is more involved to analyze.

\subsection{Limitations of perfect error correction in the Zeno regime}
\label{appendix:LimitationsZeno}

The main text argues that perfect correction of the relevant local errors is incompatible with preserving the interaction responsible for entanglement generation. For completeness, we provide here a formal proof of this statement and clarify its relation to standard fault-tolerant quantum-computing protocols.

We consider dynamics in which the coherent interaction and the local
errors are generated by the same operators,
\begin{equation}
\mathcal{L}(\rho) = -i \lambda [O_A \otimes O_B, \rho]
+ \Gamma_a \mathcal{D}[O_A](\rho)
+ \Gamma_b \mathcal{D}[O_B](\rho).
\end{equation}
As explained above, in the Zeno limit of continuous recovery, the effective error-corrected dynamics is generated by \(\mathcal{L}_{\mathrm{eff}} = R\circ \mathcal{L}\).
We show that any recovery map that \emph{perfectly} corrects the error set
generated by \(O_A\) and \(O_B\) (or equivalently, a recovery map for which $R(D[O_\alpha](\rho))= 0$) necessarily renders the unitary interaction trivial within the code space $R ([O_A \otimes O_B, \rho])=0$.

At first sight, this conclusion may appear to contradict fault-tolerant quantum
computation, where noisy entangling interactions can be used to implement
protected logical gates. There is, however, no contradiction. In fact, this conclusion is quite similar to the Eastin–Knill theorem in the infinitesimal limit~\cite{PhysRevLett.102.110502}. The present result
applies to protocols that attempt to preserve the interaction under instantaneous
(continuous) recovery, whereas fault-tolerant schemes treat the interaction as a
gate primitive and apply error correction only to the accumulated noise. In such
schemes, logical entangling operations are synthesized from sequences of short
interaction pulses with fixed interaction time (in contrast to the continuous recovery) interleaved with recovery and succeed only when the noise per
primitive remains below a finite threshold, thereby restricting attention to a
limited noise regime. Since our aim is to assess, in principle, whether error
correction can reveal gravitationally mediated entanglement across the full
parameter space of the hybrid gravity model, we do not pursue this
threshold-dependent approach here.

We also stress that this limitation is specific to \emph{perfect} error correction
in the Zeno limit. If one instead allows for imperfect recovery, the interaction
need not be completely projected out, and the signal-to-noise ratio can in
principle be improved arbitrarily by suitably tailoring the recovery map. This
alternative strategy, which does not rely on threshold assumptions but abandons
perfect correction, is discussed in App.~\ref{app:Beyond local QEC code}.

We formalize the Zeno-limit statement in the following proposition.

\addtocounter{theorem}{1}
\begin{prop}
\label{prop:KL_and_trivial_unitary}
    Given a quantum code defined on a linear subspace $\mathcal{C}\subset \mathcal{H}_A\otimes\mathcal{H}_B$, two local Hermitian operators $O_A,O_B$, and an error-correction quantum channel $R$, assume that the following conditions hold:
    \begin{itemize}
        \item the error-correction channel $R$ stabilizes code states, i.e. $R(\rho)=\rho$ for any code state $\rho\in \mathcal{L}(\mathcal{C})$;
        \item the local errors are fully corrected, i.e. $R(\mathcal{D}[O_A](\rho))=R(\mathcal{D}[O_B](\rho))=0$;
        \item the error-correction channel $R$ ensures a closed evolution in the code space, i.e. $R\circ\mathcal{L}(\rho) \in \mathcal{L}(\mathcal{C})$ for any code state $\rho\in \mathcal{L}(\mathcal{C})$.
    \end{itemize}
    Then we conclude (i) the Knill--Laflamme conditions
    \begin{align*}
        &{}_{AB}\langle \bar i|O_A|\bar j\rangle_{AB}\propto \delta_{ij}, \qquad{}_{AB}\langle \bar i|O_A^2|\bar j\rangle_{AB}\propto \delta_{ij},\qquad {}_{AB}\langle \bar i|O_B|\bar j\rangle_{AB}\propto \delta_{ij}\\
        &{}_{AB}\langle \bar i|O_B^2|\bar j\rangle_{AB}\propto \delta_{ij},\qquad {}_{AB}\langle \bar i|O_A\otimes O_B|\bar j\rangle_{AB}\propto \delta_{ij},
    \end{align*}
    and (ii) the unitary dynamics is trivial on the code after recovery, i.e. $R([O_A\otimes O_B,\rho])=0$.
\end{prop}

The third condition $R\circ\mathcal{L}(\rho) \in \mathcal{L}(\mathcal{C})$ ensures that code states remain in the code subspace under the composite infinitesimal evolution $R\circ(I+\mathcal{L} dt)$. If this condition fails, then either the code space $\mathcal{C}$ must be redefined so that the dynamics closes on it, or the leakage must be treated as an additional error channel to be corrected.

\begin{proof}
    We first use the condition $R(\rho)=\rho$ on the code. Assume without loss of generality that the channel admits a Kraus decomposition $R(\rho)=\sum_iR_i \rho R_i^\dagger$, and denote the projector onto the code subspace by $\Pi_{AB}$. Such a channel can be realized by a unitary evolution on an auxiliary system. In particular, any pure state $|\psi\rangle_{AB}$ evolves as
    \begin{align*}
        |\psi\rangle_{AB}\otimes |0\rangle_{\mathrm{aux}}\rightarrow \sum_i R_i |\psi\rangle_{AB} \otimes |i\rangle_{\mathrm{aux}}=|\psi\rangle_{AB}\otimes|\phi(\psi)\rangle_\mathrm{aux},
    \end{align*}
    where the last equality holds since $R(|\psi\rangle\langle\psi|)=|\psi\rangle\langle\psi|$ on $\mathcal{C}$. Since this holds for any $|\psi\rangle_{AB}\in\mathcal{C}$, the auxiliary state must be independent of $|\psi\rangle_{AB}$, i.e. $|\phi(\psi)\rangle_\mathrm{aux}=|\phi\rangle_\mathrm{aux}$. Equivalently,
    \begin{align*}
        R_i \Pi_{AB} \propto \Pi_{AB}, \quad \forall i.
    \end{align*}

    Next, the infinitesimal Lindbladian evolution can be cast in Kraus form with Kraus operators
    \begin{align*}
        E_0 = I - \frac{dt}{2}(O_A^2+O_B^2),\quad E_1=\sqrt{dt}O_A,\quad E_2=\sqrt{dt}O_B,
    \end{align*}
    such that
    $$\sum_{i=0}^2E_i\rho E_i^\dagger=\rho+dt(\mathcal{D}[O_A](\rho)+\mathcal{D}[O_B](\rho)+O\left(dt^{2}\right).$$
    Therefore, the two conditions are translated into
    $$R\left(\sum_{i=0}^2E_i\rho E_i^\dagger\right)=\rho+O\left(dt^{2}\right).$$
    Using the discussion above, this implies the following equation up to $O(dt)$ order
    $$R_i E_j \Pi_{AB} \propto \Pi_{AB}\Rightarrow \Pi_{AB} E_j^\dagger E_{k}\Pi_{AB}=\sum_i\Pi_{AB} E_j^\dagger R_i^\dagger R_i E_{k}\Pi_{AB}\propto \Pi_{AB},$$
    i.e. the Knill--Laflamme condition. At order $O(dt)$, the Knill--Laflamme condition reads
    \begin{align*}
        &{}_{AB}\langle \bar i|O_A|\bar j\rangle_{AB}\propto \delta_{ij}, \qquad{}_{AB}\langle \bar i|O_A^2|\bar j\rangle_{AB}\propto \delta_{ij},\qquad {}_{AB}\langle \bar i|O_B|\bar j\rangle_{AB}\propto \delta_{ij}\\
        &{}_{AB}\langle \bar i|O_B^2|\bar j\rangle_{AB}\propto \delta_{ij},\qquad {}_{AB}\langle \bar i|O_A\otimes O_B|\bar j\rangle_{AB}\propto \delta_{ij}.
    \end{align*}

    We now study the error-corrected unitary dynamics generated by $[O_A\otimes O_B,\rho]$. Using the Knill--Laflamme condition above and the assumption $\Pi_{AB}\rho\Pi_{AB}=\rho$ (hence $\rho (I-\Pi_{AB})=0$), we find
    \begin{align*}
    &\Pi_{AB} [O_A\otimes O_B,\rho]\Pi_{AB}=[\Pi_{AB} O_A\otimes O_B \Pi_{AB},\rho]\propto [\Pi_{AB},\rho]=0,\\
    &(I-\Pi_{AB})[O_A\otimes O_B ,\rho](I-\Pi_{AB})=0.
    \end{align*}
    The first equation implies that within the code space $O_A\otimes O_B$ can only generate a trivial global phase. Together with the second equation, this shows that $[O_A\otimes O_B,\rho]$ is purely off-block-diagonal with respect to the decomposition $\Pi_{AB}\oplus(I-\Pi_{AB})$. We next show that any such off-block-diagonal contribution is mapped to zero by the error correction channel. Given that the errors are fully corrected, i.e. $R(\mathcal{D}[O_A](\rho))=R(\mathcal{D}[O_B](\rho))=0$, we just need to ensure that the unitary evolution does not drive code states out of the code space. The condition $R\circ\mathcal{L}(\rho) \in \mathcal{L}(\mathcal{C})$ then implies $\Pi_{AB} R([O_A\otimes O_B,\rho])\Pi_{AB} =R([O_A\otimes O_B,\rho])$.

    Assume without loss of generality that $R_i\Pi_{AB}=r_i\Pi_{AB}$, where $r_i$ is a complex number. Then
    \begin{align*}
        0=\Pi_{AB}(I-\Pi_{AB})=\Pi_{AB}\sum_i R_i^\dagger R_i (I-\Pi_{AB})=\sum_i r_i^* \Pi_{AB}R_i (I-\Pi_{AB}),
    \end{align*}
    and therefore
    \begin{align*}
        R([O_A\otimes O_B,\rho])=&\Pi_{AB}R([O_A\otimes O_B,\rho])\Pi_{AB}\\
        =&\sum_i \Pi_{AB} R_i \Pi_{AB}[O_A\otimes O_B,\rho](I-\Pi_{AB})R_i^\dagger \Pi_{AB} +\Pi_{AB} R_i (I-\Pi_{AB})[O_A\otimes O_B,\rho]\Pi_{AB}R_i^\dagger \Pi_{AB}\\
        =&\sum_i r_i \Pi_{AB}[O_A\otimes O_B,\rho](I-\Pi_{AB})R_i^\dagger \Pi_{AB} + r_i^* \Pi_{AB} R_i (I-\Pi_{AB})[O_A\otimes O_B,\rho]\Pi_{AB}\\
        =& \Pi_{AB}[O_A\otimes O_B,\rho] \times 0+ 0\times [O_A\otimes O_B,\rho]\Pi_{AB}\\
        =&0.
    \end{align*}
\end{proof}

\subsection{Constructing an imperfect local QEC code}\label{app:Local imperfect QEC code}

Since perfect correction eliminates the desired interaction, the protocol considered in the main text relies instead on imperfect error correction. In this subsection, we provide an explicit derivation of this local QEC code and analyze how the microscopic interaction and decoherence processes are mapped onto the encoded degrees of freedom.

We show this for the model that combines Newtonian gravitational interaction and the GRW-type of noise. Other models are considered in App.~\ref{app:Effective error rates}. Keeping only the leading non-trivial contributions, the dynamics can be expanded as
\begin{align}
\label{eq: leading order grw}
    \frac{\dd}{\dd t}\rho_{AB}=&\frac{i}{\hbar}\Big[\frac{Gm_A m_B}{|d+x_A -x_B|},\rho_{AB}\Big]+\mathcal{L}_{\mathrm{GRW},A}(\rho_{AB})+\mathcal{L}_{\mathrm{GRW},B}(\rho_{AB})\nonumber\\
    =& -\frac{2iGm_A m_B}{\hbar d^3}[x_A x_B,\rho_{AB}] + \cdots \nonumber\\
    &+ \lambda_{\mathrm{GRW}} n \left( \int d^3 \boldsymbol{x}_1 d^3 \boldsymbol{x}_2  \left(1-\frac{(\boldsymbol{x}_1-\boldsymbol{x}_2)^2}{4r_\mathrm{C}^2}\right)|\boldsymbol{x}_1\rangle_A {}_A\langle \boldsymbol{x}_1|\rho |\boldsymbol{x}_2\rangle_A {}_A\langle \boldsymbol{x}_2| -\rho_{AB}\right)+\mathcal{L}_{\mathrm{GRW},B}(\rho_{AB})\nonumber\\
    =&-\frac{2iGm_A m_B}{\hbar d^3}[x_A x_B,\rho_{AB}]+ \frac{\lambda_{\mathrm{GRW}} n}{2 r_C^2} \mathcal{D}[x_A](\rho_{AB})+ \frac{\lambda_{\mathrm{GRW}} n}{2 r_C^2} \mathcal{D}[x_B](\rho_{AB}) + \cdots \nonumber\\
\end{align}
Thus, to leading order, the dynamics contains a nonlocal interaction term proportional to $x_A x_B$ and local decoherence terms associated with $x_A$ and $x_B$. In the following, we denote the unitary evolution part as
$$
\mathcal{L}_{\mathrm{NG}}(\rho_{AB})=\frac{i}{\hbar}\Big[\frac{Gm_A m_B}{|d+x_A -x_B|},\rho_{AB}\Big].
$$

Our goal is to preserve the potential entanglement generation while suppressing the collapse-induced decoherence as much as possible. However, the analysis below shows a fundamental contradiction between these two requirements for this class of error models.

We now consider a code space $\mathcal{C}=\operatorname{span}\{|\bar i\rangle_{AB}\}$ in which the interaction $x_A x_B$ induces non-trivial dynamics. This requires
\begin{equation}
    {}_{AB}\langle \bar i| x_A x_B|\bar j\rangle_{AB}\not\propto \delta_{ij},
\end{equation}
i.e. $x_A x_B$ is not proportional to the identity on $\mathcal{C}$. On the other hand, according to the Knill--Laflamme condition Eq.~\eqref{eq:qec cond} in the main text, the collapse errors $x_A$ and $x_B$ can be corrected if and only if the code satisfies
\begin{equation}
    {}_{AB}\langle \bar i| E_k^\dagger E_l |\bar j\rangle_{AB}\propto \delta_{ij}, \quad E_i\in \{\mathbbm{1},x_A, x_B\}.
\end{equation}
Therefore, these two requirements are incompatible in general. In particular, if ${}_{AB}\langle \bar 0| x_A x_B|\bar 1\rangle_{AB} \neq 0$, then the interaction $x_A x_B$ induces coherent mixing between $|\bar 0\rangle_{AB}$ and $|\bar 1\rangle_{AB}$. In this situation, an ambiguity necessarily arises between (i) $|\bar 1\rangle_{AB}$ subjected to an $x_B$ error and (ii) $|\bar 0\rangle_{AB}$ subjected to an $x_A$ error. This is manifested by the non-orthogonality condition $({}_{AB}\langle \bar 0| x_A^\dagger) (x_B|\bar 1\rangle_{AB}) \neq 0$, indicating that information can be irreversibly lost due to the stochastic and non-linear nature of the collapse process~\cite{preskill1998lecture}. More broadly, whenever one combines a Newtonian interaction with localization in the position basis (from collapse or from environmental decoherence), the local error operators cannot be cleanly distinguished from the nonlocal interaction, and an analogous contradiction persists.

We therefore focus on \emph{imperfect} QEC: we retain a non-trivial nonlocal Hamiltonian capable of generating entanglement, while protecting part of the encoded information against localization mechanisms.

We restrict to \emph{local} QEC codes of the form $\mathcal{C}=\mathcal{C}_A\otimes\mathcal{C}_B$ and \emph{local} recovery maps $R=R_A\otimes R_B$. In this setting, entanglement is not introduced by code-state preparation, measurement, or error-correction operation. Consequently, any entanglement observed in the experiment must be generated by the gravitational interaction. Further discussions beyond local QEC codes can be found in App.~\ref{app:Beyond local QEC code}.

In the following, we consider a local QEC code defined in the phononic Hilbert space of each center-of-mass motional mode, namely
$$\mathcal{C}_\alpha=\operatorname{span}\{|\bar 0\rangle_\alpha=|0\rangle_\alpha, |\bar 1\rangle_\alpha=|2N+1\rangle_\alpha\}.
$$
We denote the Fock basis by kets without overlines, e.g. $|i\rangle_\alpha$. The parameter $N$ plays the role of a ``code distance'' in the discussion below. We also define the code-space projector $\Pi=\sum_{i=0,1}|\bar i\rangle\langle \bar i|$ and introduce the logical operators $\sigma^+=|\bar 1\rangle \langle \bar 0|$, $\sigma^-=|\bar 0\rangle \langle \bar 1|$, $\sigma^x=\sigma^+ +\sigma^-$, and $\sigma^z=|\bar 0\rangle \langle \bar 0|-|\bar 1\rangle \langle \bar 1|$.

The discussion above kept only leading-order terms. In a full-order treatment of the Newtonian potential and collapse models, higher powers of the displacement operators, i.e. $x_\alpha^i$, can appear in the master equation. We now characterize how these operators act within the code space. Write $x_\alpha=x_0(a_\alpha+a_\alpha^\dagger)$, where $x_0$ is the ground-state position standard deviation, $\alpha \in \{A,B\}$, and $a_A,a_B$ are the phonon annihilation operators. Then $x_\alpha^i$ expands into monomials in $a_\alpha$ and $a_\alpha^\dagger$ of total degree $i,i-2,i-4,\cdots$, implying the following structure of matrix elements:
\begin{itemize}
    \item ${}_\alpha\langle \bar 0 | x^i_\alpha |\bar 1\rangle_\alpha\neq 0 $ only when $i\ge 2N+1$ and $i$ is odd.
    \item ${}_\alpha\langle \bar 0 | x^i_\alpha |\bar 0\rangle_\alpha={}_\alpha\langle \bar 1 | x^i_\alpha |\bar 1\rangle_\alpha= 0 $ only when $i$ is odd, and otherwise ${}_\alpha\langle \bar 0 | x^i_\alpha |\bar 0\rangle_\alpha\neq{}_\alpha\langle \bar 1 | x^i_\alpha |\bar 1\rangle_\alpha$.
\end{itemize}
According to the QEC conditions in Eq.~\eqref{eq:qec cond} in the main text and the standard discussion in~\cite{preskill1998lecture}, these matrix elements imply the following logical error structure (we omit subscripts for simplicity):
\begin{itemize}
    \item $\sigma^x$-type errors: $\langle \bar 0 | x^ix^j |\bar 1\rangle= 0 $ implies that $x^i|\bar 0\rangle$ and $x^j|\bar 1\rangle$ are perfectly distinguishable (orthogonal), hence there is no confusion between $|\bar 0\rangle$ and $|\bar 1\rangle$. This holds for all $i+j\le 2N$. By contrast, $\langle \bar 0 | x^ix^j |\bar 1\rangle\neq 0 $ implies that $x^i|\bar 0\rangle$ can be confused with $x^j|\bar 1\rangle$. For instance, $\langle \bar 0 | x^Nx^{N+1} |\bar 1\rangle\neq 0 $ implies that $x^N|\bar 0\rangle$ and $x^{N+1}|\bar 1\rangle$ are not orthogonal. If we attempt to correct an $x^N$ error using a Kraus operator $|\bar 0\rangle\langle\psi|$ with $|\psi\rangle \propto x^N|\bar 0\rangle$, then an $x^{N+1}$ error acting on $|\bar 1\rangle$ produces a component along $|\psi\rangle$, and the same Kraus operator will map part of $x^{N+1}|\bar 1\rangle$ to $|\bar 0\rangle$. This corresponds to a $\sigma^-$ error. Conversely, $\langle \bar 0 | x^{N+1}x^{N} |\bar 1\rangle\neq 0 $ corresponds to a $\sigma^+$ error. Together these contributions form a $\sigma^x$-type logical error.
    \item $\sigma^z$-type errors: $\langle \bar 0 | x^i |\bar 0\rangle\neq\langle \bar 1 | x^i |\bar 1\rangle$ implies $(\langle \bar 0|+\langle \bar 1|)x^i(|\bar 0\rangle-|\bar 1\rangle)\neq 0$, and by the same reasoning this yields an unavoidable $\sigma^z$-type logical error.
    \item nonlocal coupling: ${}_\alpha\langle \bar 0 | x^i_\alpha |\bar 1\rangle_\alpha\neq 0 $ only when $i\ge 2N+1$ implies that a nonlocal coupling between $|\bar 0\rangle_A|\bar 0\rangle_B$ and $|\bar 1\rangle_A |\bar 1\rangle_B$ generated by $x^i_A x^j_B$ appears only when $i,j\ge 2N+1$ and $i,j$ are odd.
\end{itemize}

More precisely, these errors can be characterized using Krylov spaces. For error operators $x,x^2,x^3,\cdots x^N$, define the normalized Krylov states
$|\bar i_{x^k}\rangle\propto (1-|\bar i\rangle\langle \bar i|-\sum_{j=1}^{k-1}|\bar i_{x^j}\rangle \langle \bar i_{x^j}|)x^k|\bar i\rangle$
for $k=1,2,\ldots,N$. One can verify that $\{|\bar i\rangle, |\bar i_x\rangle, \cdots, |\bar i_{x^N}\rangle\}_{i=0,1}$ are mutually orthogonal, since $\langle \bar 0|x^k|\bar 1\rangle=0$ for $k=1,2,3\cdots, 2N$. This follows iteratively: $\langle \bar 0|x^1|\bar 1\rangle=0$ implies $|\bar 0\rangle\perp|\bar 1_x\rangle$ and $|\bar 1\rangle\perp|\bar 0_x\rangle$; then $\langle \bar 0|x^2|\bar 1\rangle=0$ implies $\langle \bar 0|\bar 1_{x^2}\rangle=\langle \bar 0_x|\bar 1_x\rangle=\langle 0_{x^2}|\bar 1\rangle=0$; similarly, $\langle \bar 0|x^3|\bar 1\rangle=0$ implies $\langle \bar 0|\bar 1_{x^3}\rangle=\langle \bar 0_x|\bar 1_{x^2}\rangle=\langle \bar 0_{x^2}|\bar 1_x\rangle=\langle 0_{x^3}|\bar 1\rangle=0$; and continuing in this way, $\langle \bar 0|x^{2N}|\bar 1\rangle=0$ implies $\langle \bar 0_{x^N}|\bar 1_{x^N}\rangle=0$. One can also verify directly that $|\bar 0_{x^k}\rangle=|k\rangle$.

In this setting, we can correct these errors with the following local QEC channel:
\begin{equation}
\label{eq:recoverychannelnottracepreserving}
    R(\rho)=\Pi \rho \Pi + \sum_{k=1}^N(|\bar 0\rangle\langle \bar 0_{x^k}|+|\bar 1\rangle\langle \bar 1_{x^k}|)\rho(|\bar 0_{x^k}\rangle\langle \bar 0|+|\bar 1_{x^k}\rangle\langle \bar 1|).
\end{equation}
In general, this map is not trace preserving. However, for pedagogical purposes, we use it as a recovery operation in this appendix only. Restricting the discussion to the code subspace makes the action of the recovery channel on the code states particularly transparent, without the additional structure required to define a trace-preserving map on the full Hilbert space. And this recovery map can of course be completed to be trace-preserving. In the rigorous treatment of App.~\ref{app:Effective error rates}, we instead introduce a recovery map that is trace-preserving in the whole Hilbert space, and all the results of the main text rest on that construction. Note, nevertheless, that if $\rho$ lies within the code subspace, its population is conserved under Eq.(~\ref{eq:recoverychannelnottracepreserving}), since $\operatorname{Tr} R(\mathcal{D}x^i)=0$ for any $i\le N$. In practice, implementing the full channel may be challenging, and one may effectively obtain a finite probability of successful error correction corresponding to local post-selection. In this situation, verifying entanglement may become more demanding, while the separability classification of the underlying dynamics remains faithful. As discussed above, $R$ can correct $\sigma^x$-like errors up to the $x^N$ error operator. Concretely, for $\rho$ supported on the code space $\mathcal{C}$,
\begin{equation}
    R\circ \mathcal{D}[x^i](\rho)\propto \mathcal{D}[\sigma^z](\rho), \quad \text{for }i=1,2,\cdots, N.
\end{equation}
For higher-order errors, e.g. an $x^{N+1}$ error, $\sigma^x$-type logical errors become unavoidable. For example, $|\bar 0\rangle \langle\bar 0_{x^N}|x^{N+1}|\bar 1\rangle \neq 0$ implies that $R\circ\mathcal{D}[x^{N+1}](\rho)$ necessarily contains $\sigma^x$-type contributions. In summary, the overall error-corrected dynamics contains the following components:
\begin{itemize}
    \item $R\circ\mathcal{L}_{\mathrm{NG}}$ induces a nonlocal $\sigma^x_A\sigma^x_B$ interaction in the code space. This contribution arises from projection of the terms such as $x^{2N+1}_A x^{2N+1}_B$ (and higher orders) in the Taylor expansion of the Newtonian potential.
    \item Error-corrected decoherence terms: for example, $R\circ \mathcal{L}_{\mathrm{GRW}}$ induces local $\sigma^z$-type and $\sigma^x$-type errors acting on the two masses individually. All terms $x^i_\alpha$ contribute to $\sigma^z$-type errors, while only $x^{N+1}_\alpha$ and higher-order terms contribute to $\sigma^x$-type errors.
\end{itemize}

Note that we have so far ignored the contribution from the free Hamiltonian $H_0$. This is because it is purely local and does not affect the separability classification of the bipartite dynamics. A direct calculation yields the effective free Hamiltonian on the code,
\begin{equation}
    \Pi_A\Pi_B H_0\Pi_A\Pi_B=\hbar \omega \left(2N+2 -\frac{2N+1}{2}\sigma^z_A-\frac{2N+1}{2}\sigma^z_B\right).
\end{equation}
In general, this encoding will also introduce $\sigma^x_A\sigma^z_B,\sigma^z_A\sigma^x_B,\sigma^z_A\sigma^z_B$ nonlocal interaction, see the discussions in App.~\ref{app:Locality of the effective Lindbladian} for further considerations. These interactions are also present  from lower-order expansions $x^i_Ax^j_B$, but numerical results in App.~\ref{app:Calculation of the effective Lindbladian} will show $\sigma^x_A\sigma^x_B$ interaction is the main contribution to the non-separability.

\subsection{Effective logical error channels}\label{app:Logical error rates}

The encoded logical dynamics derived above can be expressed in terms of a small set of effective logical interactions and logical error channels. This subsection introduces a convenient parametrization of these processes that will be used in the subsequent separability analysis and in the calculation of effective error rates.

Throughout this section we write the effective corrected decoherence generator as
\begin{align}
\mathcal{L}_{\mathrm{eff}}=R\circ\mathcal{L}_{\mathrm{decoh}},
\end{align}
and we restrict attention to its action on operators supported on the code subspace.

Consider first the effective dynamics on a single encoded qubit (either party $\alpha=A$ or $B$). It is convenient to represent $\mathcal{L}_{\mathrm{eff}}$ in the operator basis $\{|\bar i\rangle\alpha{}\alpha\langle\bar j|\}_{i,j\in{0,1}}$. Using the double-ket convention
\begin{align}
|\bar i\bar j\rangle\rangle_\alpha \equiv |\bar i\rangle_\alpha{}_\alpha\langle\bar j|,
\end{align}
we will encounter generators of the form

Let us first study single-qubit errors. Without loss of generality, for the calculations in the following we will encounter effective Lindbladians in the double-ket notation as follows:
\begin{equation}
    \mathcal{L}_{\mathrm{eff}} \begin{pmatrix}
        |\bar 0 \bar 0\rangle\rangle_\alpha \\
        |\bar 0 \bar 1\rangle\rangle_\alpha \\
        |\bar 1 \bar 0\rangle\rangle_\alpha \\
        |\bar 1 \bar 1\rangle\rangle_\alpha
    \end{pmatrix} =
\begin{pmatrix}
-a & 0 & 0 & b\\
0 & -c & d & 0\\
0 & d & -c & 0\\
a & 0 & 0 & -b
\end{pmatrix}\begin{pmatrix}
        |\bar 0 \bar 0\rangle\rangle_\alpha \\
        |\bar 0 \bar 1\rangle\rangle_\alpha \\
        |\bar 1 \bar 0\rangle\rangle_\alpha \\
        |\bar 1 \bar 1\rangle\rangle_\alpha
    \end{pmatrix},
\end{equation}
where $a,b,c,d$ are real parameters determined by the microscopic model and the chosen local error correction channel $R_\alpha$.

This class of generators can be rewritten in the standard Lindblad form on the logical qubit as
\begin{align}
     \mathcal{L}_{\mathrm{eff}} (\rho)=& \Gamma_{\mathrm{eff}}^+\mathcal{D}[\sigma^+_\alpha ](\rho)+\Gamma_{\mathrm{eff}}^-\mathcal{D}[\sigma^-_\alpha ] (\rho)+\Gamma_{\mathrm{eff}}^x\mathcal{D}[\sigma^x_\alpha ](\rho)+\Gamma_{\mathrm{eff}}^z\mathcal{D}[\sigma^z_\alpha ](\rho),\nonumber\\
     \text{where } \Gamma_{\mathrm{eff}}^x =& d,\qquad \Gamma_{\mathrm{eff}}^+ = a-d,\qquad \Gamma_{\mathrm{eff}}^- = b-d,\qquad\Gamma_{\mathrm{eff}}^z = \frac{2c-a-b}{4}.
\end{align}

In addition to local terms, some microscopic models generate symmetric two-body decoherence contributions that, after the local recovery, take the structured form, see also App.~\ref{app:Effective error rates}:
\begin{align}
    \mathcal{L}_{\mathrm{eff}}(\rho)=&+\Gamma_{\mathrm{eff}}^{zz}(\mathcal{D}[\sigma^z_A-\sigma^z_B]-\mathcal{D}[\sigma^z_A]-\mathcal{D}[\sigma^z_B])(\rho)+\Gamma_{\mathrm{eff}}^{xx}(\mathcal{D}[\sigma^x_A-\sigma^x_B]-\mathcal{D}[\sigma^x_A]-\mathcal{D}[\sigma^x_B])(\rho)\nonumber\\
    &+\Gamma_{\mathrm{eff}}^{xz}(\mathcal{D}[\sigma^x_A-\sigma^z_B]-\mathcal{D}[\sigma^x_A]-\mathcal{D}[\sigma^z_B]-\mathcal{D}[\sigma^z_A-\sigma^x_B]+\mathcal{D}[\sigma^z_A]+\mathcal{D}[\sigma^x_B])(\rho),
\end{align}
This structure reflects the symmetry of the underlying decoherence Lindbladian and isolates the genuinely two-party contributions into differences of local logical operators.

A particularly useful identity is
\begin{equation}
    (\mathcal{D}[\sigma^x_A-\sigma^x_B]-\mathcal{D}[\sigma^x_A]-\mathcal{D}[\sigma^x_B])(\rho)=\sigma^x_A\otimes \sigma^x_B\rho + \rho \sigma^x_A\otimes \sigma^x_B - \sigma^x_A\rho \sigma^x_B - \sigma^x_B \rho \sigma^x_A.
\end{equation}
which makes it explicit that this term correlates the two systems even though it is generated by a single Lindblad operator of the form $\sigma_A^x-\sigma_B^x$.

Finally, it is convenient to extract the coefficients $\Gamma_{\mathrm{eff}}^{zz}$, $\Gamma_{\mathrm{eff}}^{xz}$, and $\Gamma_{\mathrm{eff}}^{xx}$ directly from matrix elements in the logical two-qubit basis. One consistent choice is
\begin{align}
    4\Gamma_{\mathrm{eff}}^{zz}=&{}_A\langle \bar 0|{}_B\langle \bar 0|\mathcal{L}_{\mathrm{eff}}(|\bar 0\rangle_A|\bar 0\rangle_B {}_A\langle \bar 0|{}_B\langle \bar 0|)|\bar 0\rangle_A|\bar 0\rangle_B+{}_A\langle \bar 1|{}_B\langle \bar 1|\mathcal{L}_{\mathrm{eff}}(|\bar 1\rangle_A|\bar 1\rangle_B {}_A\langle \bar 0|{}_B\langle \bar 0|)|\bar 0\rangle_A|\bar 0\rangle_B \nonumber\\
    &- {}_A\langle \bar 0|{}_B\langle \bar 1|\mathcal{L}_{\mathrm{eff}}(|\bar 0\rangle_A|\bar 1\rangle_B {}_A\langle \bar 0|{}_B\langle \bar 0|)|\bar 0\rangle_A|\bar 0\rangle_B - {}_A\langle \bar 1|{}_B\langle \bar 0|\mathcal{L}_{\mathrm{eff}}(|\bar 1\rangle_A|\bar 0\rangle_B {}_A\langle \bar 0|{}_B\langle \bar 0|)|\bar 0\rangle_A|\bar 0\rangle_B,\nonumber\\
    2\Gamma_{\mathrm{eff}}^{xz}=&{}_A\langle \bar 0|{}_B\langle \bar 0|\mathcal{L}_{\mathrm{eff}}(|\bar 0\rangle_A|\bar 1\rangle_B {}_A\langle \bar 0|{}_B\langle \bar 0|)|\bar 0\rangle_A|\bar 0\rangle_B- {}_A\langle \bar 0|{}_B\langle \bar 1|\mathcal{L}_{\mathrm{eff}}(|\bar 1\rangle_A|\bar 1\rangle_B {}_A\langle \bar 0|{}_B\langle \bar 0|)|\bar 0\rangle_A|\bar 0\rangle_B,\nonumber\\
    \Gamma_{\mathrm{eff}}^{xx}=&{}_A\langle \bar 0|{}_B\langle \bar 0|\mathcal{L}_{\mathrm{eff}}(|\bar 1\rangle_A|\bar 1\rangle_B {}_A\langle \bar 0|{}_B\langle \bar 0|)|\bar 0\rangle_A|\bar 0\rangle_B.
\end{align}
These relations will be used later to calculate effective logical error rates in a model-independent way and to connect them to the separability criteria discussed in App.~\ref{app:Locality of the effective Lindbladian}.

\subsection{Separability of the effective Lindbladian}\label{app:Locality of the effective Lindbladian}

The separability criterion stated in Theorem~\ref{thm:classification} of the main text is derived here in detail. Using the effective logical interaction and error channels introduced above, we determine when the resulting Lindbladian is non-separable or entanglement-generating and when it admits an equivalent separable realization. 

In general, in this section let us deal with the Lindbladian of the following form
\begin{align}
\label{eq:lindblad total}
    \mathcal{L}_{\mathrm{tot}} = & \mathcal{L}_{\mathrm{eff}}^x+\mathcal{L}_{\mathrm{eff}}^z+\mathcal{L}_{\mathrm{eff}}^{xz}+\mathcal{L}_{\mathrm{eff}}^{\pm}\nonumber\\
    \mathcal{L}_{\mathrm{eff}}^x(\rho)=&-i\lambda_\mathrm{eff}^{xx} [\sigma^x_A \otimes \sigma^x_B, \rho]+\Gamma_{\mathrm{eff}}^x(\sigma^x_A\rho\sigma^x_A-\rho)+\Gamma_{\mathrm{eff}}^x(\sigma^x_B\rho\sigma^x_B-\rho) + \Gamma_{\mathrm{eff}}^{xx}(\sigma^x_A \sigma^x_B \rho + \rho\sigma^x_A\sigma^x_B - \sigma^x_A \rho \sigma^x_B -\sigma^x_B \rho \sigma^x_A),\nonumber\\
    \mathcal{L}_{\mathrm{eff}}^z(\rho)=&-i\lambda_\mathrm{eff}^{zz} [\sigma^z_A \otimes \sigma^z_B, \rho]+\Gamma_{\mathrm{eff}}^z(\sigma^z_A\rho\sigma^z_A-\rho)+\Gamma_{\mathrm{eff}}^z (\sigma^z_B\rho\sigma^z_B-\rho) + \Gamma_{\mathrm{eff}}^{zz}(\sigma^z_A \sigma^z_B \rho + \rho\sigma^z_A\sigma^z_B - \sigma^z_A \rho \sigma^z_B -\sigma^z_B \rho \sigma^z_A)\nonumber\\
    \mathcal{L}_{\mathrm{eff}}^{xz}(\rho)=&-i\lambda_\mathrm{eff}^{xz} [\sigma^x_A \otimes \sigma^z_B-\sigma^z_A \otimes \sigma^x_B, \rho]+ \Gamma_{\mathrm{eff}}^{xz}(\sigma^x_A \sigma^z_B \rho + \rho\sigma^x_A\sigma^z_B - \sigma^x_A \rho \sigma^z_B -\sigma^z_B \rho \sigma^x_A\nonumber\\
    &\qquad-\sigma^z_A \sigma^x_B \rho - \rho\sigma^z_A\sigma^x_B + \sigma^z_A \rho \sigma^x_B +\sigma^x_B \rho \sigma^z_A),\nonumber\\
    \mathcal{L}_{\mathrm{eff}}^{\pm}(\rho) = &\sum_{\alpha=A,B}\Big(\Gamma_{\mathrm{eff}}^+\mathcal{D}[\sigma^+_\alpha ](\rho)+\Gamma_{\mathrm{eff}}^-\mathcal{D}[\sigma^-_\alpha ] (\rho)\Big),
\end{align}
which is the most general form of the effective Lindbladian in Eq.~\eqref{eq:eff lind full} and App.~\ref{app:Calculation of the effective Lindbladian}, up to local Hamiltonian terms, which does not affect the separability classification.

\subsubsection{Separability classification}\label{app:Separability classification}

Let us first prove a lemma that will be used for the proof of the theorem below. This lemma considers a bipartite system $A,B$, where its Lindbladian dynamics is governed by some Hermitian operator $O_A,O_B$, and the separability classification of this dynamics is fully classified as follows.

\begin{lemma}
\label{lemma:separability}
    Consider the following Lindbladian defined on a bipartite system:
    \begin{equation}
    \label{eq:lemma2 lindbladian}
    \mathcal{L}(\rho)= -i\lambda [O_A \otimes O_B, \rho]+\Gamma_A\mathcal{D}[O_A](\rho)+\Gamma_B\mathcal{D}[O_B](\rho) + \Gamma_{AB}(O_A O_B \rho + \rho O_A O_B - O_A \rho O_B -O_B \rho O_A),
    \end{equation}
    where $O_A,O_B$ are non-trivial Hermitian operators defined on subsystem $A,B$, respectively. Its separability classification is determined by the ratio between the effective interaction rate, $\lambda$, the effective local error rates, $\Gamma_{A}$ and $\Gamma_{B}$, as well as the effective nonlocal error rate, $\Gamma_{AB}$:
    \begin{itemize}
        \item When $\lambda^2 +\Gamma_{AB}^2>\Gamma_{A}\Gamma_{B}$, $\mathcal{L}$ is non-separable and generates entanglement.
        \item When $\lambda^2 +\Gamma_{AB}^2\le\Gamma_{A}\Gamma_{B}$, $\mathcal{L}$ is separable. Moreover, it can be realized by LOCC if $\lambda^2\le (\Gamma_A-|\Gamma_{AB}|)(\Gamma_B-|\Gamma_{AB}|)$ is also satisfied.
    \end{itemize}
\end{lemma}

Note that complete positivity of $\mathcal{L}$ requires $\Gamma_A,\Gamma_B\ge 0$ and $|\Gamma_{AB}|\le\sqrt{\Gamma_A\Gamma_B}$.

\begin{proof}
    Let us first study the non-separable side, i.e. when $\lambda^2 +\Gamma_{AB}^2>\Gamma_{A}\Gamma_{B}$. Let us consider an initial product state $|\psi\rangle_A|\phi\rangle_B$. Note that the decoherence channels in $\mathcal{L}$ are invariant under the transformation $O_A \rightarrow O_A+aI_A, O_B \rightarrow O_B + b I_B$, where $a,b$ are some constants, and the unitary part of $\mathcal{L}$ transforms up to some local unitary, without changing the separability classification. Without loss of generality, let us then assume ${}_A\langle \psi|O_A|\psi\rangle_A = {}_B\langle \phi| O_B |\phi\rangle_B=0$. And we also assume $O_A|\psi\rangle_A= |\psi_{O}\rangle_A$, $O_B|\phi\rangle_B= |\phi_O\rangle_B$, otherwise $O_A, O_B$ can be renormalized by redefining the coefficients. Then in the subspace $\operatorname{span}\{|\psi\rangle_A|\phi\rangle_B,|\psi\rangle_A|\phi_O\rangle_B,|\psi_O\rangle_A|\phi\rangle_B,|\psi_O\rangle_A|\phi_O\rangle_B\}$, the evolution in the short-time limit is
    \begin{align}
        \mathcal{L}(|\psi\rangle_A\langle\psi|\otimes |\phi\rangle_B\langle\phi|)=\begin{pmatrix}
             -\Gamma_A -\Gamma_B & -\Gamma_B\frac{{}_B\langle \phi|O_B^2|\phi_O\rangle_B}{2} & -\Gamma_A\frac{{}_A\langle \psi|O_A^2|\psi_O\rangle_A}{2} & \Gamma_{AB}+i\lambda\\
             -\Gamma_B\frac{{}_B\langle \phi_O|O_B^2|\phi\rangle_B}{2} & \Gamma_B & -\Gamma_{AB} & \\
             -\Gamma_A\frac{{}_A\langle \psi_O|O_A^2|\psi\rangle_A}{2}& -\Gamma_{AB} & \Gamma_A & \\
             \Gamma_{AB}-i\lambda & & &
        \end{pmatrix}+\cdots
    \end{align}
    Note that in the subspace $\Pi=|\psi\rangle_A\langle\psi|\otimes |\phi_O\rangle_B\langle\phi_O|+|\psi_O\rangle_A\langle\psi_O|\otimes |\phi\rangle_B\langle\phi|$, one has
    \begin{align}
        \det \Pi\times \mathcal{L}^{T_B}(|\psi\rangle_A\langle\psi|\otimes |\phi\rangle_B\langle\phi|)\times\Pi = \det \begin{pmatrix}
              \Gamma_B & \Gamma_{AB} -i\lambda\\
             \Gamma_{AB}+ i\lambda & \Gamma_A
        \end{pmatrix}= \Gamma_A \Gamma_B - \Gamma_{AB}^2 - \lambda^2
    \end{align}
    where $T_B$ is the partial transpose on subsystem $B$. This implies $\mathcal{L}$ is entangling and hence non-separable when $\lambda^2 + \Gamma_{AB}^2 > \Gamma_A\Gamma_B$, since the state in short-time limit is not partial-transpose positive. In other words, the channel $\exp \mathcal{L}t$ is entangling at least in a finite interval of time around $t>0$.

    On the other hand, let us explicitly construct a separable construction of the evolution channel. Let us consider the following Kraus operators:
    \begin{align}
        &K_{1,2}=(I_A \pm c O_A)(2I_A+2|c|^2O_A^2)^{-1/2}\otimes (I_B\pm d O_B)(2I_B+2|d|^2O_B^2)^{-1/2}, \nonumber\\
        &K_{3,4}=(I_A \pm e O_A)(2I_A+2|e|^2O_A^2)^{-1/2}\otimes (I_B \pm f O_B)(2I_B+2|f|^2O_B^2)^{-1/2}.
    \end{align}
    These Kraus operators are trace-preserving as $\sum_{i=1}^4K_{i}^\dagger K_{i}= I_{AB}$, when we take $|c|^2=|e|^2,|d|^2=|f^2|$ and $\operatorname{Re}c\operatorname{Re}d+\operatorname{Re}e\operatorname{Re}f=0$. Let us assume $c,d$ are of the order $O(\sqrt{\dd t})$ at short-time $\dd t$, then this channel read as:
    \begin{align}
        \mathcal{E}(\rho)=& \sum_{i=1}^4 K_{i}\rho K_{i}^\dagger \nonumber\\
        =& \rho + |c|^2 \mathcal{D}[O_A](\rho) + |d|^2 \mathcal{D}[O_B](\rho) + \operatorname{Re}\frac{cd+ef}{2}(O_A O_B \rho + \rho O_A O_B - O_A \rho O_B -O_B \rho O_A) \nonumber\\
        &+ i\operatorname{Im}\frac{cd+ef}{2} [O_AO_B, \rho] +i \operatorname{Im} \frac{c d^*+e f^*}{2} (O_A \rho O_B - O_B \rho O_A) + O\left(\dd t^{2}\right)
    \end{align}
    Note that the $iO_A \rho O_B - i O_B \rho O_A$ corresponds to KTM-like term, see also App.~\ref{app:Collapse models}. In this situation, let us take
    \begin{equation}
        |c|^2 = \Gamma_A \dd t, \quad|d|^2 = \Gamma_B \dd t, \quad \operatorname{Re} \frac{cd+ef}{2} = \Gamma_{AB} \dd t,\quad \operatorname{Im} \frac{cd+ ef}{2} = -\lambda \dd t,  \quad \operatorname{Im} \frac{cd^* + e f^*}{2}=0.
    \end{equation}
    More specifically, these equations have solution so long as $(\lambda^2+\Gamma_{AB}^2)\dd t^2 = |cd+ef|^2/4 \le |c|^2 |d|^2 = \Gamma_A \Gamma_B \dd t^2$. One solution is 
    \begin{align}
        &c= \sqrt{\Gamma_A \dd t} e^{i\frac{\alpha + \beta + \theta}{2}}, \quad d= \sqrt{\Gamma_B \dd t} e^{i\frac{\alpha + \beta - \theta}{2}}, \quad e= \sqrt{\Gamma_A \dd t} e^{i\frac{\alpha - \beta - \theta}{2}}, \quad f= \sqrt{\Gamma_B \dd t} e^{i\frac{\alpha - \beta + \theta}{2}},\nonumber\\
        & \text{where } \alpha = \arg \Gamma_{AB}-i \lambda, \quad \beta = \arccos \sqrt{\frac{\Gamma_{AB}^2+\lambda^2}{\Gamma_A\Gamma_B}}, \quad \theta =  \arccos  -\frac{\Gamma_{AB}}{\sqrt{\Gamma_A\Gamma_B}}
    \end{align}
    and we take $\alpha=0$ for $\Gamma_{AB}=\lambda = 0$. Therefore, we show that the separable channel $\mathcal{E}(\rho,\dd t)=\rho +\mathcal{L}(\rho)\dd t + O(\dd t^{2})$ realizes $\exp \mathcal{L} t$ in the leading order. Therefore $\exp \mathcal{L} t$ is separable since 
    \begin{equation}
        \exp \mathcal{L} t (\rho) = \lim_{N\rightarrow \infty} \mathcal{E}^N(\rho, t/N)
    \end{equation}
    can be realized by composition of separable channels.

    On the other hand, let us show this Lindbladian can also be realized in a LOCC fashion with a smaller parameter regime. Following the KTM construction in App.~\ref{app:Collapse models}, we may take $O_A$ and $O_B$ as measurement and feedback operators. The resulting LOCC Lindbladian reads
\begin{align}
\label{eq:lemma2 LOCC}
    \mathcal{L}_{\mathrm{LOCC}}(\rho)=&-i[g O_A\otimes h O_B,\rho]+\mathcal{D}[g O_A- i h O_B](\rho)\nonumber\\
    &-i[m O_A\otimes n O_B,\rho]+\mathcal{D}[n O_B- i m O_A](\rho)\nonumber\\
    =&-i(gh+mn)[O_A\otimes O_B,\rho]+(g^2+m^2) \mathcal{D}[O_A](\rho)+ (h^2+n^2) \mathcal{D}[O_B](\rho),
\end{align}
where $g,h,m,n>0$ are real numbers and we have assumed $gh=mn$. On the other hand, note the $\Gamma_{AB}$-terms can be realized in an local-operation-and-shared-randomness fashion as follows:
\begin{align}
\label{eq:lemma2 LOSR}
    &\mathbb{E}_\eta e^{i \eta O_A} e^{-i\eta O_B} \rho e^{-i \eta O_A} e^{i\eta O_B}\nonumber\\
    =& \rho + \mathbb{E}_\eta i \eta[ ( O_A - O_B) \rho -\rho (O_A -O_B)] + \mathbb{E}_\eta \eta^2 \left[( O_A - O_B) \rho (O_A -O_B) - \frac{(O_A-O_B)^2}{2} \rho- \rho\frac{(O_A-O_B)^2}{2} \right] +\cdots \nonumber\\
    =& \rho + \mathbb{E}_\eta \eta^2 \mathcal{D}[ O_A - O_B]( \rho) + O(\eta^4),
\end{align}
assuming the random variable satisfied $\mathbb{E}_\eta \eta=\mathbb{E}_\eta \eta^3=0$. Similarly one can show
\begin{equation}
\label{eq:lemma2 LOSR 2}
    \mathbb{E}_\eta e^{i \eta O_A} e^{i\eta O_B} \rho e^{-i \eta O_A} e^{-i\eta O_B}= \rho + \mathbb{E}_\eta \eta^2 \mathcal{D}[ O_A + O_B]( \rho) +O(\eta^4),
\end{equation}
Combining Eq.~\eqref{eq:lemma2 LOCC}-\eqref{eq:lemma2 LOSR 2}, and assuming $\mathbb{E}_\eta \eta^2 = p\dd t$, we conclude the following Lindbladian can be realized in an LOCC fashion:
\begin{equation}
    -i(gh+mn)[O_A\otimes O_B,\rho]+(g^2+m^2 + p) \mathcal{D}[O_A](\rho)+ (h^2+n^2+p) \mathcal{D}[O_B](\rho) \pm p (O_A O_B \rho + \rho O_A O_B - O_A \rho O_B -O_B \rho O_A).
\end{equation}
Therefore, $\mathcal{L}$ can be further realized in an LOCC fashion if
\begin{equation}
    \lambda^2 = (gh+mn)^2 \le (g^2+m^2)(h^2+n^2)= (\Gamma_A-|\Gamma_{AB}|)(\Gamma_B-|\Gamma_{AB}|),
\end{equation}
in the case $\Gamma_A,\Gamma_B\ge |\Gamma_{AB}|$. This can be solved by $g=\sqrt{\Gamma_A-|\Gamma_{AB}|}\cos \xi, m=\sqrt{\Gamma_A-|\Gamma_{AB}|}\sin \xi, h=\lambda/2g, n=\lambda/2m$, where $\sin 2\xi =|\lambda|/\sqrt{(\Gamma_A-|\Gamma_{AB}|)(\Gamma_B-|\Gamma_{AB}|)}$.  
\end{proof}

Note that there is a finite gap, where $\mathcal{L}$ is not known to be LOCC or not:
\begin{equation}
    \Gamma_A\Gamma_B - (\Gamma_A+\Gamma_B)|\Gamma_{AB}|+\Gamma_{AB}^2<\lambda^2 \le\Gamma_A \Gamma_B - \Gamma_{AB}^2,
\end{equation}
since complete positivity require $|\Gamma_{AB}|\le \sqrt{\Gamma_A\Gamma_B} \le (\Gamma_A +\Gamma_B)/2$.

Let us now consider the separability classification of $\mathcal{L}_{\mathrm{tot}}$. We first note that the Lindbladian can be rewritten as 
\begin{align}
    \mathcal{L}_{\mathrm{eff}}^x + \mathcal{L}^{\pm}_{\mathrm{eff}} =& \mathcal{L}^{x\prime}_{\mathrm{eff}}+\mathcal{L}^\perp_{\mathrm{eff}}\nonumber\\
    \mathcal{L}^{x\prime}_{\mathrm{eff}}(\rho) = & -i\lambda_\mathrm{eff}^{xx} [\sigma^x_A \otimes \sigma^x_B, \rho]+\bar{\Gamma}_{\mathrm{eff}}^x(\sigma^x_A\rho\sigma^x_A-\rho)+\bar{\Gamma}_{\mathrm{eff}}^x(\sigma^x_B\rho\sigma^x_B-\rho) + \Gamma_{\mathrm{eff}}^{xx}(\sigma^x_A \sigma^x_B \rho + \rho\sigma^x_A\sigma^x_B - \sigma^x_A \rho \sigma^x_B -\sigma^x_B \rho \sigma^x_A),\nonumber\\
    \mathcal{L}^\perp_{\mathrm{eff}}(\rho) = & \mathcal{D}[L^\perp_A](\rho)+\mathcal{D}[L^\perp_B](\rho), \qquad
    \nonumber\\
    &\text{where } \qquad \bar{\Gamma}_{\mathrm{eff}}^x=\Gamma^x_{\mathrm{eff}}+\frac{\Gamma_{\mathrm{eff}}^+\Gamma_{\mathrm{eff}}^-}{\Gamma_{\mathrm{eff}}^+ +\Gamma_{\mathrm{eff}}^-},\quad L^\perp_\alpha=\frac{i \Gamma_{\mathrm{eff}}^+ \sigma^+_\alpha -i\Gamma_{\mathrm{eff}}^- \sigma^-_\alpha}{\sqrt{\Gamma_{\mathrm{eff}}^+ +\Gamma_{\mathrm{eff}}^-}}.
\end{align}
The numerical results in App.~\ref{app:Effective error rates} shows $\Gamma^+_\mathrm{eff}<0$, and it does not violate the complete positivity, as long as $\Gamma^+_\mathrm{eff}+\Gamma^-_\mathrm{eff}>0$ and $\bar{\Gamma}^x_{\mathrm{eff}}>0$. The full complete positivity condition should take all decoherence rates into consideration with the Kossakowski matrix~\cite{gorini1976completely,lindblad1976generators}. In the case of $\Gamma^+_\mathrm{eff}+\Gamma^-_\mathrm{eff}=0$, complete positivity requires $\Gamma^+_\mathrm{eff}=\Gamma^-_\mathrm{eff}=0$, and we define $\bar{\Gamma}^x_\mathrm{eff}=\Gamma^x_\mathrm{eff}, L^\perp_\alpha=0$. In the following, we will show the term $\mathcal{L}^\perp_\mathrm{eff}$ does not affect the separability classification.

Let us restate Theorem~\ref{thm:classification} as follows:

\addtocounter{theorem}{-3}
\begin{theorem}
\label{thm:classification_app}
    Given the Lindbladian $\mathcal{L}_\mathrm{tot}$ defined in Eq.~\eqref{eq:lindblad total}, let us define
    \begin{equation}
        \eta = \left\|  \begin{pmatrix}
            \frac{\Gamma^{xx}_\mathrm{eff}-i\lambda^{xx}_\mathrm{eff}}{\bar{\Gamma}^x_\mathrm{eff}} & -\frac{\Gamma^{xz}_\mathrm{eff}-i\lambda^{xz}_\mathrm{eff}}{\sqrt{\Gamma^z_\mathrm{eff}\bar{\Gamma}^x_\mathrm{eff}}} \\
            \frac{\Gamma^{xz}_\mathrm{eff}-i\lambda^{xz}_\mathrm{eff}}{\sqrt{\Gamma^z_\mathrm{eff}\bar{\Gamma}^x_\mathrm{eff}}} & \frac{\Gamma^{zz}_\mathrm{eff}-i\lambda^{zz}_\mathrm{eff}}{\Gamma^z_\mathrm{eff}}\\
        \end{pmatrix}\right\|_2.
    \end{equation}
    The separability classification is determined as follows:
    \begin{itemize}
        \item When $\eta >1$, $\mathcal{L}_\mathrm{tot}$ is non-separable.
        \item When $\eta\le 1$, $\mathcal{L}_\mathrm{tot}$ is separable. A set of equivalent scalar conditions are given in Eq.~\eqref{eq: separable scalar conditions}.
    \end{itemize}
    Therefore, $\eta>1$ serves as a necessary and sufficient condition for non-separability.
\end{theorem}
\addtocounter{theorem}{2}

Consider the simplified situation of $\Gamma^{xz}_\mathrm{eff}=\lambda^{xz}_\mathrm{eff}=\lambda^{zz}_\mathrm{eff}=\Gamma^{zz}_\mathrm{eff}=0$, the separability parameter reduces to $\eta = |\Gamma^{xx}_\mathrm{eff}-i\lambda^{xx} |/\bar{\Gamma}^x_\mathrm{eff}$. And $\eta>1$ also implies the Lindbladian generates entanglement, see also Prop.~\ref{prop:entangling}. As we will see from the numerical results in App.~\ref{app:Effective error rates}, in the non-separable regime we have $|\Gamma^{xz}_\mathrm{eff}-i\lambda^{xz}_\mathrm{eff}|/\sqrt{\Gamma^z_\mathrm{eff}\bar{\Gamma}^x_\mathrm{eff}},  |\Gamma^{zz}_\mathrm{eff}-i\lambda^{zz}_\mathrm{eff}|/\Gamma^z_\mathrm{eff}  \ll  |\Gamma^{xx}_\mathrm{eff}-i\lambda^{xx}_\mathrm{eff}|/\bar{\Gamma}^x_\mathrm{eff}$ and hence the non-separable side $\eta$ is mainly controlled by this ratio. In this case, the separability parameter can be expanded as follows:
\begin{align}
    \eta =& |a| + \frac{|b|^2}{|a|} -\operatorname{Re}\frac{c^* b^2}{a|a|}  + O\left(\max\left(\frac{|b|}{|a|},\frac{|c|}{|a|}\right)^4\right),\nonumber\\
    \text{where}\quad a = & \frac{\Gamma^{xx}_\mathrm{eff}-i\lambda^{xx}_\mathrm{eff}}{\bar{\Gamma}^x_\mathrm{eff}}, \quad b = \frac{\Gamma^{xz}_\mathrm{eff}-i\lambda^{xz}_\mathrm{eff}}{\sqrt{\Gamma^z_\mathrm{eff}\bar{\Gamma}^x_\mathrm{eff}}},\quad c = \frac{\Gamma^{zz}_\mathrm{eff}-i\lambda^{zz}_\mathrm{eff}}{\Gamma^z_\mathrm{eff}} 
\end{align}
Therefore, it is well approximated by 
\begin{equation}
    \eta \approx \frac{\sqrt{(\Gamma^{xx}_\mathrm{eff})^2+(\lambda^{xx}_\mathrm{eff})^2}}{\bar{\Gamma}^x_\mathrm{eff}}.
\end{equation}

On the other hand, in general we have 
\begin{equation}
    \eta \ge \eta_x\equiv  \frac{\sqrt{(\Gamma^{xx}_\mathrm{eff})^2+(\lambda^{xx}_\mathrm{eff})^2}}{\bar{\Gamma}^x_\mathrm{eff}}.
\end{equation}
Hence $\sqrt{(\Gamma^{xx}_\mathrm{eff})^2+(\lambda^{xx})^2}>{\bar{\Gamma}^x_\mathrm{eff}}$ already implies non-separability. This can also be read from the violation of Eq.~\eqref{eq: separable scalar conditions}. Therefore, it is sufficient to consider the rough scaling analysis as in the main text, to account for the non-separability of the effective dynamics. In other words, it is sufficient to claim non-separability of the error-corrected dynamics given $\lambda^{xx}_{\mathrm{eff}}\gg \bar{\Gamma}^x_\mathrm{eff}$.

\begin{proof}
    In the following proof, for simplicity let us omit the $\mathrm{eff}$ subscripts. Let us first consider the non-separable side. The non-separability of the Lindbladian $\mathcal{L}_{\mathrm{tot}}$ can be directly inferred from the partial-transpose negativity of the Choi state. More specifically, let us denote the maximally entangled state $|\Phi\rangle = (|00\rangle+|11\rangle)/\sqrt{2}$, and $|\Psi\rangle_{A_1 A_2; B_1 B_2}= |\Phi\rangle_{A_1 A_2}\otimes |\Phi\rangle_{B_1 B_2}$. Similar to the proof in Lemma~\ref{lemma:separability}, let us consider the six-dimensional subspace $\operatorname{span}\{\sigma^x_{A_1}|\Psi\rangle,\sigma^y_{A_1}|\Psi\rangle,\sigma^z_{A_1}|\Psi\rangle,\sigma^x_{
    B_1}|\Psi\rangle,\sigma^y_{
    B_1}|\Psi\rangle,\sigma^z_{
    B_1}|\Psi\rangle\}$ with corresponding projection operator $\Pi_6$. Then the corresponding matrix elements of the partial-transposed Choi state read as follows\footnote{In the case of $\Gamma^+_\mathrm{eff}+\Gamma^-_\mathrm{eff}=0$, complete positivity requires $\Gamma^+_\mathrm{eff}=\Gamma^-_\mathrm{eff}=0$. $\sigma^y_\alpha$ sectors are then decoupled.}:
    \begin{align}
        G\equiv\Pi_6 \mathcal{L}_{\mathrm{tot},A_1B_1} (|\Psi\rangle_{A_1 A_2; B_1 B_2}\langle \Psi|)^{T_{B_1 B_2}} \Pi_6= \begin{pmatrix}
            \Gamma^x + \frac{\Gamma^++\Gamma^-}{4}& -i\frac{\Gamma^+-\Gamma^-}{4} & & \Gamma^{xx}-i\lambda^{xx} & &  \Gamma^{xz}-i \lambda^{xz} \\
            i\frac{\Gamma^+-\Gamma^-}{4} & \frac{\Gamma^++\Gamma^-}{4} & & & & \\
            & & \Gamma^z & -\Gamma^{xz}+i\lambda^{xz} & & \Gamma^{zz}-i\lambda^{zz} \\
            \Gamma^{xx}+i\lambda^{xx}& & -\Gamma^{xz}-i\lambda^{xz} & \Gamma^x+\frac{\Gamma^++\Gamma^-}{4} & i\frac{\Gamma^+-\Gamma^-}{4} & \\
            & & & -i\frac{\Gamma^+-\Gamma^-}{4} & \frac{\Gamma^++\Gamma^-}{4} & \\
            \Gamma^{xz}+i\lambda^{xz}& & \Gamma^{zz}+i\lambda^{zz} & & & \Gamma^z \\
        \end{pmatrix}
    \end{align}
    Note that complete positivity requires $\Gamma^+ + \Gamma^- \ge 0$, then we can rewrite $G$ as $4\oplus 2$ blocks, i.e.
    \begin{align}
        G\cong \begin{pmatrix}
            \Gamma^x + \frac{\Gamma^++\Gamma^-}{4}& \Gamma^{xz}-i \lambda^{xz}  & & \Gamma^{xx}-i\lambda^{xx} & &   -i\frac{\Gamma^+-\Gamma^-}{4}\\
            \Gamma^{xz}+i\lambda^{xz} & \Gamma^z  & \Gamma^{zz}+i\lambda^{zz} & & & \\
            & \Gamma^{zz}-i\lambda^{zz} & \Gamma^z & -\Gamma^{xz}+i\lambda^{xz} & &  \\
            \Gamma^{xx}+i\lambda^{xx}& & -\Gamma^{xz}-i\lambda^{xz} & \Gamma^x+\frac{\Gamma^++\Gamma^-}{4} & i\frac{\Gamma^+-\Gamma^-}{4} & \\
            & & & -i\frac{\Gamma^+-\Gamma^-}{4} & \frac{\Gamma^++\Gamma^-}{4} & \\
            i\frac{\Gamma^+-\Gamma^-}{4}& &  & & & \frac{\Gamma^++\Gamma^-}{4} \\
        \end{pmatrix}
    \end{align}
    The positivity of $G$ is equivalent to its Schur complement, i.e. the matrix
    \begin{align}
        &\begin{pmatrix}
            \Gamma^x + \frac{\Gamma^++\Gamma^-}{4}& \Gamma^{xz}-i \lambda^{xz}  & & \Gamma^{xx}-i\lambda^{xx} \\
            \Gamma^{xz}+i\lambda^{xz} & \Gamma^z  & \Gamma^{zz}+i\lambda^{zz} &\\
            & \Gamma^{zz}-i\lambda^{zz} & \Gamma^z & -\Gamma^{xz}+i\lambda^{xz}\\
            \Gamma^{xx}+i\lambda^{xx}& & -\Gamma^{xz}-i\lambda^{xz} & \Gamma^x+\frac{\Gamma^++\Gamma^-}{4}\\
        \end{pmatrix} - \begin{pmatrix}
            &  -i\frac{\Gamma^+-\Gamma^-}{4}\\
            & \\
            & \\
            i\frac{\Gamma^+-\Gamma^-}{4} &
        \end{pmatrix}\frac{4 I_2}{\Gamma^++\Gamma^-} \begin{pmatrix}
             & & & -i\frac{\Gamma^+-\Gamma^-}{4}\\
             i\frac{\Gamma^+-\Gamma^-}{4} & & &
        \end{pmatrix}\nonumber\\
        &= \begin{pmatrix}
            \bar{\Gamma}^x & \Gamma^{xz}-i \lambda^{xz}  & & \Gamma^{xx}-i\lambda^{xx} \\
            \Gamma^{xz}+i\lambda^{xz} & \Gamma^z  & \Gamma^{zz}+i\lambda^{zz} &\\
            & \Gamma^{zz}-i\lambda^{zz} & \Gamma^z & -\Gamma^{xz}+i\lambda^{xz}\\
            \Gamma^{xx}+i\lambda^{xx}& & -\Gamma^{xz}-i\lambda^{xz} & \bar{\Gamma}^x\\
        \end{pmatrix}.
    \end{align}
    Note that this last 4-by-4 matrix can be rewritten as a $2\oplus 2$ block as follows:
    \begin{equation}
       \begin{pmatrix}
            \bar{\Gamma}^x &  & \Gamma^{xx}+i\lambda^{xx}  & \Gamma^{xz}+i \lambda^{xz}  \\
             & \Gamma^z  & -\Gamma^{xz}-i\lambda^{xz} & \Gamma^{zz}+i\lambda^{zz}\\
             \Gamma^{xx}-i\lambda^{xx} & -\Gamma^{xz}+i\lambda^{xz} & \bar{\Gamma}^x & \\
             \Gamma^{xz}-i\lambda^{xz} & \Gamma^{zz}-i\lambda^{zz} &  &  \Gamma^z\\
        \end{pmatrix}.
    \end{equation}
    Therefore, the separability of this dynamics is controlled by the rewritten $\mathcal{L}^{x\prime}$ and is independent of the $\mathcal{L}^\perp$ term. Without loss of generality, let us assume $\bar{\Gamma}^x,\Gamma^z\neq0$. Otherwise it returns to the single-axis case in Lemma~\ref{lemma:separability}. Then its positivity is also equivalent to the positivity of the following matrix, given $\bar{\Gamma}^x,\Gamma^z> 0$ imposed by the complete positivity:
    \begin{align}
    \label{eq:block 2 positivity}
        &\begin{pmatrix}
            \bar{\Gamma}^x &\\
            & \Gamma^z
        \end{pmatrix} - \begin{pmatrix}
            \Gamma^{xx}+i\lambda^{xx} & \Gamma^{xz}+i\lambda^{xz} \\
            -\Gamma^{xz}-i\lambda^{xz} & \Gamma^{zz}+i\lambda^{zz}\\
        \end{pmatrix} \begin{pmatrix}
            \frac{1}{\bar{\Gamma}^x} &\\
            & \frac{1}{\Gamma^z}
        \end{pmatrix}\begin{pmatrix}
            \Gamma^{xx}-i\lambda^{xx} & -\Gamma^{xz}+i\lambda^{xz} \\
            \Gamma^{xz}-i\lambda^{xz} & \Gamma^{zz}-i\lambda^{zz}\\
        \end{pmatrix} \nonumber\\
        =&  \begin{pmatrix}
            \bar{\Gamma}^x-\frac{(\Gamma^{xx})^2+(\lambda^{xx})^2}{\bar{\Gamma}^x} - \frac{(\Gamma^{xz})^2+(\lambda^{xz})^2}{\Gamma^z} & \frac{(\Gamma^{xx}+i\lambda^{xx})(\Gamma^{xz}-i\lambda^{xz})}{\bar{\Gamma}^x}-\frac{(\Gamma^{xz}+i\lambda^{xz})(\Gamma^{zz}-i\lambda^{zz})}{\Gamma^z}\\
            \frac{(\Gamma^{xx}-i\lambda^{xx})(\Gamma^{xz}+i\lambda^{xz})}{\bar{\Gamma}^x}-\frac{(\Gamma^{xz}-i\lambda^{xz})(\Gamma^{zz}+i\lambda^{zz})}{\Gamma^z}& \Gamma^z-\frac{(\Gamma^{xz})^2+(\lambda^{xz})^2}{\bar{\Gamma}^x}-\frac{(\Gamma^{zz})^2+(\lambda^{zz})^2}{\Gamma^z}
        \end{pmatrix} \ge 0
    \end{align}
    In other words, we have the following separability necessary condition:
    \begin{align}
    \label{eq: separable scalar conditions}
        &\frac{(\Gamma^{xx})^2+(\lambda^{xx})^2}{(\bar{\Gamma}^x)^2} + \frac{(\Gamma^{xz})^2+(\lambda^{xz})^2}{\bar{\Gamma}^x\Gamma^z} \le 1, \qquad \frac{(\Gamma^{xz})^2+(\lambda^{xz})^2}{\bar{\Gamma}^x\Gamma^z}+\frac{(\Gamma^{zz})^2+(\lambda^{zz})^2}{(\Gamma^z)^2}\le 1,\nonumber\\
        & \Bigg|\frac{(\Gamma^{xx}+i\lambda^{xx})(\Gamma^{xz}-i\lambda^{xz})}{\bar{\Gamma}^x}-\frac{(\Gamma^{xz}+i\lambda^{xz})(\Gamma^{zz}-i\lambda^{zz})}{\Gamma^z}\Bigg|^2 \nonumber\\
        \le&  \left(\bar{\Gamma}^x-\frac{(\Gamma^{xx})^2+(\lambda^{xx})^2}{\bar{\Gamma}^x} - \frac{(\Gamma^{xz})^2+(\lambda^{xz})^2}{\Gamma^z}\right)\left(\Gamma^z-\frac{(\Gamma^{xz})^2+(\lambda^{xz})^2}{\bar{\Gamma}^x}-\frac{(\Gamma^{zz})^2+(\lambda^{zz})^2}{\Gamma^z}\right).
    \end{align}
    Equivalently, Eq.~\eqref{eq:block 2 positivity} can also be rewritten as
    \begin{align}
    \label{eq:positivity with J}
        0\le I - \begin{pmatrix}
            \frac{1}{\sqrt{\bar{\Gamma}^x}} &\\
            & \frac{1}{\sqrt{\Gamma^z}}
        \end{pmatrix}\begin{pmatrix}
            \Gamma^{xx}+i\lambda^{xx} & \Gamma^{xz}+i\lambda^{xz} \\
            -\Gamma^{xz}-i\lambda^{xz} & \Gamma^{zz}+i\lambda^{zz}\\
        \end{pmatrix} \begin{pmatrix}
            \frac{1}{\bar{\Gamma}^x} &\\
            & \frac{1}{\Gamma^z}
        \end{pmatrix}\begin{pmatrix}
            \Gamma^{xx}-i\lambda^{xx} & -\Gamma^{xz}+i\lambda^{xz} \\
            \Gamma^{xz}-i\lambda^{xz} & \Gamma^{zz}-i\lambda^{zz}\\
        \end{pmatrix}\begin{pmatrix}
            \frac{1}{\sqrt{\bar{\Gamma}^x}} &\\
            & \frac{1}{\sqrt{\Gamma^z}}
        \end{pmatrix}.
    \end{align}
    Therefore, we can define the 2-norm, i.e. the largest singular value of $\sqrt{J^\dagger J}$, as the separability parameter and rewrite the separability condition as
    \begin{equation}
        \eta = ||J||_2 \le 1, \quad \text{where } J =  \begin{pmatrix}
            \frac{\Gamma^{xx}-i\lambda^{xx}}{\bar{\Gamma}^x} & -\frac{\Gamma^{xz}-i\lambda^{xz}}{\sqrt{\Gamma^z\bar{\Gamma}^x}} \\
            \frac{\Gamma^{xz}-i\lambda^{xz}}{\sqrt{\Gamma^z\bar{\Gamma}^x}} & \frac{\Gamma^{zz}-i\lambda^{zz}}{\Gamma^z}\\
        \end{pmatrix}.
    \end{equation}
    This is because Eq.~\eqref{eq:positivity with J} is equivalent to $I-J^\dagger J \ge 0$. Therefore, we conclude $\eta>1$ implies $\mathcal{L}_{\mathrm{tot}}$ is not partial-transpose positive, hence $\mathcal{L}_\mathrm{eff}$ is non-separable.

    Then let us consider the separable side. Note that the Choi state above is a $4\times 4$ system, hence partial-transpose positivity does not necessarily imply separability. Let us explicitly construct the separable decomposition of $\exp \mathcal{L}_{\mathrm{tot}}t$. For the convenience of discussion, let us consider $\mathcal{L}'=\mathcal{L}^{x\prime}+\mathcal{L}^z+\mathcal{L}^{xz}$ instead, ignoring $\mathcal{L}^\perp$. This is because $\mathcal{L}^\perp$ itself composes of local Lindblad operator and can be realized with a separable evolution. According to Lemma~\ref{lemma:separability}, it's sufficient to construct $\mathcal{L}'$ from combination of Lindbladian of the form in Eq.~\eqref{eq:lemma2 lindbladian}. Let us consider the following ansatz of $\mathcal{L}'$:
    \begin{align}
        \mathcal{L}'=&\sum_j \mathcal{L}_j^{O_{A,j},O_{B,j}}(\rho),\nonumber\\
        \mathcal{L}_j^{O_{A,j},O_{B,j}}=& -i\lambda_j [O_{A,j} \otimes O_{B,j}, \rho]+\Gamma_{A,j}\mathcal{D}[O_{A,j}](\rho)+\Gamma_{B,j}\mathcal{D}[O_{B,j}](\rho)\nonumber\\
        &+ \Gamma_{AB,j}(O_{A,j} O_{B,j} \rho + \rho O_{A,j} O_{B,j} - O_{A,j} \rho O_{B,j} -O_{B,j} \rho O_{A,j}),\nonumber\\
        \text{where } O_{A,j}\equiv& a_j\cdot \sigma_A^T =  a_j^1 \sigma^x_A + a^2_j \sigma^z_A, \qquad O_{B,j}\equiv b_j\cdot \sigma_B^T= b_j^1 \sigma^x_B + b^2_j \sigma^z_B,
    \end{align}
    where we denote $a_j=(a_j^1,a_j^2), b_j=(b_j^1,b_j^2)$ and $\sigma_\alpha = (\sigma^x_\alpha,\sigma^z_\alpha)$, and the coefficient vectors $a_j,b_j$ are real. By comparing between those coefficients, we obtain the following equations:
    \begin{align}
        \sum_j (\Gamma_{AB,j}-i\lambda_j) a_j^T b_j = \begin{pmatrix}
            \Gamma^{xx}-i\lambda^{xx}& \Gamma^{xz}-i\lambda^{xz}\\
            -\Gamma^{xz}+i\lambda^{xz} & \Gamma^{zz}-i\lambda^{zz}
        \end{pmatrix}, \qquad \sum_j \Gamma_{A,j} a^T_j a_j = \sum_j \Gamma_{B,j} b^T_j b_j = \begin{pmatrix}
            \bar{\Gamma}^x & \\
             & \Gamma^z
        \end{pmatrix}.
    \end{align}
    For simplicity, let us denote $Q = \operatorname{diag}(\bar{\Gamma}^x,\Gamma^z)$. Equivalently, since $\bar{\Gamma}^x,\Gamma^z> 0$\footnote{Otherwise, the separable condition $\eta\le 1$ would imply the interaction rate and non-local error rate are also zero, and reduce to the single-axis situation, which can be tackled with a single $\mathcal{L}_j$ as in Lemma~\ref{lemma:separability}. For instance, $\Gamma^z=0$ implies $\Gamma^{zz}=\Gamma^{xz}=\lambda^{xz}=\lambda^{zz}=0$.}, let us define $c_j = \sqrt{\Gamma_{A,j}} a_j Q^{-1/2}$ and $d_j = \sqrt{\Gamma_{B,j}}b_j Q^{-1/2} $, then the equations above reduce to 
    \begin{equation}
    \label{eq:J separable construction}
        \sum_j \frac{\Gamma_{AB,j}-i\lambda_j}{\sqrt{\Gamma_{A,j}\Gamma_{B,j}}} d_j^T c_j = Q^{-1/2}\begin{pmatrix}
            \Gamma^{xx}-i\lambda^{xx}& -\Gamma^{xz}+i\lambda^{xz}\\
            \Gamma^{xz}-i\lambda^{xz} & \Gamma^{zz}-i\lambda^{zz}
        \end{pmatrix}Q^{-1/2} = J, \qquad \sum_j c^T_j c_j = \sum_j d^T_j d_j = \begin{pmatrix}
            1 & \\
             & 1
        \end{pmatrix}
    \end{equation}
    Assuming there exists a solution of $c,d$ for this equation, then for any two-dimensional complex vector $e,f$, one reads from Cauchy-Schwarz inequality that
    \begin{align}
        |eJ f^\dagger| =& \Bigg|\sum_j \frac{\Gamma_{AB,j}-i\lambda_j}{\sqrt{\Gamma_{A,j}\Gamma_{B,j}}} (e\cdot d_j^T) (c_j \cdot f^\dagger)\Bigg|\nonumber\\
        \le&\sum_j  \Big|(e\cdot d_j^T)\Big|\Big| (c_j \cdot f^\dagger)\Big|\nonumber\\
        \le& \sqrt{\sum_{j,k}  |e\cdot d_j^T|^2 |c_k \cdot f^\dagger|^2}\nonumber\\
        =& \sqrt{\sum_j e\cdot (d_j^T d_j)\cdot e^\dagger}\sqrt{\sum_k f\cdot (c_k^T c_k) f^\dagger}\nonumber\\
        =& |e| |f|
    \end{align}
    For the first inequality we have used $\Gamma_{AB,j}^2+\lambda_j^2 \le \Gamma_{A,j}\Gamma_{B,j}$, see also Lemma~\ref{lemma:separability}. This implies the separability condition $||J||_2\le 1$. 
    
    Then let us consider the sufficiency of this condition. Let us assume the number of $j$ over summation is 2, anddenote two matrices $O_1 = (d_1^T,d_2^T), O_2 = (c_1^T, c_2^T)$, then we have
    \begin{equation}
        \begin{pmatrix}
            1 & \\
             & 1 \\
        \end{pmatrix}=\sum_j c_j^T c_j =O_2 O_2^T = \sum_j d_j^T d_j = O_1 O_1^T,
    \end{equation}
    Therefore we can take $O_1, O_2$ to be orthogonal matrices under certain conditions, discussed below. Then the first equation in Eq.~\eqref{eq:J separable construction} can be rewritten as
    \begin{equation}
        J = O_1 \begin{pmatrix}
            z_1 & \\
             & z_2
        \end{pmatrix} O_2^T,\quad \text{ where } z_j =  \frac{\Gamma_{AB,j}-i\lambda_j}{\sqrt{\Gamma_{A,j}\Gamma_{B,j}}}.
    \end{equation}
    For the convenience of discussion, let us first consider the boundary case that both singular value of $J$ is 1, i.e. $J$ is a unitary and then reduce the general case to summation of unitary $J$. This implies 
    \begin{align}
        &\frac{|\Gamma^{xx}-i\lambda^{xx}|}{\bar{\Gamma}^{x}} = \frac{|\Gamma^{zz}-i\lambda^{zz}|}{{\Gamma}^{z}}, \qquad \frac{|\Gamma^{xx}-i\lambda^{xx}|^2}{(\bar{\Gamma}^{x})^2} + \frac{|\Gamma^{xz}-i\lambda^{xz}|^2}{\bar{\Gamma}^{x}{\Gamma}^{z}}=1, \nonumber\\
        &\arg (\Gamma^{xx}-i\lambda^{xx})+\arg(\Gamma^{zz}-i\lambda^{zz}) - 2\arg(\Gamma^{xz}-i\lambda^{xz}) \equiv 0 \pmod{2\pi}.
    \end{align} 
    Let us denote
    \begin{equation}
        g = \frac{|\Gamma^{xx}-i\lambda^{xx}|}{\bar{\Gamma}^x} = \sqrt{1-h^2}, \quad \theta = -\arg (\Gamma^{xz}-i\lambda^{xz}), \quad \gamma = \theta + \arg (\Gamma^{xx}-i\lambda^{xx}), \quad J= e^{-i\theta}\begin{pmatrix}
            g e^{i\gamma} & -h \\
            h & ge^{-i\gamma}
        \end{pmatrix}.
    \end{equation}
    This equation can be solved in two-fold ways. Firstly, let us solve
    \begin{align}
        J^T J = & O_2\begin{pmatrix}
            z_1^2 & \\  
             & z_2^2
        \end{pmatrix} O_2^T = e^{-2i\theta-i\pi/2}\begin{pmatrix}
            ih^2 + i g^2 e^{2i\gamma} & 2gh\sin\gamma\\
            2gh\sin\gamma & i h^2 +i g^2 e^{-2i\gamma}
        \end{pmatrix} = e^{-2i\theta-i\pi/2}\begin{pmatrix}
            m e^{i\beta} & n\\
            n & -m e^{-i\beta}
        \end{pmatrix},\nonumber\\
        &\text{where } m = \sqrt{(h^2 + g^2 \cos 2\gamma)^2 +g^4 \sin^2 2\gamma}, \beta = \arg ih^2 + i g^2 e^{2i\gamma}, n = 2gh\sin\gamma
    \end{align}
    This equation is readily solved as follows:
    \begin{align}
        c_1 =& \left(x,s\sqrt{1-x^2}\right),\quad c_2= \left(s\sqrt{1-x^2}, -x\right),\quad x= \sqrt{\frac{1+\frac{m\cos \beta}{\sqrt{m^2 \cos^2\beta+n^2}}}{2}}, \quad s = \frac{n}{|n|}, \nonumber\\
         z_1^2 =& e^{-2i\theta - i\pi/2}(\sqrt{m^2\cos^2\beta+n^2}+im\sin \beta), \quad z_2^2 = e^{-2i\theta - i\pi/2}(-\sqrt{m^2\cos^2\beta+n^2}+im\sin \beta)
    \end{align}
    Note that here $|z_1|=|z_2|=1$, which implies corresponding $\mathcal{L}_j^{O_{A,j},O_{B,j}}$ are also at the separable boundary.

    On the other hand, we have 
    \begin{align}
        J J^T = & O_1\begin{pmatrix}
            z_1^2 & \\  
             & z_2^2
        \end{pmatrix} O_1^T = e^{-2i\theta-i\pi/2}\begin{pmatrix}
            ih^2 + i g^2 e^{2i\gamma} & -2gh\sin\gamma\\
            -2gh\sin\gamma & i h^2 +i g^2 e^{-2i\gamma}
        \end{pmatrix} = e^{-2i\theta-i\pi/2}\begin{pmatrix}
            m e^{i\beta} & -n\\
            -n & -m e^{-i\beta}
        \end{pmatrix},
    \end{align}
    where only the sign of off-diagonal element is altered. Therefore it is solved by
    \begin{equation}
        d_1 = \left(x,-s\sqrt{1-x^2}\right),\quad d_2= \left(-s\sqrt{1-x^2}, -x\right)
    \end{equation}

     More specifically, this $\mathcal{L}'$ is realized as follows:
    \begin{align}
        \mathcal{L}' = &\mathcal{L}_1^{O_{A,1},O_{B,1}}+\mathcal{L}_2^{O_{A,2},O_{B,2}},\nonumber\\
        O_{A,1} = &\frac{x\sqrt{\bar{\Gamma}^{x}}\sigma^x_A+s\sqrt{(1-x^2){\Gamma}^{z}}\sigma^z_A}{\sqrt{\Gamma_{A,1}}},\quad O_{A,2} = \frac{s\sqrt{(1-x^2)\bar{\Gamma}^{x}}\sigma^x_A-x\sqrt{{\Gamma}^{z}}\sigma^z_A}{\sqrt{\Gamma_{A,2}}}, \nonumber\\
        O_{B,1} = &\frac{x\sqrt{\bar{\Gamma}^{x}}\sigma^x_B-s\sqrt{(1-x^2){\Gamma}^{z}}\sigma^z_B}{\sqrt{\Gamma_{B,1}}},\quad O_{B,2} = \frac{-s\sqrt{(1-x^2)\bar{\Gamma}^{x}}\sigma^x_B-x\sqrt{{\Gamma}^{z}}\sigma^z_B}{\sqrt{\Gamma_{B,2}}}, \nonumber\\
        \Gamma_{AB,1}-i\lambda_1 =& \sqrt{\Gamma_{A,1}\Gamma_{B,1}}z_1, \quad  \Gamma_{AB,2}-i\lambda_2 = \sqrt{\Gamma_{A,2}\Gamma_{B,2}}z_2
    \end{align}
    and $\Gamma_{\alpha,j}>0$ can be arbitrarily assigned. 
    
    In general, $J$ is not an unitary operator. However, since $||J||_2\le 1$, we can write the singular value decomposition of $J$ as
    \begin{equation}
        J=U \begin{pmatrix}
            s_1 & \\
             & s_2 
        \end{pmatrix} V^\dagger,
    \end{equation}
    where $U$ and $V$ are some unitary matrices, and singular values $1\ge s_1,s_2\ge 0$. Then $J$ can be rewritten as a sum of two unitaries as follows:
    \begin{equation}
        J = \frac{1}{2}U \begin{pmatrix}
            s_1+i\sqrt{1-s_1^2} & \\
             & s_2 + i\sqrt{1-s_2^2}
        \end{pmatrix} V^\dagger +\frac{1}{2} U \begin{pmatrix}
            s_1-i\sqrt{1-s_1^2} & \\
             & s_2 - i\sqrt{1-s_2^2}
        \end{pmatrix} V^\dagger.
    \end{equation}
    Therefore, similar to the discussions above, using product of the unitary and its transpose, we can write down a decomposition of the first unitary as $z_3 c_3^T d_3 + z_4 c^T_4 d_4$, and a decomposition of the second unitary as $z_5 d_5^T c_5 + z_6 d^T_6 c_6$, such that
    \begin{equation}
        z_3 \frac{c_3^T d_3}{2} + z_4 \frac{c^T_4 d_4}{2} +z_5 \frac{c_5^T d_5}{2} + z_6 \frac{c^T_6 d_6}{2} = J, \qquad \sum_{j=3}^6 \frac{c_j^T c_j}{2}=\sum_{j=3}^6 \frac{d_j^T d_j}{2}=\begin{pmatrix}
            1 &\\
             & 1
        \end{pmatrix}.
    \end{equation}
    Therefore, $\mathcal{L}'$ can be rewritten as a summation over four separable Lindbladians, and hence $\mathcal{L}_\mathrm{tot}=\mathcal{L}'+\mathcal{L}^\perp$ is separable, since $\mathcal{L}^\perp$ is just local decoherence channel.

    In summary, we have shown that $\eta = ||J||_2$ can serve as a separability classification parameter for $\mathcal{L}_\mathrm{tot}$ defined in Eq.~\eqref{eq:lindblad total}, such that: (i) $\mathcal{L}_{\mathrm{tot}}$ is separable when $\eta\le 1$, since we give rise to an explicit separable construction according to Lemma.~\ref{lemma:separability}; $\mathcal{L}_{\mathrm{tot}}$ is  non-separable when $\eta>1$, since it violates partial-transpose positivity.   
    
\end{proof}

\subsubsection{Entangling condition}

On the other hand, non-separability does not necessarily leads to entanglement. In general, the evolution channels could be $\operatorname{SEPP\backslash SEP}$, i.e. separability-preserving but non-separable channels. Let us also consider the (non-)separability-preserving, or (non-)entangling criterion. In general, this is equivalent to apply the evolution channel to arbitrary product state and detect whether entanglement is generated; or equivalently, whether the output state has negative partial-transpose~\cite{benatti2008environment}.

\begin{prop}
\label{prop:entangling}
    Given the Lindbladian in Eq.~\eqref{eq:lindblad total}, it is entangling/not separability-preserving at $t=0^+$ leading-order iff for initial product state $|\psi\rangle_A\otimes|\phi\rangle_B$, we have
    \begin{equation}
        \inf_{|\psi\rangle_A,|\phi\rangle_B}(h C h^\dagger)(g C g^\dagger)- |g^*(-D+iH) h^\dagger|^2<0.
    \end{equation}
    The definition of these symbols are presented below. As a simple sufficient condition, if we take $|\psi\rangle_A\otimes|\phi\rangle_B=|00\rangle_{AB}$ or $|11\rangle_{AB}$,  it reduces to 
    \begin{equation}
        (\lambda^{xx})^2 + (\Gamma^{xx})^2 > \min(\Gamma^x+\Gamma^+,\Gamma^x+\Gamma^-)^2
    \end{equation}
\end{prop}

In the situation that only $\lambda^{zz}_\mathrm{eff},\Gamma^{zz},\Gamma^z\neq 0$, it reduces to $(\lambda^{zz}_\mathrm{eff})^2+(\Gamma^{zz}_\mathrm{eff})^2 > (\Gamma^z_\mathrm{eff})^2$.

\begin{proof}
    In general, if the channel $\exp\mathcal{L}_\mathrm{tot} t$ is not separability-preserving for all time, then there exists some pure state such that at $t=0^+$ $\mathcal{L}_\mathrm{tot}$ generates entanglement. In the following proof, let us omit the subscript $\mathrm{eff}$ for simplicity.
    
    Therefore, it is sufficient to consider a product initial state $|\psi\rangle_A|\phi\rangle_B$ and determine if entanglement can be generated with it. Let us denote their orthogonal bases as $|\psi_\perp\rangle_A$ and $|\phi_\perp\rangle_B$. At the leading $O(\dd t)$ order, the relevant component used for entanglement detection is the partial-transposed subspace $\operatorname{span}\{|\psi_\perp\rangle_A|\phi^*\rangle_B,|\psi\rangle_A|\phi_\perp^*\rangle_B\}$, which is denote with corresponding projection operator $\Pi_\perp$. Let us rewrite the Lindbladian with the following Kossakowski representation\cite{gorini1976completely,lindblad1976generators}
    \begin{align}
        \mathcal{L}_\mathrm{tot}(\rho) = & -i[\sum_{ij}H_{ij} \sigma^i_A \otimes \sigma_B^j,\rho] + \sum_{ij} \Bigg(C_{ij} \left(\sigma^i_A\rho \sigma^j_A -\frac{1}{2}\{\sigma^j_A \sigma^i_A,\rho\}\right)+D_{ij} \left(\sigma^i_A\rho \sigma^j_B -\frac{1}{2}\{\sigma^i_A \sigma^j_B,\rho\}\right)\nonumber\\
        &+D_{ji}^* \left(\sigma^i_B\rho \sigma^j_A -\frac{1}{2}\{\sigma^i_B \sigma^j_A,\rho\}\right)+E_{ij} \left(\sigma^i_B\rho \sigma^j_B -\frac{1}{2}\{\sigma^j_B \sigma^i_B,\rho\}\right)\Bigg),\nonumber\\
        \text{where } H =& \begin{pmatrix}
            \lambda^{xx} &  &\lambda^{xz} \\
            & & \\
           -\lambda^{xz} & & \lambda^{zz}\\
        \end{pmatrix}, \quad C= E=  \begin{pmatrix}
            \Gamma^x + \frac{\Gamma^++\Gamma^-}{4}& i\frac{\Gamma^+-\Gamma^-}{4} & \\
            -i\frac{\Gamma^+-\Gamma^-}{4} & \frac{\Gamma^++\Gamma^-}{4} & \\
            & & \Gamma^z \\
        \end{pmatrix}, \quad  D =  \begin{pmatrix}
            -\Gamma^{xx} & &  -\Gamma^{xz} \\
            & & \\
            \Gamma^{xz}& & -\Gamma^{zz} \\
        \end{pmatrix}, 
    \end{align}
    where we have taken $\sigma^1=\sigma^x,\sigma^2=\sigma^y,\sigma^3=\sigma^z$. Let us write the relevant matrix elements as  ${}_A\langle\psi_\perp|\sigma^i_A|\psi_A\rangle=g^i,{}_B\langle \phi_\perp|\sigma^i_B|\phi\rangle_B=h^i$. Then the entanglement generation is detected by the partial-transpose negativity of
    \begin{align}
        \Pi_\perp \mathcal{L}_\mathrm{tot}(|\psi\rangle_A\langle\psi|\otimes |\phi\rangle_B\langle\phi|)^{T_B}\Pi_\perp = & \begin{pmatrix}
            h E h^\dagger & g^*(-\operatorname{Re}D+iH) h^\dagger\\
            g(-\operatorname{Re}D - iH)h^T & g C g^\dagger
        \end{pmatrix}\nonumber\\
        = & \begin{pmatrix}
            h C h^\dagger & g^*(-D+iH) h^\dagger\\
            g(-D - iH)h^T & g C g^\dagger
        \end{pmatrix}.
    \end{align}
    Since $C=E \ge0$, this matrix is non-negative iff
    \begin{equation}
        (h C h^\dagger)(g C g^\dagger)\ge |g^*(-D+iH) h^\dagger|^2
    \end{equation}
    and therefore $\mathcal{L}_\mathrm{tot}$ is entangling at $t=0^+$ iff
    \begin{equation}
        \inf_{|\psi\rangle_A,|\phi\rangle_B}(h C h^\dagger)(g C g^\dagger)- |g^*(-D+iH) h^\dagger|^2<0.
    \end{equation}
    On the other hand, note that this criterion is only at the leading order. If $\Pi_\perp \mathcal{L}_\mathrm{tot}(|\psi\rangle_A\langle\psi|\otimes |\phi\rangle_B\langle\phi|)^{T_B}\Pi_\perp$ has zero eigenvalue, then one needs to look at higher-order term to determine if it generates entanglement.

    As an example, let us take $|\psi\rangle_A = |0\rangle, |\phi\rangle_B = |0\rangle_B$, hence $g=h=(1,i,0)$. In this situation, direct evaluation shows
    \begin{align}
        (h C h^\dagger)(g C g^\dagger)- |g^*(-D+iH) h^\dagger|^2= (\Gamma^x + \Gamma^+)^2 - (\lambda^{xx})^2 - (\Gamma^{xx})^2.
    \end{align}
    Similarly, if we take $|\psi\rangle_A = |1\rangle, |\phi\rangle_B = |1\rangle_B$, then $g=h=(1,-i,0)$ implies 
    \begin{align}
        (h C h^\dagger)(g C g^\dagger)- |g^*(-D+iH) h^\dagger|^2= (\Gamma^x + \Gamma^-)^2 - (\lambda^{xx})^2 - (\Gamma^{xx})^2.
    \end{align}
    Hence we conclude $\mathcal{L}_{\mathrm{tot}}$ generates entanglement if
    \begin{equation}
        (\lambda^{xx})^2 + (\Gamma^{xx})^2 > \min(\Gamma^x+\Gamma^+,\Gamma^x+\Gamma^-)^2
    \end{equation}
\end{proof}

In summary, for the separability classification of the Lindbladian dynamics $\mathcal{L}_\mathrm{tot}$ in Eq.~\eqref{eq:lindblad total}, which is the general form of the error-corrected effective dynamics in App.~\ref{app:Calculation of the effective Lindbladian}, in Theorem~\ref{thm:classification_app} we define 
\begin{equation}
    \eta = \left\|  \begin{pmatrix}
            \frac{\Gamma^{xx}_\mathrm{eff}-i\lambda^{xx}_\mathrm{eff}}{\bar{\Gamma}^x_\mathrm{eff}} & -\frac{\Gamma^{xz}_\mathrm{eff}-i\lambda^{xz}_\mathrm{eff}}{\sqrt{\Gamma^z_\mathrm{eff}\bar{\Gamma}^x_\mathrm{eff}}} \\
            \frac{\Gamma^{xz}_\mathrm{eff}-i\lambda^{xz}_\mathrm{eff}}{\sqrt{\Gamma^z_\mathrm{eff}\bar{\Gamma}^x_\mathrm{eff}}} & \frac{\Gamma^{zz}_\mathrm{eff}-i\lambda^{zz}_\mathrm{eff}}{\Gamma^z_\mathrm{eff}}\\
        \end{pmatrix}\right\|_2,
\end{equation}
such that $\mathcal{L}_\mathrm{tot}$ is separable if $\eta \le 1$, and $\mathcal{L}_\mathrm{tot}$ is non-separable if $\eta > 1$. On the other hand, Prop.~\ref{prop:entangling} give rises to a sufficient condition for $\mathcal{L}_\mathrm{tot}$ to be entangling as follows:
\begin{equation}
        \frac{\sqrt{(\lambda^{xx}_\mathrm{eff})^2 + (\Gamma^{xx}_\mathrm{eff})^2}}{\min(\Gamma^x_\mathrm{eff}+\Gamma^+_\mathrm{eff},\Gamma^x_\mathrm{eff}+\Gamma^-_\mathrm{eff})} > 1.
\end{equation}

\subsection{Beyond local QEC codes}\label{app:Beyond local QEC code}

In the previous sections we analyzed an \emph{imperfect} \emph{local} QEC protocol, where both the code space and the recovery operations are defined locally, ensuring that entanglement is not introduced by the QEC setting itself. More generally, one may consider \emph{separable} recovery operations that also do not introduce entanglement. In this section, we investigate whether one can design a better imperfect QEC code equipped with a general separable recovery map.

In App.~\ref{app:Local imperfect QEC code} we considered a QEC code that encodes two logical qubits. We start from the leading-order master equation in Eq.~\eqref{eq: leading order grw}, which contains local $x_A$ and $x_B$ errors. It can be written in the form
\begin{equation}
    \frac{\dd}{\dd t} \rho =\mathcal{L}_{x_A,x_B}(\rho) = -i \lambda[ x_A\otimes x_B, \rho] + \Gamma \big(\mathcal{D}[x_A\otimes I_B] + \mathcal{D}[I_A\otimes x_B]\big)(\rho),
\end{equation}
If we choose a code with distance $N=0$, then the error-correction conditions in Eq.~\eqref{eq:qec cond} in the main text reduce to
\begin{align}
    &\Pi_\alpha x_\alpha \Pi_\alpha  = x_0 \sigma^x_\alpha , \quad \Pi_\alpha x^2_\alpha \Pi_\alpha  =2x_0^2\Pi_\alpha -x^2_0\sigma^z_\alpha , \nonumber\\
     &\Pi_A\otimes\Pi_B\times x_A\otimes x_B\times \Pi_A\otimes\Pi_B =x_0^2 \sigma^x_A \sigma^x_B,
\end{align}
i.e. none of the error-correction conditions are satisfied. This failure is already enforced by the requirement that $x_A\otimes x_B$ generate a non-trivial nonlocal interaction on the code. If the system is initialized in the product state $|\bar 0\rangle_A|\bar 0\rangle_B$, this interaction drives Rabi oscillations between $|\bar 0\rangle_A|\bar 0\rangle_B$ and $|\bar 1\rangle_A|\bar 1\rangle_B$, and the dynamics is effectively confined to a two-dimensional subspace.

We now consider a more general QEC code in which the error-correction conditions can be satisfied \emph{except} for the condition involving $x_A\otimes x_B$. Since the nonlocal unitary dynamics is described by Rabi oscillations in a two-dimensional subspace, it suffices to consider a two-dimensional code space $\mathcal{C}=\operatorname{span}\{|\bar i\rangle_{AB}\}_{i=0,1}\subset\mathcal{H}_A\otimes\mathcal{H}_B$, which may involve entangled states. This would not be an obstruction as long as one can initialize the system in a product state within $\mathcal{C}$ and implement the recovery operation in a separable manner. We define $\Pi=\sum_i |\bar i\rangle \langle \bar i|$ and $\sigma^x=|\bar 0\rangle\langle \bar 1|+|\bar 1\rangle \langle \bar 0|$.

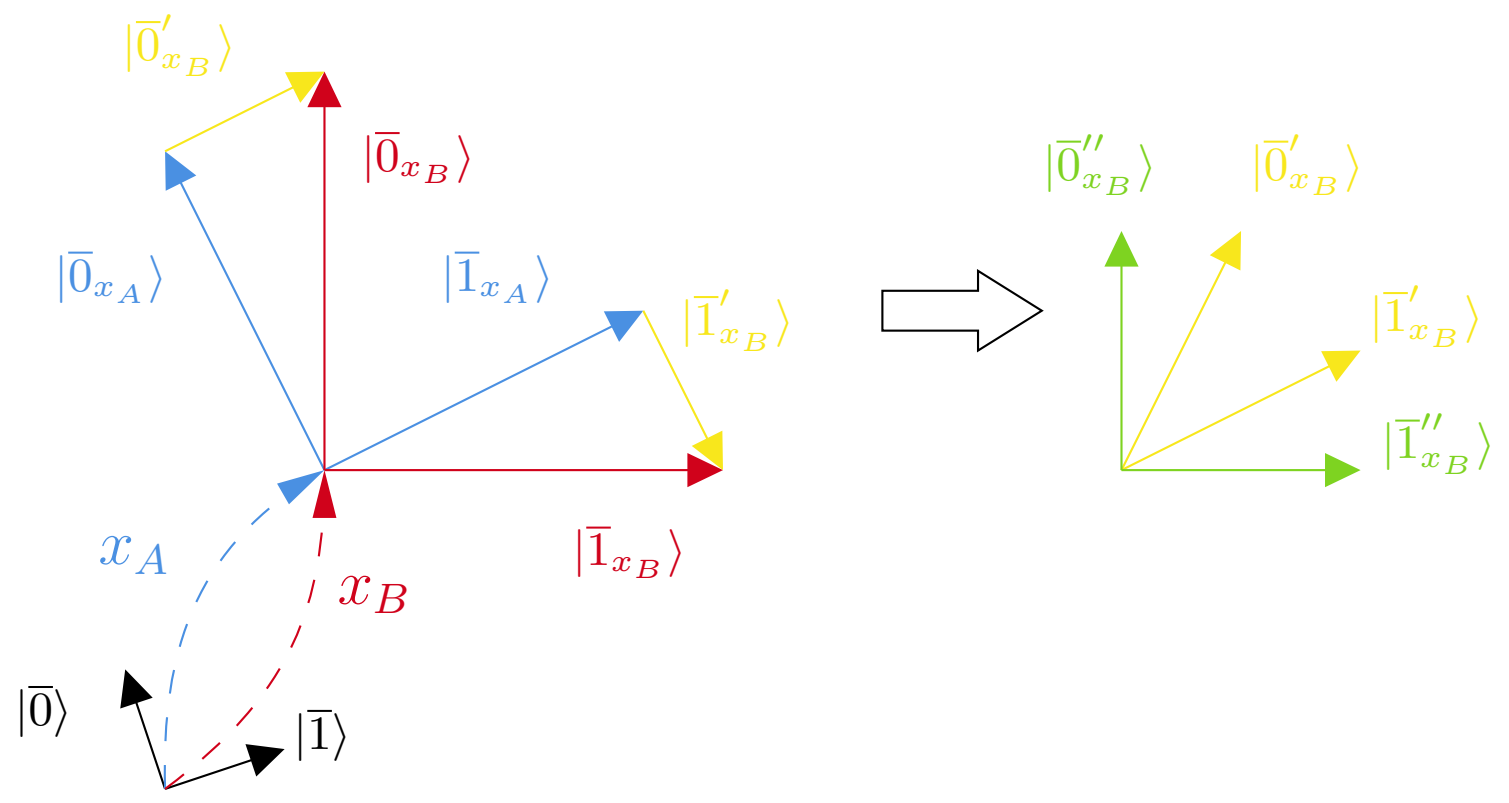
\begin{figure}
    \centering
    \begin{tikzpicture}[x=0.75pt,y=0.75pt,yscale=-1,xscale=1]

\draw    (70,200) -- (60.95,172.85) ;
\draw [shift={(60,170)}, rotate = 71.57] [fill={rgb, 255:red, 0; green, 0; blue, 0 }  ][line width=0.08]  [draw opacity=0] (8.93,-4.29) -- (0,0) -- (8.93,4.29) -- cycle    ;
\draw    (70,200) -- (97.15,190.95) ;
\draw [shift={(100,190)}, rotate = 161.57] [fill={rgb, 255:red, 0; green, 0; blue, 0 }  ][line width=0.08]  [draw opacity=0] (8.93,-4.29) -- (0,0) -- (8.93,4.29) -- cycle    ;
\draw [color={rgb, 255:red, 74; green, 144; blue, 226 }  ,draw opacity=1 ]   (110,120) -- (71.34,42.68) ;
\draw [shift={(70,40)}, rotate = 63.43] [fill={rgb, 255:red, 74; green, 144; blue, 226 }  ,fill opacity=1 ][line width=0.08]  [draw opacity=0] (8.93,-4.29) -- (0,0) -- (8.93,4.29) -- cycle    ;
\draw [color={rgb, 255:red, 74; green, 144; blue, 226 }  ,draw opacity=1 ]   (110,120) -- (187.32,81.34) ;
\draw [shift={(190,80)}, rotate = 153.43] [fill={rgb, 255:red, 74; green, 144; blue, 226 }  ,fill opacity=1 ][line width=0.08]  [draw opacity=0] (8.93,-4.29) -- (0,0) -- (8.93,4.29) -- cycle    ;
\draw [color={rgb, 255:red, 208; green, 2; blue, 27 }  ,draw opacity=1 ]   (110,120) -- (110,23) ;
\draw [shift={(110,20)}, rotate = 90] [fill={rgb, 255:red, 208; green, 2; blue, 27 }  ,fill opacity=1 ][line width=0.08]  [draw opacity=0] (8.93,-4.29) -- (0,0) -- (8.93,4.29) -- cycle    ;
\draw [color={rgb, 255:red, 208; green, 2; blue, 27 }  ,draw opacity=1 ]   (110,120) -- (207,120) ;
\draw [shift={(210,120)}, rotate = 180] [fill={rgb, 255:red, 208; green, 2; blue, 27 }  ,fill opacity=1 ][line width=0.08]  [draw opacity=0] (8.93,-4.29) -- (0,0) -- (8.93,4.29) -- cycle    ;
\draw [color={rgb, 255:red, 248; green, 231; blue, 28 }  ,draw opacity=1 ]   (70,40) -- (107.32,21.34) ;
\draw [shift={(110,20)}, rotate = 153.43] [fill={rgb, 255:red, 248; green, 231; blue, 28 }  ,fill opacity=1 ][line width=0.08]  [draw opacity=0] (8.93,-4.29) -- (0,0) -- (8.93,4.29) -- cycle    ;
\draw [color={rgb, 255:red, 248; green, 231; blue, 28 }  ,draw opacity=1 ]   (190,80) -- (208.66,117.32) ;
\draw [shift={(210,120)}, rotate = 243.43] [fill={rgb, 255:red, 248; green, 231; blue, 28 }  ,fill opacity=1 ][line width=0.08]  [draw opacity=0] (8.93,-4.29) -- (0,0) -- (8.93,4.29) -- cycle    ;
\draw [color={rgb, 255:red, 248; green, 231; blue, 28 }  ,draw opacity=1 ]   (310,120) -- (338.66,62.68) ;
\draw [shift={(340,60)}, rotate = 116.57] [fill={rgb, 255:red, 248; green, 231; blue, 28 }  ,fill opacity=1 ][line width=0.08]  [draw opacity=0] (8.93,-4.29) -- (0,0) -- (8.93,4.29) -- cycle    ;
\draw [color={rgb, 255:red, 248; green, 231; blue, 28 }  ,draw opacity=1 ]   (310,120) -- (367.32,91.34) ;
\draw [shift={(370,90)}, rotate = 153.43] [fill={rgb, 255:red, 248; green, 231; blue, 28 }  ,fill opacity=1 ][line width=0.08]  [draw opacity=0] (8.93,-4.29) -- (0,0) -- (8.93,4.29) -- cycle    ;
\draw [color={rgb, 255:red, 126; green, 211; blue, 33 }  ,draw opacity=1 ]   (310,120) -- (310,63) ;
\draw [shift={(310,60)}, rotate = 90] [fill={rgb, 255:red, 126; green, 211; blue, 33 }  ,fill opacity=1 ][line width=0.08]  [draw opacity=0] (8.93,-4.29) -- (0,0) -- (8.93,4.29) -- cycle    ;
\draw [color={rgb, 255:red, 126; green, 211; blue, 33 }  ,draw opacity=1 ]   (310,120) -- (367,120) ;
\draw [shift={(370,120)}, rotate = 180] [fill={rgb, 255:red, 126; green, 211; blue, 33 }  ,fill opacity=1 ][line width=0.08]  [draw opacity=0] (8.93,-4.29) -- (0,0) -- (8.93,4.29) -- cycle    ;
\draw [color={rgb, 255:red, 74; green, 144; blue, 226 }  ,draw opacity=1 ] [dash pattern={on 4.5pt off 4.5pt}]  (70,200) .. controls (69.02,164.95) and (77.73,139.19) .. (108.58,120.83) ;
\draw [shift={(110,120)}, rotate = 150.09] [fill={rgb, 255:red, 74; green, 144; blue, 226 }  ,fill opacity=1 ][line width=0.08]  [draw opacity=0] (12,-3) -- (0,0) -- (12,3) -- cycle    ;
\draw [color={rgb, 255:red, 208; green, 2; blue, 27 }  ,draw opacity=1 ] [dash pattern={on 4.5pt off 4.5pt}]  (70,200) .. controls (102.34,175.9) and (109.71,153.33) .. (109.99,121.93) ;
\draw [shift={(110,120)}, rotate = 90] [fill={rgb, 255:red, 208; green, 2; blue, 27 }  ,fill opacity=1 ][line width=0.08]  [draw opacity=0] (12,-3) -- (0,0) -- (12,3) -- cycle    ;
\draw   (250,75) -- (274,75) -- (274,70) -- (290,80) -- (274,90) -- (274,85) -- (250,85) -- cycle ;

\draw (31,172.4) node [anchor=north west][inner sep=0.75pt]    {$| \overline{0} \rangle $};
\draw (101,178.4) node [anchor=north west][inner sep=0.75pt]    {$| \overline{1} \rangle $};
\draw (52,135.4) node [anchor=north west][inner sep=0.75pt]  [font=\large,color={rgb, 255:red, 74; green, 144; blue, 226 }  ,opacity=1 ]  {$x_{A}$};
\draw (112,145.4) node [anchor=north west][inner sep=0.75pt]  [font=\large,color={rgb, 255:red, 208; green, 2; blue, 27 }  ,opacity=1 ]  {$x_{B}$};
\draw (41,63.4) node [anchor=north west][inner sep=0.75pt]  [color={rgb, 255:red, 74; green, 144; blue, 226 }  ,opacity=1 ]  {$| \overline{0}_{x_{A}} \rangle $};
\draw (138,63.4) node [anchor=north west][inner sep=0.75pt]  [color={rgb, 255:red, 74; green, 144; blue, 226 }  ,opacity=1 ]  {$| \overline{1}_{x_{A}} \rangle $};
\draw (118,33.4) node [anchor=north west][inner sep=0.75pt]  [color={rgb, 255:red, 208; green, 2; blue, 27 }  ,opacity=1 ]  {$| \overline{0}_{x_{B}} \rangle $};
\draw (171,132.4) node [anchor=north west][inner sep=0.75pt]  [color={rgb, 255:red, 208; green, 2; blue, 27 }  ,opacity=1 ]  {$| \overline{1}_{x_{B}} \rangle $};
\draw (58,4.4) node [anchor=north west][inner sep=0.75pt]  [color={rgb, 255:red, 248; green, 231; blue, 28 }  ,opacity=1 ]  {$| \overline{0} '_{x_{B}} \rangle $};
\draw (198,73.4) node [anchor=north west][inner sep=0.75pt]  [color={rgb, 255:red, 248; green, 231; blue, 28 }  ,opacity=1 ]  {$| \overline{1} '_{x_{B}} \rangle $};
\draw (371,72.4) node [anchor=north west][inner sep=0.75pt]  [color={rgb, 255:red, 248; green, 231; blue, 28 }  ,opacity=1 ]  {$| \overline{1} '_{x_{B}} \rangle $};
\draw (341,34.4) node [anchor=north west][inner sep=0.75pt]  [color={rgb, 255:red, 248; green, 231; blue, 28 }  ,opacity=1 ]  {$| \overline{0} '_{x_{B}} \rangle $};
\draw (289,34.4) node [anchor=north west][inner sep=0.75pt]  [color={rgb, 255:red, 126; green, 211; blue, 33 }  ,opacity=1 ]  {$| \overline{0} ''_{x_{B}} \rangle $};
\draw (374,104.4) node [anchor=north west][inner sep=0.75pt]  [color={rgb, 255:red, 126; green, 211; blue, 33 }  ,opacity=1 ]  {$| \overline{1} ''_{x_{B}} \rangle $};

\end{tikzpicture}
    \caption{Schematics of the change of bases $|\bar i\rangle$ under the action of $x_A$ and $x_B$. Black vectors represent the code states $|\bar 0\rangle$ and $|\bar 1\rangle$. Under $x_A$ or $x_B$, the code states $|\bar i\rangle$ are mapped to $|\bar i_{x_A}\rangle$ (blue) or $|\bar i_{x_B}\rangle$ (red), respectively. We further define $|\bar i'_{x_B}\rangle$ (yellow) as the component of $|\bar i_{x_B}\rangle$ orthogonal to $|\bar 0_{x_A}\rangle$ and $|\bar 1_{x_A}\rangle$. Finally, since $|\bar 0'_{x_B}\rangle$ and $|\bar 1'_{x_B}\rangle$ are not orthogonal, we perform a symmetric orthogonalization to obtain $|\bar 0''_{x_B}\rangle$ and $|\bar 1''_{x_B}\rangle$ (green).}
    \label{fig:imperfect correction bases}
\end{figure}

Assume that the following error conditions hold:
\begin{align}
\label{eq:bipartite cond}
    \Pi x_A\otimes I\Pi =a_1\Pi, \quad \Pi x_A^2\otimes I\Pi =(a_1^2 +a_2^2) \Pi, \nonumber\\
     \quad \Pi I\otimes x_B\Pi =b_1 \Pi, \quad \Pi I\otimes x_B^2\Pi =(b_1^2 +b_2^2) \Pi, 
\end{align}
where $a_1,a_2,b_1,b_2$ are real numbers and we assume $a_2,b_2\neq 0$ for convenience. Note that $a_2=\sqrt{\langle x_A^2\rangle-\langle x_A\rangle^2}=\Delta x_A$ is just the standard deviation, and similarly we have $b_2=\Delta x_B$. The first two conditions imply that if we define the normalized basis
$|\bar i_{x_A}\rangle\propto (I-|\bar i\rangle\langle \bar i|)x_A|\bar i\rangle$,
then $\langle \bar i| \bar j_{x_A}\rangle=0$ and $\langle \bar i_{x_A} | \bar j_{x_A}\rangle=\delta_{ij}$, and moreover
$x_A|\bar i\rangle = a_1|\bar i\rangle + a_2|\bar i_{x_A}\rangle$.
We thus define the Krylov space $\mathcal{C}_{x_A}=\{|\bar i_{x_A}\rangle\}$, see also Fig.~\ref{fig:imperfect correction bases}. This subspace is orthogonal to the code space $\mathcal{C}$, and the states $\{|\bar i_{x_A}\rangle\}$ are mutually orthogonal. These properties imply that the action of an $x_A$ error transfers the code state into $\mathcal{C}_{x_A}$ in a distinguishable manner, so the error can be corrected.

Similarly, define
$|\bar i_{x_B}\rangle\propto (I-|\bar i\rangle\langle \bar i|)x_B|\bar i\rangle$
and $\mathcal{C}_{x_B}=\{|\bar i_{x_B}\rangle\}$. The last two conditions imply $\langle \bar i| \bar j_{x_B}\rangle=0$, $\langle \bar i_{x_B} | \bar j_{x_B}\rangle=\delta_{ij}$ and
$x_B|\bar i\rangle = b_1|\bar i\rangle + b_2|\bar i_{x_B}\rangle$,
see also Fig.~\ref{fig:imperfect correction bases}. What remains undetermined is the relationship between $\mathcal{C}_{x_A}$ and $\mathcal{C}_{x_B}$. We therefore introduce the projectors $\Pi_{x_A}=\sum_i |\bar i_{x_A}\rangle \langle \bar i_{x_A}|$ and $\Pi_{x_B}=\sum_i |\bar i_{x_B}\rangle \langle \bar i_{x_B}|$.

To ensure that $x_A\otimes x_B$ drives non-trivial dynamics, $\mathcal{C}_{x_A}$ and $\mathcal{C}_{x_B}$ cannot be mutually orthogonal. We therefore study their overlap. Without loss of generality, we assume
\begin{align}
\label{eq:O1O2 mat ele imperfect nonlocal}
     \Pi x_A\otimes x_B\Pi = (a_1 b_1 +a_2 c_1 ) \Pi+ a_2 c_2 \sigma^x,
\end{align}
where $c_1,c_2$ are real numbers. Otherwise, one may rotate the basis of $\mathcal{C}$ so that the matrix elements take this form. The presence of the $\sigma^x$ term implies that $|\bar i_{x_A}\rangle$ and $|\bar j_{x_B}\rangle$ with $i\neq j$ are not mutually orthogonal, hence $|\bar i_{x_A}\rangle$ can be misidentified as $|\bar j_{x_B}\rangle$. This is the origin of the logical error discussed below. We first determine the relation between $\mathcal{C}_{x_A}$ and $\mathcal{C}_{x_B}$.

Note that $x_A\otimes I$ and $I\otimes x_B$ commute, implying
\begin{equation}
    0=\langle \bar i| [x_A\otimes I, I\otimes x_B]| \bar j\rangle = a_2 b_2(\langle \bar i_{x_A}| \bar j_{x_B}\rangle -\langle \bar i_{x_B}| \bar j_{x_A}\rangle).
\end{equation}
Define the component of $\mathcal{C}_{x_B}$ that is orthogonal to $\mathcal{C}_{x_A}$ by
$|\bar i_{x_B}'\rangle\propto (I-|\bar i\rangle\langle \bar i|-\Pi_{x_A})x_B|\bar i\rangle$,
see also Fig.~\ref{fig:imperfect correction bases}. One can then solve
\begin{equation}
    x_B|\bar i\rangle =b_1 |\bar i\rangle +b_2 |\bar i_{x_B}\rangle= b_1 |\bar i\rangle +c_1 |\bar i_{x_A}\rangle +c_2 |\overline{i\oplus 1}_{x_A}\rangle +c_3 |\bar i_{x_B}'\rangle,
\end{equation}
where $c_3=\sqrt{b_2^2-c_1^2-c_2^2}$. Without loss of generality, we consider $c_1,c_2,c_3\ge 0$. However, $|\bar 0_{x_B}'\rangle$ and $|\bar 1_{x_B}'\rangle$ are not orthogonal: enforcing that $|\bar 0_{x_B}\rangle$ is orthogonal to $|\bar 1_{x_B}\rangle$ implies
$\langle \bar 0_{x_B}'| \bar 1_{x_B}'\rangle = -z= -2 c_1 c_2 / c_3^2\le 0$.
We therefore introduce a symmetric orthogonal basis $|\bar i_{x_B}''\rangle$, see also Fig.~\ref{fig:imperfect correction bases}:
\begin{align}
    & |\bar 0_{x_B}'\rangle = \sqrt{\frac{1+\sqrt{1-z^2}}{2}}|\bar 0_{x_B}''\rangle-\sqrt{\frac{1-\sqrt{1-z^2}}{2}}|\bar 1_{x_B}''\rangle,\\
    & |\bar 1_{x_B}'\rangle =- \sqrt{\frac{1-\sqrt{1-z^2}}{2}}|\bar 0_{x_B}''\rangle+\sqrt{\frac{1+\sqrt{1-z^2}}{2}}|\bar 1_{x_B}''\rangle.
\end{align}

Both $\mathcal{C}_{x_A}=\{|\bar i_{x_A}\rangle\}$ and $\mathcal{C}_{x_B}=\{|\bar i_{x_B}\rangle\}$ form orthogonal bases, suggesting that $x_A$ and $x_B$ errors can, in principle, be corrected individually by mapping $|\bar i_{x_A}\rangle$ or $|\bar i_{x_B}\rangle$ back to $|\bar i\rangle$. However, the non-orthogonality induced by the $\sigma^x$ term in Eq.~\eqref{eq:O1O2 mat ele imperfect nonlocal} prevents both errors from being corrected unambiguously within a single recovery procedure. Concretely: (i) $\Pi_{x_A}x_B|\bar 0\rangle=c_1|\bar 0_{x_A}\rangle + c_2 |\bar 1_{x_A}\rangle$ and $\Pi_{x_A}x_B|\bar 1\rangle=c_1|\bar 1_{x_A}\rangle + c_2 |\bar 0_{x_A}\rangle$ are not orthogonal; (ii) if an $x_B$ error occurs and we apply a recovery that maps $|\bar i_{x_A}\rangle\mapsto|\bar i\rangle$, then the term proportional to $c_2$ produces a $\sigma^x$ logical error. Equivalently, one cannot distinguish whether $|\bar 0_{x_A}\rangle$ originated from $x_A|\bar 0\rangle$ or from $x_B|\bar 1\rangle$, and a misidentification leads to an incorrect recovery. Similarly, $|\bar 0_{x_B}'\rangle$ and $|\bar 1_{x_B}'\rangle$ are not orthogonal and therefore cannot be corrected unambiguously. Conversely, if one attempts to correct $x_B$ perfectly, then $x_A$ cannot be corrected perfectly. This tradeoff is a direct consequence of requiring $\langle \bar i|x_A\otimes x_B|\bar j\rangle \not \propto \delta_{ij}$, i.e. that $x_A \otimes x_B$ generate a non-trivial unitary dynamics. In this regime, the $\sigma^x$ logical error is unavoidable.

Given the orthogonality structure above, we introduce the recovery map
\begin{equation}
\label{eq: nonlocal EC}
    R(\rho)=\sum_{i=0,1,2} R_i \rho R_i^\dagger, \text{ where } R_0=\Pi, R_1= \sum_{i=0,1}|\bar i\rangle\langle \bar i_{x_A}|, R_2=\sum_{i=0,1}|\bar i\rangle\langle \bar i''_{x_B}|.
\end{equation}
Note that this recovery map is actually not trace-preserving, also only for pedagogical purpose. In principle, one should take the recovery map supplemented with additional Kraus operators, such that it is trace-preserving. But $\mathrm{Tr}R(\mathcal{L}(\rho))=0$, so $R$ here won't cause any leaking of population, thus it does not affect our discussion here without those additional Kraus operators. A direct calculation yields the corresponding effective error-corrected dynamics
\begin{align}
\label{eq:imperfect nonlocal ec effL}
    R(\mathcal{L}_{x_A,x_B}(\rho))=& -i \lambda_{\mathrm{eff}}[  \sigma^x, \rho] + \Gamma_{\mathrm{eff}}(\sigma^x \rho \sigma^x -\rho),
\end{align}
where $\lambda_{\mathrm{eff}}=\lambda a_2 c_2$ and $\Gamma_{\mathrm{eff}}=\Gamma(c_2^2+c_3^2(1-\sqrt{1-z^2})/2)$. We therefore obtain a damped Rabi oscillator on the logical qubit: the coherent oscillation reflects the presence of the nonlocal interaction, while the damping originates from the residual logical $\sigma^x$ error.

The resulting channel can be written explicitly as
\begin{align}
    e^{R\circ \mathcal{L}T}(\rho)=&\frac{1+e^{-2\Gamma_{\mathrm{eff}}T}\cos(2\lambda_{\mathrm{eff}}T)}{2}\rho + \frac{1-e^{-2\Gamma_{\mathrm{eff}}T}\cos(2\lambda_{\mathrm{eff}}T)}{2}\sigma^x\rho\sigma^x\\
    & - \frac{i}{2} e^{-2\Gamma_{\mathrm{eff}}T}\sin(2\lambda_{\mathrm{eff}}T) \sigma^x \rho + \frac{i}{2} e^{-2\Gamma_{\mathrm{eff}}T}\sin(2\lambda_{\mathrm{eff}}T) \rho \sigma^x 
\end{align}
Therefore, $\lambda_{\mathrm{eff}}$ must be sufficiently larger than $\Gamma_{\mathrm{eff}}$ for the coherent Rabi flopping to remain visible. The ratio of these two rates is
\begin{align}
    \frac{\Gamma_{\mathrm{eff}}}{\lambda_{\mathrm{eff}}}=&\frac{\Gamma}{\lambda}\frac{c_2^2+c_3^2 \frac{1-\sqrt{1-z^2}}{2} }{a_2 c_2}\nonumber\\
    =&\frac{\Gamma}{\lambda}\frac{c_2^2+\frac{2c_1 c_2}{z} \frac{1-\sqrt{1-z^2}}{2} }{a_2 c_2}\nonumber\\
    =&\frac{\Gamma}{\lambda}\frac{c_2+\frac{1}{z} \frac{-c_2 + \sqrt{c_2^2(1-z^2)+b_2^2 z^2}}{z} (1-\sqrt{1-z^2}) }{a_2 }\nonumber\\
    =&\frac{\Gamma}{\lambda}\frac{\frac{c_2}{b_2}+ \frac{-\frac{c_2}{b_2} + \sqrt{\frac{c_2^2}{b_2^2}(1-z^2)+z^2}}{z^2} (1-\sqrt{1-z^2}) }{\frac{a_2}{b_2}}
\end{align}
where we used $z= 2c_1 c_2 /c_3^2$ and $c_1^2 + c_2^2 + c_3^2 = b_2^2$ to solve for $c_1$. One can verify that for $0\le z \le 1$ and $0\le c_2/b_2 \le 1$, the numerator is a monotonically increasing function ranging from $0$ (attained at $z=c_2/b_2=0$, i.e. $c_2=c_3=0$, $c_1=b_2$, so that $x_B$ has the same effect on the codewords as $x_A$) to $1$. In summary,
\begin{equation}
    0\le \frac{\Gamma_{\mathrm{eff}}}{\lambda_{\mathrm{eff}}}\le \frac{\Gamma b_2}{\lambda a_2} = \frac{\Gamma  \Delta x_B}{\lambda \Delta x_A}.
\end{equation}
Note also that when $\Gamma_{\mathrm{eff}}/\lambda_{\mathrm{eff}}$ approaches $0$, the interaction rate $\lambda_{\mathrm{eff}}\propto c_2$ also approaches $0$.\footnote{Note that the upper bound is not symmetric between $x_A$ and $x_B$. This asymmetry arises because we correct the $x_A$ error completely while correcting the $x_B$ error only partially. For our purposes, this upper bound is less important than the possibility of suppressing the error ratio.} To leading order, $\lambda_{\mathrm{eff}}\propto c_2$ while $\Gamma_{\mathrm{eff}}\propto c_2^2$, so the error rate decreases faster than the interaction strength. \textit{In conclusion, for a nonlocal code space (potentially equipped with nonlocal recovery), the ratio of error rate to interaction strength can be made arbitrarily small, at the price of simultaneously weakening the interaction.} In other words, the signal-to-noise ratio can be magnified arbitrarily with this approach.

A natural question is whether such a code can be realized in a local fashion. This question has two aspects: whether code-state initialization can be implemented locally, and whether the recovery operation can be implemented as a separable channel.

First, suppose the code basis states are product states, i.e. $|\bar 0\rangle=|\alpha_0\rangle_A\otimes |\beta_0\rangle_B$ and $|\bar 1\rangle=|\alpha_1\rangle_A\otimes |\beta_1\rangle_B$. A linear combination of two product states is a product state if and only if $|\alpha_0\rangle_A=|\alpha_1\rangle_A$ or $|\beta_0\rangle_B=|\beta_1\rangle_B$. Assume the latter. Then the code space is a tensor product of a qubit supported on $\mathcal{H}_A$ and a single fixed state $|\beta_0\rangle_B$ on $\mathcal{H}_B$. In this case, as discussed in App.~\ref{app:Local imperfect QEC code}, the four conditions in Eq.~\eqref{eq:bipartite cond} already imply $\langle \bar i|x_A\otimes x_B|\bar j\rangle\propto \delta_{ij}$, i.e. the effective dynamics is trivial on the code space. Therefore, for the matrix elements in Eq.~\eqref{eq:bipartite cond} and Eq.~\eqref{eq:O1O2 mat ele imperfect nonlocal} to hold while retaining non-trivial dynamics, the code subspace $\mathcal{C}$ must contain entangled states. Moreover, if a code state is a product state (e.g. $|\bar 0\rangle$), then $\langle \bar 0|x_A\otimes x_B|\bar 0\rangle = \langle \bar 0|x_A\otimes I|\bar 0\rangle \langle \bar 0|I\otimes x_B|\bar 0\rangle$ implies $c_1=0$, and hence the covariance vanishes for that state: $\langle x_A\otimes x_B\rangle-\langle x_A\rangle \langle x_B\rangle=0$. This already implies $\Gamma_{\mathrm{eff}}/\lambda_{\mathrm{eff}}=\Gamma c_2/(\lambda a_2)$.

Second, we consider whether the recovery operation can be implemented by a separable channel. As discussed above, in the limit $c_2\rightarrow 0$ the coherent Rabi oscillation dominates over the residual $\sigma^x$ errors. If the states in $\mathcal{C}$ do not all carry the same amount of entanglement, then the Rabi oscillation produces an oscillatory entanglement dynamics that necessarily includes an entanglement growth process. In this case, even if one starts from a separable Lindbladian (e.g. $\Gamma>\lambda$), the generation of entanglement implies that the recovery channel $R$ cannot be separable.

In summary, given the operator matrix elements in Eq.~\eqref{eq:bipartite cond} and Eq.~\eqref{eq:O1O2 mat ele imperfect nonlocal}, and since $\mathcal{C}$ cannot consist solely of separable states, we are left with two possibilities: (i) $\mathcal{C}$ contains at least one separable state, so that initialization can be implemented locally, but the presence of other entangled states implies that the effective dynamics can generate entanglement, hence the recovery map $R$ is non-separable; (ii) all code states in $\mathcal{C}$ have non-zero entanglement.  Then the initialization must be non-separable because all code states are entangled.   Therefore, the error-correction protocol in this section cannot be implemented in a fully separable manner.

\subsection{Concatenated QEC for suppressing logical $\sigma_z$ errors}\label{app:Alternative QEC toolkit}

The local code discussed above generally produces both logical $\sigma_x$ and $\sigma_z$ contributions. In parameter regimes where the $\sigma_z$ sector dominates, it is useful to consider additional layers of encoding, for a clearer separability classification. As an illustrative example, we show how a repetition-code construction distributed across multiple masses can suppress logical $\sigma_z$ dynamics while preserving the logical $\sigma_x$ interaction and the associated separability classification.

As discussed above, for the Fock-state code, $\sigma^z$-type errors are typically stronger than the $\sigma^x$-type interactions and errors. Here we consider introducing multiple masses and dividing them into two parties $A$ and $B$. We then define a repetition code across all masses in $A$ and across all masses in $B$. We will show that this repetition code can eliminate the $\sigma^z$-like interactions and errors, while preserving the $\sigma^x$-like nonlocal interactions between $A$ and $B$ and keeping the local $\sigma^x$ errors intact.

Consider a bipartition of $2N$ masses into parties $A$ and $B$, with each mass in a party labeled by $i$. Each mass is encoded into a qubit, for instance using the Fock-state encoding above, with codewords $\{|\bar 0\rangle _{A,i},|\bar 1\rangle _{A,i}\}$ and $\{|\bar 0\rangle _{B,i},|\bar 1\rangle _{B,i}\}$. We also define $\sigma^x=|\bar 1\rangle \langle \bar0|+|\bar 0\rangle \langle \bar 1|$ and $\sigma^z=|\bar 0\rangle \langle \bar 0|-|\bar 1\rangle \langle \bar 1|$. Without loss of generality, consider the following Lindbladian, which contains nonlocal $\sigma^z_A \otimes \sigma^z_B$ and $\sigma^x_A\otimes \sigma^x_B$ interactions together with local $\sigma^z$ and $\sigma^x$ errors:
\begin{align}
    \mathcal{L}(\rho)=&-i\sum_{i,j} \lambda_x [\sigma^x_{A,i}\otimes \sigma^x_{B,j},\rho]-i\sum_{i,j} \lambda_z [\sigma^z_{A,i}\otimes \sigma^z_{B,j},\rho]\nonumber\\
    &+ \Gamma_{x}\sum_i (\mathcal{D}[\sigma^x_{A,i}]+\mathcal{D}[\sigma^x_{B,i}])(\rho)+ \Gamma_{z}\sum_i (\mathcal{D}[\sigma^z_{A,i}]+\mathcal{D}[\sigma^z_{B,i}])(\rho).
\end{align}
We focus on the regime in which the $\sigma^z$ interaction rate and $\sigma^z$ error rate are of the same order, $\Gamma_{z}\sim \lambda_z$, and both are much larger than the corresponding $\sigma^x$ rates, i.e. $\Gamma_{z},\lambda_z \gg \Gamma_{x},\lambda_x$. We further assume that the $\sigma^x$ interaction rate is much larger than the $\sigma^x$ error rate, i.e. $\lambda_x \gg \Gamma_{x}$. In this situation, if one can correct $\sigma^z$ errors (and hence render the $\sigma^z$ interactions effectively trivial within the logical subspace), then it may be possible to observe entanglement generated by the $\sigma^x$ nonlocal interactions. In the following we show that this can be achieved.

In principle, $\sigma^z$ is orthogonal to $\sigma^x$ under the Frobenius inner product, $\operatorname{Tr} \sigma^z(\sigma^x)^\dagger=0$, so it should be possible to preserve the $\sigma^x$ dynamics while eliminating the $\sigma^z$ error. Let $|\overline{\pm}\rangle =(|\bar 0\rangle\pm|\bar 1\rangle )/\sqrt{2}$. This can be realized by the following quantum error-correction code, taking party $A$ as an example:
\begin{equation}
    |\bar{\bar{0}}\rangle_A = \frac{1}{\sqrt{2}}\left(\bigotimes_{i=1}^N |\overline{+}\rangle_{A,i}+\bigotimes_{i=1}^N |\overline{-}\rangle_{A,i}\right), \quad |\bar{\bar{1}}\rangle_A = \frac{1}{\sqrt{2}}\left(\bigotimes_{i=1}^N |\overline{+}\rangle_{A,i}-\bigotimes_{i=1}^N |\overline{-}\rangle_{A,i}\right).
\end{equation}
That is, we encode one logical qubit into the $N$ qubits defined on the $N$ masses in party $A$. We define the projector onto the code space as $\Pi_A=|\bar{\bar{0}}\rangle_A{}_A\langle \bar{\bar{0}}|+|\bar{\bar{1}}\rangle_A{}_A\langle \bar{\bar{1}}|$ and the recovery operation as
\begin{equation}
    R_A(\rho)=\Pi_A\rho \Pi_A + \sum_{i=1}^N \Pi_A \sigma^z_{A,i}\rho \sigma^z_{A,i} \Pi_A.
\end{equation}
For the $N$ masses in party $B$, the codewords and recovery operations are defined symmetrically. Note the recovery operation here is not trace-preserving, practically one should supplement it with additional Kraus operators, but it makes no difference for the effective dynamics here.

Suppose the state is initially supported on the code space, i.e. $\rho=\Pi_A\otimes \Pi_B \rho \Pi_A\otimes \Pi_B$. The corresponding error-corrected effective dynamics can then be computed as
\begin{align}
\label{eq:eff L multimass}
   & R_A\otimes R_B \circ\mathcal{L}(\mathcal{\rho})\nonumber\\
   =& -i\sum_{j} N\lambda_x R_B\left( [X_A\otimes \sigma^x_{B,j},\rho]  \right) +\Gamma_{x} R_B\left(  (N\mathcal{D}[X_A]+\sum_i\mathcal{D}[\sigma^x_{B,i}])(\rho) \right)\nonumber\\
    & +\Gamma_{z} R_B\left( -N\rho  + \sum_i\mathcal{D}[\sigma^z_{B,i}](\rho)  +  \sum_i\sum_{j=1}^N \delta_{ij}\rho\right)\nonumber\\
    =& -i N^2\lambda_x  [X_A\otimes X_B,\rho]  +N\Gamma_{x}  (\mathcal{D}[X_A]+\mathcal{D}[X_B])(\rho).
\end{align}
Therefore, with this second layer of error correction, it is sufficient to retain only the $\sigma^x_{A,i}\otimes \sigma^x_{B,j}$ nonlocal interaction and the local $\sigma^x$ errors. Moreover, if there are interactions among masses within party $A$, for instance a Hamiltonian term $\sigma^x_{A,i}\otimes \sigma^x_{A,j}$, then it becomes trivial under this encoding because $\sigma^x_{A,i}\otimes \sigma^x_{A,j}|\bar{\bar{i}}\rangle_A=|\bar{\bar{i}}\rangle$ is invariant. Beyond these errors, operators such as $\sigma^+=|\bar 1\rangle \langle \bar 0|=(1-\sigma^z)\sigma^x/2$ (see also App.~\ref{app:Effective error rates}) are also mapped to $X$ errors, as shown by
\begin{align}
    &R_A\left(\sigma^+_{A,i}\rho (\sigma^+_{A,i})^\dagger-\frac{1}{2}\{|\bar 0\rangle_{A,i}{}_{A,i}\langle \bar 0|,\rho\}\right)\displaybreak[1]\nonumber\\
    =& R_A\left(\frac{(1-\sigma^z_{A,i})\sigma^x_{A,i}}{2}\rho \frac{\sigma^x_{A,i}(1-\sigma^z_{A,i})}{2}-\frac{1}{2}\Big\{\frac{1+\sigma^z_{A,i}}{2},\rho\Big\}\right)\nonumber\\
     =& \frac{1}{2}(X_A \rho X_A -\rho).
\end{align}

We now study the separability classification of $\mathcal{L}$. Note that it can be rewritten as
\begin{align}
    \mathcal{L}(\rho)=&\sum_{i,j}  -i[\lambda_x\sigma^x_{A,i}\otimes \sigma^x_{B,j}+ \lambda_z\sigma^z_{A,i}\otimes \sigma^z_{B,j},\rho]+ \frac{\Gamma_{x}}{N} (\mathcal{D}[\sigma^x_{A,i}]+\mathcal{D}[\sigma^x_{B,j}])(\rho)+ \frac{\Gamma_{z}}{N} (\mathcal{D}[\sigma^z_{A,i}]+\mathcal{D}[\sigma^z_{B,j}])(\rho).
\end{align}
By Lemma~\ref{lemma:separability}, the $\sigma^x$ part of each term is non-separable if and only if $\lambda_x> \Gamma_x/N$. On the other hand, the error-corrected master equation in Eq.~\eqref{eq:eff L multimass} is non-separable if and only if $\lambda_x N^2 > \Gamma_x N$. Therefore, this QEC protocol is separable and preserves the separability of the underlying Lindbladian, while it may reduce the degree of non-separability of an initially non-separable Lindbladian. Consequently, to determine whether the system is entanglement-generating, it suffices to compare the rate of the $\sigma^x$-like nonlocal interaction with the rate of the $\sigma^x$-like local errors.

Applying this protocol to the decoherence channels in App.~\ref{app:Effective error rates}, for instance CSL decoherence channel in Eq.~\eqref{eq:LeffCSL}, we obtain a clean $\sigma^x$-axis-only Lindbladian
\begin{align}
    R_A\otimes R_B \circ\mathcal{L}_{\mathrm{eff,CSL}}(\mathcal{\rho})= N\left(\Gamma^{x}_{\mathrm{eff}}+\frac{\Gamma^+_{\mathrm{eff}}+\Gamma^-_{\mathrm{eff}}}{2}\right)  (\mathcal{D}[X_A]+\mathcal{D}[X_B])(\rho)+\sum_{ij}\Gamma^{xx}_{\mathrm{eff},ij}  (\mathcal{D}[X_A-X_B]-\mathcal{D}[X_A]-\mathcal{D}[X_B])(\rho),
\end{align}
with the price of increasing logical error rates, since 
$$
\Gamma^{x}_{\mathrm{eff}}+\frac{\Gamma^+_{\mathrm{eff}}+\Gamma^-_{\mathrm{eff}}}{2}\ge\Gamma^{x}_{\mathrm{eff}}+\frac{\Gamma^+_{\mathrm{eff}}\Gamma^-_{\mathrm{eff}}}{\Gamma^+_{\mathrm{eff}}+\Gamma^-_{\mathrm{eff}}}.
$$
This protocol can be a cleaner alternative for separability classification in App.~\ref{app:Separability classification}. Moreover, the separability classification  in Theorem~\ref{thm:classification_app} of this effective dynamics with concatenated QEC with $\eta>1$ or $\eta \le 1$ also coincides with the entangling criterion in Prop.~\ref{prop:entangling}. However, the separability classification is not faithfully kept, since this concatenated layer of QEC introduces more noise.

\newpage
\section{Application of the QEC framework to collapse models}\label{app:Calculation of the effective Lindbladian}

This appendix provides technical and supplementary details on the application of the quantum error-correction framework developed in App.~\ref{app:Local imperfect QEC code} to each of the collapse models considered in the main text. We begin by deriving the logical gravitational interaction generated within the encoded subspace and discuss the validity of the approximations used in its evaluation. We then construct model-specific recovery procedures, derive the corresponding effective logical dynamics, and analyze their separability properties. Since the different models can largely be treated using the same methodology, we collect the corresponding derivations here rather than repeating similar calculations in the main text. For completeness, we also discuss a small extension of the separability analysis that proves useful for analyzing one of the effective dynamics encountered below. Then we compare the resulting classifications across the different collapse models. Finally, we analytically prove the entanglement recoverability with respect to GRW/CSL model in Thm.~\ref{thm:recoverability}.

\subsection{Effective gravitational interaction}

Before analyzing the various collapse models, we determine the logical interaction generated by the Newtonian potential within the encoded subspace introduced in App.~\ref{app:Local imperfect QEC code}. This interaction constitutes the coherent component of the effective dynamics and serves as a common reference point for all recovery procedures considered below.

As discussed in App.~\ref{app:Local imperfect QEC code}, the QEC recovery channel has the form
\begin{equation}
    R(\rho_{AB})=R_A\otimes R_B(\rho_{AB}), R_\alpha(\rho_\alpha)=\sum_i R^i_\alpha\rho_\alpha R^{i\dagger}_\alpha,
\end{equation}
where $R_\alpha ^0=\Pi_\alpha$ is the projector onto the code space $\mathcal{C}_\alpha$. By direct calculation, the effective error-corrected unitary dynamics reads as follows, assuming that $\rho=\Pi^A\Pi^B\rho \Pi^A\Pi^B$ is supported on the code space:
\begin{equation}
    R\circ\mathcal{L}_{\mathrm{NG}}(\rho)=\frac{i}{\hbar}\Big[\Pi^A\Pi^B\frac{Gm_A m_B}{|d+x_A -x_B|}\Pi^A\Pi^B,\rho_{AB}\Big].
\end{equation}
Therefore, this part of the master equation is determined by the matrix elements of the Newtonian potential.

The Newtonian potential
$$
-\frac{G m_A m_B}{|d+x_A-x_B|}
$$
is singular when the positions of the two masses coincide, i.e. when $|d+x_A-x_B|=0$. Consequently, any nonzero wavefunction overlap can render the corresponding matrix elements ill-defined. Moreover, when the two particles approach each other at distances comparable to their size, one expects additional short-range interactions that are not captured by the Newtonian potential alone. To consistently isolate the long-range gravitational coupling, we therefore restrict our analysis to configurations in which the centers of mass of the two particles remain separated by at least twice their radius and introduce the following regularized Newtonian potential:
\begin{equation}
    f_{\mathrm{NG}}(|d+x_A-x_B|)=\begin{cases}
        -\frac{G m_A m_B}{|d+x_A-x_B|},\quad &|d+x_A-x_B|>2r_0,\\
        0,\quad &|d+x_A-x_B|\le 2r_0.
    \end{cases}
\end{equation}
We then compute the nonlocal matrix elements of this regularized potential. By symmetry of the potential and of the wavefunctions under exchange of the two masses, the following four off-diagonal matrix elements are equal:
\begin{align}
   \hbar \lambda^{xx}_{\mathrm{eff}}=&{}_A\langle \bar 0|{}_B\langle \bar 0|f_{\mathrm{NG}}(|d+x_A-x_B|)|\bar 1\rangle_A|\bar 1\rangle_B\nonumber\\
    =&{}_A\langle \bar 0|{}_B\langle \bar 1|f_{\mathrm{NG}}(|d+x_A-x_B|)|\bar 1\rangle_A|\bar 0\rangle_B\nonumber\\
    =&{}_A\langle \bar 1|{}_B\langle \bar 0|f_{\mathrm{NG}}(|d+x_A-x_B|)|\bar 0\rangle_A|\bar 1\rangle_B\nonumber\\
    =&{}_A\langle \bar 1|{}_B\langle \bar 1|f_{\mathrm{NG}}(|d+x_A-x_B|)|\bar 0\rangle_A|\bar 0\rangle_B,
\end{align}
where the superscript indicates that this matrix element contributes to the effective $\sigma^x_A\sigma^x_B$ interaction. Similarly, we also have $\sigma^x_A\sigma^z_B,\sigma^z_A\sigma^x_B$ and $\sigma^z_A\sigma^z_B$ interaction terms, and their interaction rates are determined as follows:
\begin{align}
    2\hbar \lambda^{xz}_{\mathrm{eff}}=&{}_A\langle \bar 0|{}_B\langle \bar 0|f_{\mathrm{NG}}(|d+x_A-x_B|)|\bar 1\rangle_A|\bar 0\rangle_B-{}_A\langle \bar 0|{}_B\langle \bar 1|f_{\mathrm{NG}}(|d+x_A-x_B|)|\bar 1\rangle_A|\bar 1\rangle_B\nonumber = -2\hbar \lambda^{zx}_{\mathrm{eff}}\\
    4\hbar \lambda^{zz}_{\mathrm{eff}}=&{}_A\langle \bar 0|{}_B\langle \bar 0|f_{\mathrm{NG}}(|d+x_A-x_B|)|\bar 0\rangle_A|\bar 0\rangle_B-{}_A\langle \bar 0|{}_B\langle \bar 1|f_{\mathrm{NG}}(|d+x_A-x_B|)|\bar 0\rangle_A|\bar 1\rangle_B\nonumber\\
    & -{}_A\langle \bar 1|{}_B\langle \bar 0|f_{\mathrm{NG}}(|d+x_A-x_B|)|\bar 1\rangle_A|\bar 0\rangle_B+{}_A\langle \bar 1|{}_B\langle \bar 1|f_{\mathrm{NG}}(|d+x_A-x_B|)|\bar 1\rangle_A|\bar 1\rangle_B
\end{align}

Consequently, the effective error-corrected unitary dynamics can be written as
\begin{equation}
\label{eq:NG lind}
  \mathcal{L}_{\mathrm{eff,NG}}(\rho)=  R\circ \mathcal{L}_{\mathrm{NG}}(\rho)=-i[\lambda_{\mathrm{eff}}^{xx}\sigma^x_A\otimes\sigma^x_B+\lambda_{\mathrm{eff}}^{xz}\sigma^x_A\otimes\sigma^z_B+\lambda_{\mathrm{eff}}^{zx}\sigma^z_A\otimes\sigma^x_B+\lambda_{\mathrm{eff}}^{zz}\sigma^z_A\otimes\sigma^z_B,\rho]+\cdots
\end{equation}
where we have omitted additional local Hamiltonian contributions, since they are irrelevant for the separability classification. We defer the numerical evaluation of $\lambda_{\mathrm{eff}}^{xx}, \lambda_{\mathrm{eff}}^{xz},\lambda_{\mathrm{eff}}^{zx}$ and $\lambda_{\mathrm{eff}}^{zz}$ to the subsection after the next one, where we directly compare it with the effective error rates.

\subsection{Validity of the non-contact approximation}

The effective interaction calculation above relies on a regularized Newtonian potential and assumes that the two masses never contact each other throughout the evolution. In this subsection, we quantify the contact probability between the particles and identify the parameter regime in which the regularized interaction provides a consistent description.

Without loss of generality, consider the two masses prepared in Fock states $|n\rangle_A$ and $|m\rangle_B$. The wavefunction of $|m\rangle_B$ is
\begin{equation}
    {}_B\langle m|x_B\rangle_B=\psi_m(x_B)=\frac{1}{\sqrt{ m!} (2\pi x_0^2)^\frac{1}{4}} e^{-\frac{x_B^2}{4 x_0^2}} H_{e_m}\left(\frac{x_B}{x_0}\right),
\end{equation}
where $H_{e_p}$ is the probabilist's Hermite polynomial and $x_0$ is the delocalization length of the ground state of the harmonic trap. Similarly, the wavefunction of $|n\rangle_A$ is ${}_A\langle n|x_A\rangle_A=\psi_n(x_A)$.

The probability that the two masses are separated by a distance no more than $\delta d$ can then be computed as
\begin{equation}
\label{eq:app overlap}
    P(\delta d)=\int |\psi_m(x_B)|^2|\psi_n(x_A)|^2 \Theta(\delta d-|d+x_A-x_B|)\dd x_B \dd x_A = {}_A\langle n|{}_B\langle m| \mathbbm{1}_{|d+x_A-x_B|\le \delta d}|n\rangle_A |m\rangle_B,
\end{equation}
where $\Theta(x)=1$ for $x>0$ and $\Theta(x)=0$ for $x\le 0$. Without loss of generality, we set $\delta d$ to be twice the mass radius, i.e. $\delta d=2r_0$.

\begin{figure}
    \centering
    \includegraphics[width=0.6\textwidth]{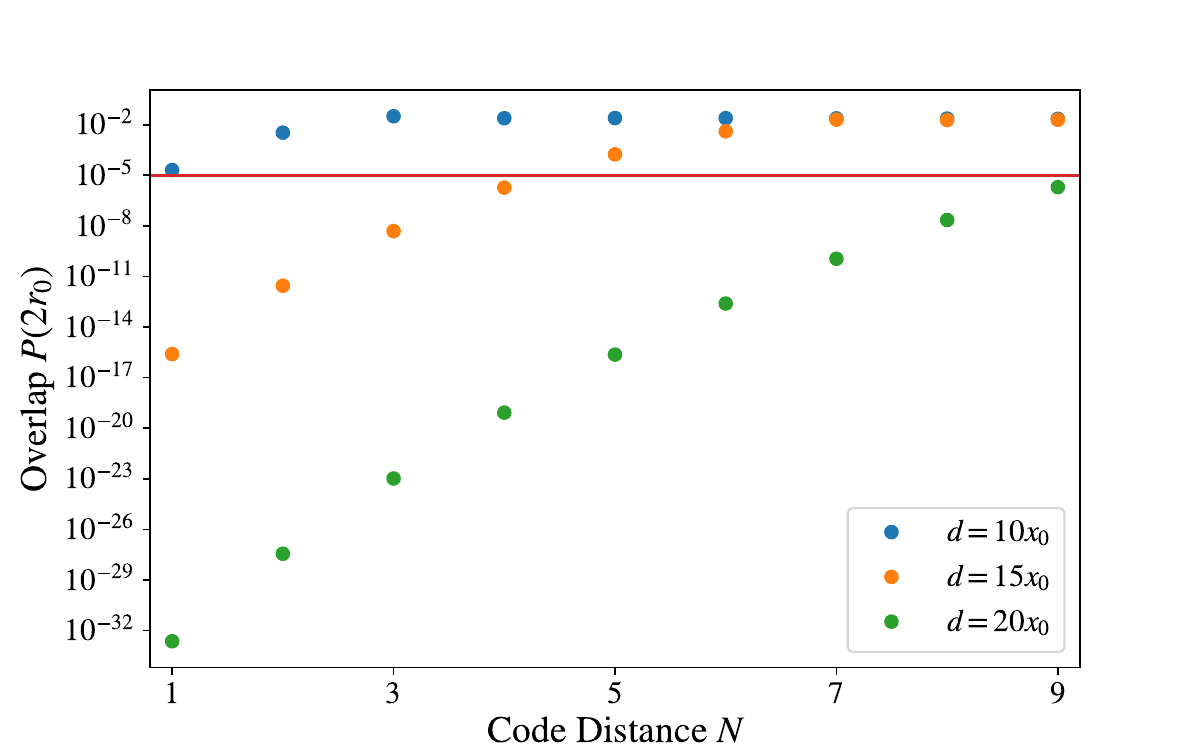}
    \caption{The contact probability $P(2 r_0)$ between the wavefunctions of the two masses, prepared in the state $|\bar 1\rangle_A|\bar 1\rangle_B=| 2N+1\rangle _A| 2N+1\rangle_B$. We impose the bound $P(2 r_0)\le 10^{-5}$ to determine up to which code distance $N$ the two wavefunctions remain effectively non-contacting.}
    \label{fig:overlap}
\end{figure}

In the numerics, we consider separations $d=10x_0$, $15x_0$, and $20x_0$. We take the radius to be $r_0\approx 0.238 x_0$, which will be used throughout the following calculations. We consider the two masses prepared in the state $|\bar 1\rangle_A|\bar 1\rangle_B=| 2N+1\rangle _A| 2N+1\rangle_B$. We impose the conservative bound $P(2r_0)\le 10^{-5}$ and identify the values of the code distance $N$ for which the two masses remain effectively non-contacting. The results are shown in Fig.~\ref{fig:overlap}. In particular, for $d=20 x_0$ we find that the code distance must satisfy $N\le 9$. For proofs relying on the asymptotic limit $N\rightarrow\infty$, $d$ and $x_0$ must be adapted as $N$ increases to keep the contact probability below the prescribed threshold. In App.~\ref{app:Rigorous proof of the entanglement recoverability}, we provide a formal derivation of this requirement.

\subsection{Recovery procedures and effective dynamics for different collapse models}\label{app:Effective error rates}

We now apply the quantum error-correction framework to the collapse and decoherence models considered in the main text and in App.~\ref{app:Local imperfect QEC code}. For each model, we construct an appropriate recovery procedure, derive the resulting effective logical dynamics, and analyze the corresponding separability properties with respect to the separability parameter $\eta$, see Thm.~\ref{thm:classification_app} in App.~\ref{app:Locality of the effective Lindbladian}.

To that end, we explicitly calculate the effective logical error rates for the various models introduced in App.~\ref{app:Collapse models}. These effective error rates can be extracted from the matrix elements of the effective error-corrected Lindbladian, see App.~\ref{app:Logical error rates}.

Specifically, the error-corrected Lindbladian is
\begin{equation}
   \mathcal{L}_\mathrm{eff,err}(\rho)= R\circ\mathcal{L}_{\mathrm{err}}(\rho)=\sum_{i,j}R_A^i R_B^j \mathcal{L}_{\mathrm{err}}(\rho)R^{i\dagger}_A R^{j\dagger}_B.
\end{equation}
Since the recovery channel $R$ maps error states back into the code subspace $\mathcal{C}=\mathcal{C}_A\otimes \mathcal{C}_B$, it suffices to analyze the effective decoherence generator within the two-qubit logical code space.

\subsubsection{KTM model}

We begin with the KTM model. Since this dynamics is constructed from local measurement and feedback, it provides a useful benchmark for validating both the recovery procedure and the separability framework developed in the main text.

Consider the recovery procedure for $\mathcal{L}_{\mathrm{KTM}}$ introduced in Eq.~\eqref{eq:KTM} in App.~\ref{app:Collapse models}. Here we choose the local Fock code space with code distance $0$, i.e. $\mathcal{C}=\mathcal{C}_A\otimes \mathcal{C}_B$ with $\mathcal{C}_\alpha =\{|\bar 0\rangle_\alpha =|0\rangle_\alpha ,|\bar 1\rangle_\alpha =|1\rangle_\alpha \}$. We denote the projector onto the code space by $\Pi_\alpha =\sum_{i=0,1}|\bar i\rangle_\alpha {}_\alpha \langle \bar i|$. A direct computation gives
\begin{equation}
    x_A|\bar 0\rangle_A = x_0|\bar 1\rangle, \quad x_A |\bar 1\rangle_A = x_0 (|\bar 0\rangle_A + \sqrt{2}|2\rangle_A).
\end{equation}
Thus, the codeword $|\bar 1\rangle_A$ is mapped partly to $|2\rangle_A$, which lies outside $\mathcal{C}_A$. We therefore introduce a recovery operation that corrects this leakage:
\begin{equation}
    R_\alpha(\rho)=R^0_\alpha\rho R^{0\dagger}_\alpha+ R^1_\alpha \rho R^{1\dagger}_\alpha= \Pi_\alpha \rho \Pi_\alpha +|\bar 1\rangle_\alpha{}_\alpha\langle 2| \rho |2\rangle_\alpha {}_\alpha\langle \bar 1|,
\end{equation}
where $\alpha \in \{A,B\}$.  Note the recovery operation here is not trace-preserving just for the convenience of discussion, only introduced in this subsection for pedagogical purpose. Practically one should supplement it with additional Kraus operators, but it makes no difference for the discussion of effective dynamics here. The corresponding effective error-corrected Lindbladian is then
\begin{align}
    R\circ \mathcal{L}_\mathrm{KTM}(\rho)  =&-i\frac{2Gm_Am_B x_0^2}{\hbar d^3}[\sigma^x_A\otimes \sigma^x_B, \rho] +\frac{2Gm_Am_B x_0^2}{\hbar d^3}(\mathcal{D}[\sigma^x_A] + \frac{1}{2} \mathcal{D}[\sigma^z_A]+\mathcal{D}[\sigma^x_B] + \frac{1}{2} \mathcal{D}[\sigma^z_B])(\rho)
    \label{eq: L1eff}
\end{align}
where we have imposed that the state is supported on the code space, i.e. $\Pi_A\otimes \Pi_B \rho \Pi_A\otimes \Pi_B=\rho$, and
\begin{equation}
    \sigma^x_\alpha =|\bar 0\rangle_\alpha {}_\alpha \langle \bar 1|+|\bar 1\rangle_\alpha {}_\alpha \langle \bar 0|, \quad \sigma^z_\alpha =|\bar 0\rangle_\alpha {}_\alpha \langle \bar 0|-|\bar 1\rangle_\alpha {}_\alpha \langle \bar 1|.
\end{equation}
We also assume that the KTM error rate is set to its minimum. Note that although the leakage component has been corrected, logical errors within the code space remain.

We now determine the separability classification of the error-corrected generator. Direct application of  Thm.~\ref{thm:classification_app} implies that the effective Lindbladian is separable, since its corresponding $\eta =1$. This conclusion is consistent with the fact that $\mathcal{L}_\mathrm{KTM}$ is, by construction, an LOCC dynamics.

\subsubsection{GRW model}

We next analyze the effective logical dynamics generated by the GRW collapse model. The primary objective is to determine how the effective interaction and logical error rates scale with the code distance and whether error correction can enhance the visibility of non-separable dynamics.

As discussed in App.~\ref{app:Local imperfect QEC code}, the QEC Kraus operators $R_\alpha ^i$ are defined such that errors of the form $x^i_\alpha$ can be corrected. Consider a code space with code distance $N$, i.e. $\mathcal{C}_\alpha =\{|\bar 0\rangle_\alpha =|0\rangle_\alpha ,|\bar 1\rangle_\alpha =|2N+1\rangle_\alpha \}$. We explicitly define the local Kraus operators as follows:
\begin{align}
\label{eq: app recovery GRW}
    R^0= &\Pi = |\bar 0\rangle \langle \bar 0|+ |\bar 1\rangle \langle \bar 1|, \quad R^i = \begin{cases}
        |\bar 0\rangle\langle i|+|\bar 1\rangle (c_i\langle 2N+1-i|+d_i\langle 2N+1+i|), &1\le i\le N,\\
        |\bar 1\rangle (d_{i-N}\langle 3N+1-i|-c_{i-N}\langle N+1+i|), &N+1\le i \le 2N,\\
        |\bar 1\rangle \langle  i+N+1|, &i>2N,
    \end{cases}\\
    \text{where }c_k=&\frac{\langle k|\mathcal{L}_{\mathrm{GRW}}(|\bar 0\rangle \langle \bar 1|)|2N+1-k\rangle }{\sqrt{(\langle k|\mathcal{L}_{\mathrm{GRW}}(|\bar 0\rangle \langle \bar 1|)|2N+1-k\rangle)^2+(\langle k|\mathcal{L}_{\mathrm{GRW}}(|\bar 0\rangle \langle \bar 1|)|2N+1+k\rangle)^2}}\nonumber\\
    d_k=&\frac{\langle k|\mathcal{L}_{\mathrm{GRW}}(|\bar 0\rangle \langle \bar 1|)|2N+1+k\rangle }{\sqrt{(\langle k|\mathcal{L}_{\mathrm{GRW}}(|\bar 0\rangle \langle \bar 1|)|2N+1-k\rangle)^2+(\langle k|\mathcal{L}_{\mathrm{GRW}}(|\bar 0\rangle \langle \bar 1|)|2N+1+k\rangle)^2}}.
    \label{eq: GRW ck dk}
\end{align}
Note that for $i>2N$, $R^i$ ensures that the QEC channel is trace-preserving and maps the full Fock space back into the code subspace.

We then calculate the matrix elements of the effective Lindbladian for a single mass:
\begin{align}
\label{eq: app GRW transition}
&\langle \bar i|R\circ\mathcal{L}_{\mathrm{GRW}}(|\bar j\rangle \langle \bar k|)|\bar l\rangle =n\times\lambda_{\mathrm{GRW}} \left( -\delta_{ij}\delta_{kl}+\sum_m\int d x_1 d x_2  e^{-\frac{(x_1-x_2)^2}{4r_\mathrm{C}^2}}\langle \bar i|R_m|x_1\rangle \langle x_1|\bar j\rangle \langle \bar k|x_2\rangle \langle x_2|R_m^\dagger|\bar l\rangle \right),
\end{align}
by performing the integration over the wavefunctions. These integrals can be evaluated analytically because they involve only Gaussian factors multiplied by polynomials. We define the logical operators $\sigma^+=|\bar 1\rangle\langle \bar 0|$, $\sigma^-=|\bar 0\rangle \langle \bar 1|$, $\sigma^x=\sigma^+ +\sigma^-$, and $\sigma^z=|\bar 0\rangle \langle \bar 0|-|\bar 1\rangle \langle \bar 1|$. Using the symmetry of the matrix elements, the effective error dynamics for a single mass can be written, for an initial state $\rho$ supported on the code subspace, as
\begin{align}
\label{eq:effLind GRW}
   \mathcal{L}_{\text{eff,GRW}}^\alpha (\rho)= R_\alpha \circ\mathcal{L}_{\mathrm{GRW}}^\alpha (\rho)= \Gamma_{\mathrm{eff}}^+\mathcal{D}[\sigma^+_\alpha ](\rho)+\Gamma_{\mathrm{eff}}^-\mathcal{D}[\sigma^-_\alpha ] (\rho)+\Gamma_{\mathrm{eff}}^x\mathcal{D}[\sigma^x_\alpha ](\rho)+\Gamma_{\mathrm{eff}}^z\mathcal{D}[\sigma^z_\alpha ](\rho).
\end{align}

\begin{figure}
    \centering
    \includegraphics[width=\textwidth]{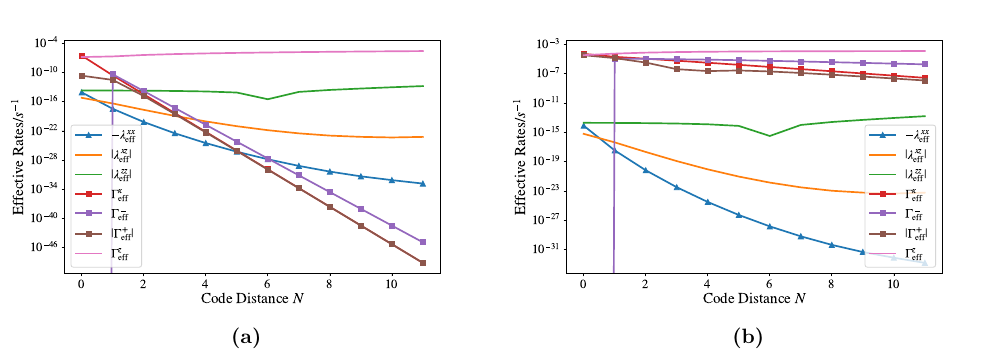}
    \caption{The effective interaction rate $\lambda_{\mathrm{eff}}^{xx}$, $\lambda^{xz}_\mathrm{eff}, \lambda^{zz}_\mathrm{eff}$ and effective error rates $\Gamma_{\mathrm{eff}}^x,\Gamma_{\mathrm{eff}}^-,\Gamma_{\mathrm{eff}}^+,\Gamma_{\mathrm{eff}}^z$ of the GRW model as a function of the code distance $N$. (a) Results for a large localization length $r_C$. We observe that $\lambda_{\mathrm{eff}}^{xx}$ decreases with $N$, but more slowly than $\Gamma_{\mathrm{eff}}^x$, $\Gamma_{\mathrm{eff}}^-$, and $\Gamma_{\mathrm{eff}}^+$. Consequently, the effective dynamics is non-separable for $N\ge 5$. (b) Results for a small localization length $r_C$.}
    \label{fig:GRW}
\end{figure}

We now illustrate these effective interaction and error rates with a concrete example. We first take $r_\mathrm{C}=2\times 10^{-6}\mathrm{m}$ and $\lambda_{\mathrm{GRW}}=10^{-8}\mathrm{s^{-1}}$ for the GRW model (this parameter choice is already excluded experimentally \cite{vinante2020narrowing,carlesso2022present}). We set $x_0=2\times10^{-8}\mathrm{m}$, $m_A=m_B=10^{-21}\mathrm{kg}$, and $d=4\times 10^{-7}\mathrm{m}$. We further assume that the mass is made of silica with density $\rho \approx 2.2\times 10^3 \mathrm{kg/m^3}$. The corresponding radius is $r_0=(3m_A/4\pi \rho)^{1/3}\approx 4.8\times 10^{-9}\mathrm{m}$. The mass of a silicon dioxide molecule is $9.98\times 10^{-26}\mathrm{kg}$, which implies a constituent number $n\approx 6\times 10^5$, since each silicon dioxide molecule contains sixty nucleons. This yields a relative strength $(4 r_C^2Gm_A m_B)/(\hbar \lambda_{\mathrm{GRW}}n d_{\min}^3 ) \approx 1.9\times 10^{-3}$, where $d_{\mathrm{min}}=2r_0$ is the minimal allowed distance between the two masses, implying that the error rate is much larger than the interaction scale, see also App.~\ref{app:Gravity-induced entanglement with spatially-superposed states}. For these parameters, the effective interaction rate and error rates are computed as a function of the code distance $N$, as shown in Fig.~\ref{fig:GRW}a. We observe that the $\sigma^x$-type error rates decrease exponentially faster with $N$ than the effective $\sigma^x_A\otimes\sigma^x_B$ interaction rate. According to Thm.~\ref{thm:classification_app}, 
the separability of the effective dynamics is determined by
\begin{equation}
    \eta_\mathrm{GRW} = \left\|  \begin{pmatrix}
            \frac{\lambda^{xx}_\mathrm{eff}}{\bar{\Gamma}^x_\mathrm{eff}} & -\frac{\lambda^{xz}_\mathrm{eff}}{\sqrt{\Gamma^z_\mathrm{eff}\bar{\Gamma}^x_\mathrm{eff}}} \\
            \frac{\lambda^{xz}_\mathrm{eff}}{\sqrt{\Gamma^z_\mathrm{eff}\bar{\Gamma}^x_\mathrm{eff}}} & \frac{\lambda^{zz}_\mathrm{eff}}{\Gamma^z_\mathrm{eff}}\\
        \end{pmatrix}\right\|_2,
\end{equation}
where $ \bar{\Gamma}_{\mathrm{eff}}^x =\Gamma^x_{\mathrm{eff}}+\Gamma_{\mathrm{eff}}^+\Gamma_{\mathrm{eff}}^-/(\Gamma_{\mathrm{eff}}^+ +\Gamma_{\mathrm{eff}}^-)$. According to the numerical results, the $\sigma^z_\alpha$-sector Lindbladian is less relevant of the separability classification since $|\lambda^{xz}_\mathrm{eff}|/\sqrt{\Gamma^z_\mathrm{eff}\bar{\Gamma}^x_\mathrm{eff}}, |\lambda^{zz}_\mathrm{eff}|/\Gamma^z_\mathrm{eff}\ll |\lambda^{xx}_\mathrm{eff}|/\bar{\Gamma}^x_\mathrm{eff}$.  Then the separability classification of the effective master equation combining $\mathcal{L}_\mathrm{eff,NG}$ and $\mathcal{L}_{\mathrm{eff,GRW}}^A+\mathcal{L}_{\mathrm{eff,GRW}}^B$, which is determined by $\eta_\mathrm{GRW}$, is well approximated by $|\lambda^{xx}_\mathrm{eff}|/\bar{\Gamma}^x_\mathrm{eff}$. The non-separability condition $\eta>1$ condition is satisfied for $N\ge 5$ for the parameters used here, and $\eta<1$ for $N\le 4$ implies separability, see also Fig.~\ref{fig:classification_app}. Moreover, when $N\ge 5$ the sufficient condition for entanglement generation in Prop.~\ref{prop:entangling} is also satisfied, i.e. $|\lambda^{xx}_\mathrm{eff}|>\Gamma^x_\mathrm{eff}+\Gamma^+_\mathrm{eff}$.

For the parameters above, note that $r_C=100 x_0$ and $d=20 x_0$. This choice of $r_C$ allows coherence to survive over the spatial support of the wavefunction, thereby permitting the overall dynamics to become non-separable. For sufficiently small $r_C$ and fixed $d,x_0$, however, the single-particle wavefunction itself collapses, and entanglement cannot be established even with QEC. To illustrate this regime, we set $r_C=x_0$, comparable to the delocalization length of the single-body wavefunction, and present the results in Fig.~\ref{fig:GRW}b. In this case, the error rates still decrease exponentially with $N$, but now more slowly than the effective interaction rate. Notably, the effective interaction rate decreases in a convex manner, suggesting that it could still exceed the effective error rates for sufficiently large $N$. However, in that regime the contact probability between the two masses becomes non-negligible, and the contribution from collision interactions can no longer be ignored.

Note that, in contrast to analyses that apply a leading-order separability condition directly to the uncorrected master equation, the separability classification in our setting is governed by how the effective rates scale with the code distance $N$. For example, if one considers only the leading-order composite dynamics in Eq.~\eqref{eq: leading order grw}, then it is non-separable if $\lambda_{\mathrm{GRW}} n /2 r_C^2< 2Gm_Am_B/\hbar d^3$, which would impose an upper bound on $\lambda_\mathrm{GRW}$. In our analysis, increasing $\lambda_\mathrm{GRW}$ primarily shifts the curves in Fig.~\ref{fig:GRW}a upward without changing their scaling with $N$. Consequently, the separability classification in this error-corrected setting is determined mainly by the parameter $r_C$ of the GRW model. 

\subsubsection{CSL model}

We now perform the same analysis for the CSL model. In contrast to the GRW case, the recovered dynamics contains additional nonlocal-like contributions, leading to a richer effective logical structure and additional entries in the separability parameter.

We first simplify the CSL model in a coarse-grained description. We consider two solid spherical masses of equal radius $r_0$, composed of identical constituents with mass $m_c$. The relevant Hilbert space for mass $A$ is expanded by the basis
\begin{equation}
    \mathcal{H}_A= \operatorname{span}\{|x_A\rangle\equiv \prod_i|1_{x_A+x_i, y_i, z_i}\rangle=\prod_i a^\dagger(x_A+x_i, y_i, z_i)|\operatorname{vac}\rangle\}_{x_A},
\end{equation}
where $\boldsymbol{x}_i=(x_i,y_i,z_i)$ denotes the relative coordinate of the $i$th atom with respect to the center of mass position $\boldsymbol{x}_A=(x_A,0,0)$, and $a^\dagger(x,y,z)$ is the lattice atom creation operator. The Hilbert space for mass $B$ is defined analogously but displaced by a distance $d$ from mass $A$. The matrix element of the mass density operator $M(\boldsymbol{x})$ can then be computed, under coarse graining, as
\begin{align}
    M(\boldsymbol{x})|x_A\rangle|x_B\rangle=&\sum_{k} m_k \psi^\dagger_k(\boldsymbol{x})\psi_k(\boldsymbol{x})\prod_i|1_{x_A+x_i, y_i, z_i}\rangle \prod_j|1_{x_B+x_j, y_j, z_j}\rangle\displaybreak[1]\nonumber\\
    =& m_c \left(\sum_{i}\delta(\boldsymbol{x}-d \hat x-\boldsymbol{x}_A-\boldsymbol{x}_i)+ \sum_{j}\delta(\boldsymbol{x}-\boldsymbol{x}_B-\boldsymbol{x}_j)\right)|x_A\rangle|x_B\rangle\displaybreak[1]\nonumber\\
    \approx & \rho_0(\rho_{x_A}(\boldsymbol{x})+\rho_{x_B}(\boldsymbol{x}))|x_A\rangle|x_B\rangle,\displaybreak[1]\\
    \text{where } \rho_{x_A}(\boldsymbol{x})=&  \begin{cases}
        1, & |\boldsymbol{x}-d\hat x-\boldsymbol{x}_A|\le r_0,\\
        0,& \text{otherwise,}\\
    \end{cases}, \quad
    \rho_{x_B}(\boldsymbol{x})=  \begin{cases}
        1, & |\boldsymbol{x}-\boldsymbol{x}_B|\le r_0,\\
        0,& \text{otherwise,}
    \end{cases}
\end{align}
where $\rho_0$ is the mass density of the two spheres.

With this setting, we compute the CSL Lindbladian matrix elements in the positional basis of $\mathcal{H}_A\otimes\mathcal{H}_B$ by evaluating the spatial integrals. The CSL master equation takes the form
\begin{align}
\label{eq: CSL computed}
    \mathcal{L}_{\mathrm{CSL}}(\rho)=&-\frac{\lambda_{\mathrm{CSL}}\pi^2 \rho_0^2 r_\mathrm{C}^6}{m_0^2}\int \dd  x_A \dd  x_B \dd y_A \dd y_B \rho_{x_A x_B, y_Ay_B} |x_A\rangle \langle y_A| \otimes |x_B\rangle\langle y_B|\nonumber\\
    &\times\Bigg[2f_{r_0/r_\mathrm{C}}\left(0\right)-f_{r_0/r_\mathrm{C}}\left(\frac{|x_A-y_A|}{r_\mathrm{C}}\right)-f_{r_0/r_\mathrm{C}}\left(\frac{|x_B-y_B|}{r_\mathrm{C}}\right)+f_{r_0/r_\mathrm{C}}\left(\frac{|d+x_A-x_B|}{r_\mathrm{C}}\right)\nonumber\\
    &+f_{r_0/r_\mathrm{C}}\left(\frac{|d+y_A-y_B|}{r_\mathrm{C}}\right)-f_{r_0/r_\mathrm{C}}\left(\frac{|d+x_A-y_B|}{r_\mathrm{C}}\right)-f_{r_0/r_\mathrm{C}}\left(\frac{|d+y_A-x_B|}{r_\mathrm{C}}\right)\Bigg],
\end{align}
where the function $f_b(a)$ is defined as
\begin{align}
\label{eq:fba_app}
    f_b(a)=&\frac{1}{3 a} \Bigg(4 a e^{-\frac{a^2}{4}} \left(a^2-12 b^2+10\right)-4 e^{-\frac{a^2}{4}-b^2} \left(a \left(a^2-8 b^2+10\right) \cosh (a b)+2 \left(a^2+6\right) b \sinh (a b)\right)\nonumber\\
    &+\sqrt{\pi } \left(a^4-12 \left(a^2+2\right) b^2+12 a^2+12\right) \left(2 \operatorname{Erf}\left(\frac{a}{2}\right)-\operatorname{Erf}\left(\frac{a}{2}-b\right)-\operatorname{Erf}\left(\frac{a}{2}+b\right)\right)\nonumber\\
    &-\sqrt{\pi } 16 a b^3\left(\operatorname{Erf}\left(\frac{a}{2}-b\right)-\operatorname{Erf}\left(\frac{a}{2}+b\right)\right)\Bigg),
\end{align}
and $\operatorname{Erf}(x)=2/\sqrt{\pi}\int_0^x e^{-t^2}dt$ is the error function. When $a,b\approx 0$, one has $f_b(a)\sim b^6$ so the leading scaling in the self-decoherence regime is $\rho_0^2 r_\mathrm{C}^6 (r_0/r_\mathrm{C})^6\sim \rho_0^2 r_0^6$, which is consistent with a mass-squared scaling. The function $f_b(a)$ is even, $f_b(-a)=f_b(a)$. Numerically, $f_b(a)$ is well approximated by a Gaussian. Compared with the standard results in the literature, note that in the limit $a\ll 1$, the leading-order term $f''_b(0)a^2/2$ gives rise to the geometric factor in Ref.~\cite{PhysRevLett.113.020405}. We therefore introduce a Gaussian approximation $f_b'(a)$ with the same peak and variance:
\begin{align}
\label{eq:fba'}
    f'_b(a)=&\frac{32}{3} \left(\sqrt{\pi } b^3 \operatorname{Erf}(b)-3 b^2+\left(b^2-2\right) e^{-b^2}+2\right) \exp \left[-a^2{\left(\frac{4 b^6}{3 \left(\sqrt{\pi } b^3 \text{Erf}(b)-3 b^2+\left(b^2-2\right) e^{-b^2}+2\right)}\right)^{-2/3}}\right].
\end{align}
Using this approximation, the matrix elements of the approximated CSL Lindbladian $\mathcal{L}'_{\mathrm{CSL}}(\rho)$ can be evaluated in the same manner as for the GRW model. A key difference is that terms such as $f'_{r_0/r_\mathrm{C}}\left(\frac{|d+x_A-x_B|}{r_\mathrm{C}}\right)$ can generate nonlocal-like contributions. However, since the integrals are centered near $f'_{r_0/r_\mathrm{C}}(d/r_\mathrm{C})$, these nonlocal-like terms are suppressed by a factor of the form $\exp(-O(d^2/r_\mathrm{C}^2))$.

As in the GRW case, we define local Kraus operators as
\begin{align}
\label{eq:CSL_QEC_protocol}
    R_0= &\Pi = |\bar 0\rangle \langle \bar 0|+ |\bar 1\rangle \langle \bar 1|, \quad R_i= \begin{cases}
        |\bar 0\rangle\langle i|+|\bar 1\rangle (c_i\langle 2N+1-i|+d_i\langle 2N+1+i|), &1\le i\le N,\\
        |\bar 1\rangle (d_{i-N}\langle 3N+1-i|-c_{i-N}\langle N+1+i|), &N+1\le i \le 2N,\\
        |\bar 1\rangle \langle  i+N+1|, &i>2N,
    \end{cases}\\
    \text{where }c_k\propto&\int \dd x_1 \dd  x_2  f'_{r_0/r_\mathrm{C}}\left(\frac{|x_1-x_2|}{r_\mathrm{C}}\right)\langle k|x_1\rangle \langle x_1|\bar 0\rangle \langle \bar 1|x_2\rangle \langle x_2|2N+1-k\rangle ,\nonumber\\
    d_k\propto&\int \dd x_1 \dd  x_2  f'_{r_0/r_\mathrm{C}}\left(\frac{|x_1-x_2|}{r_\mathrm{C}}\right)\langle k|x_1\rangle \langle x_1|\bar 0\rangle \langle \bar 1|x_2\rangle \langle x_2|2N+1+k\rangle ,\nonumber\\
    c_k^2+d_k^2=&1.
\end{align}
We then compute the matrix elements of the effective error-corrected Lindbladian in the full two-mass space:
\begin{equation}
    {}_A\langle \bar i|{}_B\langle \bar j|R\circ\mathcal{L}_{\mathrm{CSL}}'(|\bar k\rangle_A|\bar l\rangle_B {}_A\langle \bar m|{}_B\langle\bar n|)|\bar p\rangle_A|\bar q\rangle_B. 
\end{equation}
These integrals again reduce to Gaussian-polynomial integrals and can be computed analytically. By symmetry, the effective error dynamics for the two masses takes the form, for $\rho$ supported on the code subspace,
\begin{align}
\label{eq:LeffCSL}
  \mathcal{L}_{\text{eff,CSL}}(\rho)= &R\circ\mathcal{L}_{\mathrm{CSL}}(\rho)\nonumber\\
    =&\Gamma_{\mathrm{eff}}^+(\mathcal{D}[\sigma^+_A]+\mathcal{D}[\sigma^+_B])(\rho)+\Gamma_{\mathrm{eff}}^-(\mathcal{D}[\sigma^-_A]+\mathcal{D}[\sigma^-_B])(\rho)+\Gamma_{\mathrm{eff}}^x(\mathcal{D}[\sigma^x_A]+\mathcal{D}[\sigma^x_B])(\rho)+\Gamma_{\mathrm{eff}}^z(\mathcal{D}[\sigma^z_A]+\mathcal{D}[\sigma^z_B])(\rho)\nonumber\\
    &+\Gamma_{\mathrm{eff}}^{zz}(\mathcal{D}[\sigma^z_A-\sigma^z_B]-\mathcal{D}[\sigma^z_A]-\mathcal{D}[\sigma^z_B])(\rho)+\Gamma_{\mathrm{eff}}^{xx}(\mathcal{D}[\sigma^x_A-\sigma^x_B]-\mathcal{D}[\sigma^x_A]-\mathcal{D}[\sigma^x_B])(\rho)\nonumber\\
    &+\Gamma_{\mathrm{eff}}^{xz}(\mathcal{D}[\sigma^x_A-\sigma^z_B]-\mathcal{D}[\sigma^x_A]-\mathcal{D}[\sigma^z_B]-\mathcal{D}[\sigma^z_A-\sigma^x_B]+\mathcal{D}[\sigma^z_A]+\mathcal{D}[\sigma^x_B])(\rho).
\end{align}
This structure is distinct from the term arising from $\mathcal{D}[\sigma^x_A-i\sigma^x_B]$ in the KTM construction, see App.~\ref{app:Collapse models}. According to Thm.~\ref{thm:classification_app}, the separability classification is determined by
\begin{equation}
    \eta_{\mathrm{CSL}} = \left\|  \begin{pmatrix}
            \frac{\Gamma^{xx}_\mathrm{eff}-i\lambda^{xx}_\mathrm{eff}}{\bar{\Gamma}^x_\mathrm{eff}} & -\frac{\Gamma^{xz}_\mathrm{eff}-i\lambda^{xz}_\mathrm{eff}}{\sqrt{\Gamma^z_\mathrm{eff}\bar{\Gamma}^x_\mathrm{eff}}} \\
            \frac{\Gamma^{xz}_\mathrm{eff}-i\lambda^{xz}_\mathrm{eff}}{\sqrt{\Gamma^z_\mathrm{eff}\bar{\Gamma}^x_\mathrm{eff}}} & \frac{\Gamma^{zz}_\mathrm{eff}-i\lambda^{zz}_\mathrm{eff}}{\Gamma^z_\mathrm{eff}}\\
        \end{pmatrix}\right\|_2.
\end{equation}

\begin{figure}
    \centering
    \includegraphics[width=\textwidth]{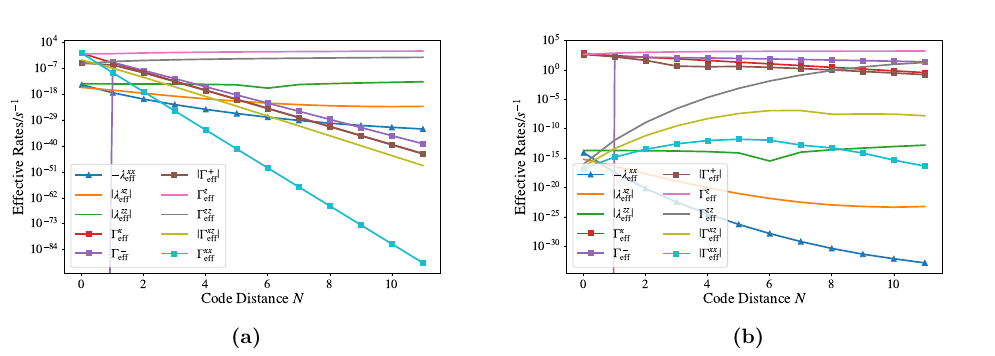}
    \caption{The effective error rates $\Gamma_{\mathrm{eff}}^x,\Gamma_{\mathrm{eff}}^-,\Gamma_{\mathrm{eff}}^+,\Gamma_{\mathrm{eff}}^z,\Gamma_{\mathrm{eff}}^{zz},\Gamma_{\mathrm{eff}}^{xx},\Gamma_{\mathrm{eff}}^{xz}$ of the CSL model and the effective interaction rate $\lambda_{\mathrm{eff}}^{xx}$, $\lambda^{xz}_\mathrm{eff}, \lambda^{zz}_\mathrm{eff}$ as functions of the code distance $N$. (a) For a large localization length $r_C$, $\lambda_{\mathrm{eff}}^{xx}$ decreases with $N$ but more slowly than the $\sigma^x$-type error rates, so the overall effective dynamics becomes non-separable for $N\ge 8$. (b) For a sufficiently small localization length $r_C$, $\lambda_{\mathrm{eff}}^{xx}$ decreases faster than the relevant effective error rates, and the dynamics does not exhibit non-separable characteristics.}
    \label{fig:CSL}
\end{figure}

As an example, we take $r_\mathrm{C}=2\times 10^{-6}\mathrm{m}$, $\lambda_{\mathrm{CSL}}=10^{-8}\mathrm{s^{-1}}$, and proton mass $m_0=1.6\times 10^{-27}\mathrm{kg}$ as the reference mass for the CSL model (this parameter choice is already excluded experimentally \cite{vinante2020narrowing,carlesso2022present}). We choose $x_0=2\times10^{-8}\mathrm{m}$, $m_A=m_B=10^{-21}\mathrm{kg}$, and $d=4\times 10^{-7}\mathrm{m}$. We again assume an silica density $\rho \approx 2.2\times 10^3 \mathrm{kg/m^3}$ so that $r_0=(3m_A/4\pi \rho)^{1/3}\approx 4.8\times 10^{-9}\mathrm{m}$. This setting gives a rough relative strength $(4 r_C^2Gm_A m_B m_0^2)/(\hbar \lambda_{\mathrm{CSL}}m_A m_B d_{\min}^3 ) \approx 3.0\times 10^{-9}$, where $d_\mathrm{min}=2r_0$ is the minimal allowed distance between the two masses, as an analogy to the GRW case in App.~\ref{app:Gravity-induced entanglement with spatially-superposed states}. This ratio is even many orders of magnitude smaller than that in the GRW case.  With these parameters, the effective interaction and error rates as functions of the code distance $N$ are shown in Fig.~\ref{fig:CSL}a. We observe that the $\sigma^x$-type error rates, including both local and nonlocal-like contributions, decrease exponentially faster than the effective interaction rate. Hence the combined effective master equation $\mathcal{L}_\mathrm{eff,NG}+\mathcal{L}_{\text{eff,CSL}}$ is non-separable when $\eta_\mathrm{CSL} >1$. Numerically, in the non-separable regime, the major contribution to $\eta_\mathrm{CSL}$ is from the $\sigma^x$-sector, or in other words the following sufficient condition is satisfied: 
\begin{equation}
    \sqrt{(\lambda_{\mathrm{eff}}^{xx})^2+(\Gamma_{\mathrm{eff}}^{xx})^2}>\bar{\Gamma}_{\mathrm{eff}}^x =\Gamma^x_{\mathrm{eff}}+\frac{\Gamma_{\mathrm{eff}}^+\Gamma_{\mathrm{eff}}^-}{\Gamma_{\mathrm{eff}}^+ +\Gamma_{\mathrm{eff}}^-}.
\end{equation}
This is satisfied for $N\ge 8$ for the parameters used here. When $N\le7$, the separability of the effective error-corrected dynamics is ensured by the condition $\eta < 1$, see also Fig.~\ref{fig:classification_app}. Note that for the parameters here, when $N=0$, $\eta_\mathrm{CSL}$ is majorly contributed by $\Gamma^{xx}_\mathrm{eff}/\bar{\Gamma}^x_\mathrm{eff}$, since $\lambda^{xx}_\mathrm{eff}\ll \Gamma^{xx}_\mathrm{eff}$. Although it seems $\eta_\mathrm{eff}\approx 0.97$ is quite close to $1$, the complete positivity of $\mathcal{L}_\mathrm{eff,CSL}$ requires $\Gamma^{xx}_\mathrm{eff}\le\bar{\Gamma}^x_\mathrm{eff}$ and hence it is not possible to be non-separable at $N=0$. The flat curve of $\eta_\mathrm{CSL}$ for $1\le N\le 6$ is caused the saturation of $\sigma^z$-sector contribution, i.e. $|\Gamma^{zz}_\mathrm{eff}-i\lambda^{zz}_\mathrm{eff}|/{\Gamma^z_\mathrm{eff}}$. Moreover, when $N\ge 8$ the sufficient condition for entanglement generation in Prop.~\ref{prop:entangling} is also satisfied, i.e. $\sqrt{(\lambda^{xx}_\mathrm{eff})^2+(\Gamma^{xx}_\mathrm{eff})^2}>\Gamma^x_\mathrm{eff}+\Gamma^+_\mathrm{eff}$.

As in the GRW discussion, error correction becomes ineffective if the localization length $r_C$ is sufficiently small so that single-body coherence is lost. We therefore also consider $r_C=x_0=2\times 10^{-8}\mathrm{m}$, see Fig.~\ref{fig:CSL}b. In this regime, the interaction rate $\lambda_{\mathrm{eff}}^{xx}$ and the nonlocal-like error rate $\Gamma_{\mathrm{eff}}^{xx}$ decrease much faster with $N$ than the local $\sigma^x$-type error rate $\Gamma_{\mathrm{eff}}^{x}$, and the overall dynamics remains separable, i.e. $\eta_\mathrm{CSL}\ll1$.

Compared with the GRW model, the CSL model introduces an additional nonlocal-like contribution proportional to $\Gamma_{\mathrm{eff}}^{xx}$, which can enhance the nonlocality of the effective dynamics and can even be comparable to the Newtonian contribution, see Fig.~\ref{fig:CSL}b. This does not mean the CSL collapse mechanism itself is fundamentally nonlocal. Positivity of the CSL generator implies $|\Gamma_{\mathrm{eff}}^{xx}|\le\Gamma_{\mathrm{eff}}^{x}$, which is consistent with the structure of Eq.~\eqref{eq: CSL computed}. The CSL model is therefore separable by itself, but it can enhance the non-separable features introduced by the Newtonian interaction.

\subsubsection{DP model}

Finally, we investigate the DP model within the same framework. As will become apparent, the DP model differs qualitatively from the GRW and CSL cases because the effective interaction and the dominant error channels originate from the same underlying kernel.

Without loss of generality, we assume the motion is effectively one-dimensional along the relative axis connecting the two equilibrium positions, and we denote the displacements from equilibrium by operators $x_A$ and $x_B$ for the two masses. Following the same coarse-graining spirit as in the CSL derivation and carrying out the integrals over the (smeared) mass densities, the DP master equation in Eq.~\eqref{eq:dp} reduces to the following form in the $x$-basis:
\begin{align}
\label{eq: DP Lind x basis}
    \mathcal{L}_\mathrm{DP}(\rho)=& \frac{G}{\hbar} \int \dd x_A\dd x_B\dd y_A\dd y_B\rho_{x_A x_B;y_A y_B}|x_Ax_B\rangle \langle y_A y_B|\nonumber\\
    & \times \Bigg( \frac{m_A^2}{|x_A-y_A|}  \operatorname{Erf}\left(\frac{|x_A-y_A|}{2\sigma}\right)+\frac{m_B^2}{|x_B-y_B|}  \operatorname{Erf}\left(\frac{|x_B-y_B|}{2\sigma}\right)-\frac{ m_A^2}{\sqrt{\pi } \sigma}-\frac{ m_B^2}{\sqrt{\pi } \sigma}\nonumber\\
    & +\frac{m_A m_B}{|d+x_A-y_B|}  \operatorname{Erf}\left(\frac{|d+x_A-y_B|}{2\sigma}\right)+\frac{m_A m_B}{|d+y_A-x_B|}  \operatorname{Erf}\left(\frac{|d+y_A-x_B|}{2\sigma}\right)\nonumber\\
    &  -\frac{(1-i)m_A m_B}{|d+x_A-x_B|}  \operatorname{Erf}\left(\frac{|d+x_A-x_B|}{2\sigma}\right)  -\frac{(1+i)m_A m_B}{|d+y_A-y_B|}\operatorname{Erf}\left(\frac{|d+y_A-y_B|}{2\sigma}\right)\Bigg).
\end{align}
In the original DP formulation, the unitary interaction term also includes the regime where the two masses contact, so we keep this expression as the starting point for our effective dynamics analysis.

As recently noted in \cite{PhysRevD.111.L121101}, the DP model can display nonlocal characteristics in the large-smearing regime $d< 0.850872\sigma$. In the following we show that this can occur  for the Fock code here and large-smearing regime examined numerically, but only for $N=0$, where our protocol reduces almost to a projection rather than a non-trivial correction map. The reason is that in Eq.~\eqref{eq: DP Lind x basis} the nonlocal interaction and the dominant errors share the same distance-dependent kernel. Consequently, increasing the code distance $N$ suppresses the effective nonlocal coupling and the effective error channels, and the non-trivial nonlocal contribution cannot parametrically surpass the effective errors. In this situation, QEC can in fact worsen the visibility of nonlocality, in sharp contrast to the GRW and CSL cases where the relevant nonlocal coupling is suppressed more slowly than the correctable error contributions.

We now compute the effective error-corrected dynamics for the DP model. We define the local Kraus operators similarly to the CSL construction:
\begin{align}
    R_0= &\Pi = |\bar 0\rangle \langle \bar 0|+ |\bar 1\rangle \langle \bar 1|, \quad R_i= \begin{cases}
        |\bar 0\rangle\langle i|+|\bar 1\rangle (c_i\langle 2N+1-i|+d_i\langle 2N+1+i|), &1\le i\le N,\\
        |\bar 1\rangle (d_{i-N}\langle 3N+1-i|-c_{i-N}\langle N+1+i|), &N+1\le i \le 2N,\\
        |\bar 1\rangle \langle  i+N+1|, &i>2N,
    \end{cases}\\
    \text{where }c_k\propto&\int \dd  x_1 \dd  x_2  \frac{ \operatorname{Erf}\left(\frac{|{x}_1-{x}_2|}{2\sigma}\right)}{|x_1-x_2|} \langle k|x_1\rangle \langle x_1|\bar 0\rangle \langle \bar 1|x_2\rangle \langle x_2|2N+1-k\rangle ,\nonumber\\
    d_k\propto&\int  \dd  x_1 \dd  x_2  \frac{ \operatorname{Erf}\left(\frac{|{x}_1-{x}_2|}{2\sigma}\right)}{|x_1-x_2|}\langle k|x_1\rangle \langle x_1|\bar 0\rangle \langle \bar 1|x_2\rangle \langle x_2|2N+1+k\rangle ,\nonumber\\
    c_k^2+d_k^2=&1.
\end{align}
We then compute the matrix elements of the effective error-corrected Lindbladian in the full two-mass space. By the same symmetry arguments as in the CSL case, the effective two-mass dynamics takes the form, for $\rho$ supported on the code subspace,
\begin{align}
\label{eq:LeffDP}
  \mathcal{L}_{\text{eff,DP}}(\rho)= &R\circ\mathcal{L}_{\mathrm{DP}}(\rho)\nonumber\\
    =&-i[\lambda_{\mathrm{eff}}^{xx}\sigma^x_A\otimes\sigma^x_B+\lambda_{\mathrm{eff}}^{xz}\sigma^x_A\otimes\sigma^z_B-\lambda_{\mathrm{eff}}^{xz}\sigma^z_A\otimes\sigma^x_B+\lambda_{\mathrm{eff}}^{zz}\sigma^z_A\otimes\sigma^z_B,\rho]+\Gamma_{\mathrm{eff}}^+(\mathcal{D}[\sigma^+_A]+\mathcal{D}[\sigma^+_B])(\rho)\nonumber\\
    &+\Gamma_{\mathrm{eff}}^-(\mathcal{D}[\sigma^-_A]+\mathcal{D}[\sigma^-_B])(\rho)+\Gamma_{\mathrm{eff}}^x(\mathcal{D}[\sigma^x_A]+\mathcal{D}[\sigma^x_B])(\rho)+\Gamma_{\mathrm{eff}}^z(\mathcal{D}[\sigma^z_A]+\mathcal{D}[\sigma^z_B])(\rho)+\Gamma_{\mathrm{eff}}^{zz}(\mathcal{D}[\sigma^z_A-\sigma^z_B]\nonumber\\
    &-\mathcal{D}[\sigma^z_A]-\mathcal{D}[\sigma^z_B])(\rho)+\Gamma_{\mathrm{eff}}^{xx}(\mathcal{D}[\sigma^x_A-\sigma^x_B]-\mathcal{D}[\sigma^x_A]-\mathcal{D}[\sigma^x_B])(\rho)+\Gamma_{\mathrm{eff}}^{xz}(\mathcal{D}[\sigma^x_A-\sigma^z_B]-\mathcal{D}[\sigma^x_A]-\mathcal{D}[\sigma^z_B]\nonumber\\
    &-\mathcal{D}[\sigma^z_A-\sigma^x_B]+\mathcal{D}[\sigma^z_A]+\mathcal{D}[\sigma^x_B])(\rho).
\end{align}
In the DP case, the effective coherent coupling is tied to the same kernel responsible for the nonlocal-like $\sigma^x_A-\sigma^x_B$ contribution, and one finds
\begin{equation}
    \lambda_{\mathrm{eff}}^{xx}=\Gamma_{\mathrm{eff}}^{xx}, \qquad \lambda_{\mathrm{eff}}^{xz}=\Gamma_{\mathrm{eff}}^{xz},\qquad \lambda_{\mathrm{eff}}^{zz}=\Gamma_{\mathrm{eff}}^{zz}.
\end{equation}
Following Thm.~\ref{thm:classification_app}, the non-separability is determined by the separability parameter
\begin{equation}
    \eta_{\mathrm{DP}} = \left\| (1-i) \begin{pmatrix}
            \frac{\lambda^{xx}_\mathrm{eff}}{\bar{\Gamma}^x_\mathrm{eff}} & -\frac{\lambda^{xz}_\mathrm{eff}}{\sqrt{\Gamma^z_\mathrm{eff}\bar{\Gamma}^x_\mathrm{eff}}} \\
            \frac{\lambda^{xz}_\mathrm{eff}}{\sqrt{\Gamma^z_\mathrm{eff}\bar{\Gamma}^x_\mathrm{eff}}} & \frac{\lambda^{zz}_\mathrm{eff}}{\Gamma^z_\mathrm{eff}}\\
        \end{pmatrix}\right\|_2=\sqrt{2} \left\| \begin{pmatrix}
            \frac{\lambda^{xx}_\mathrm{eff}}{\bar{\Gamma}^x_\mathrm{eff}} & -\frac{\lambda^{xz}_\mathrm{eff}}{\sqrt{\Gamma^z_\mathrm{eff}\bar{\Gamma}^x_\mathrm{eff}}} \\
            \frac{\lambda^{xz}_\mathrm{eff}}{\sqrt{\Gamma^z_\mathrm{eff}\bar{\Gamma}^x_\mathrm{eff}}} & \frac{\lambda^{zz}_\mathrm{eff}}{\Gamma^z_\mathrm{eff}}\\
        \end{pmatrix}\right\|_2.
\end{equation}
A simple sufficient condition for non-separability is $\eta_x= \sqrt{(\lambda_{\mathrm{eff}}^{xx})^2+(\Gamma_{\mathrm{eff}}^{xx})^2}/\bar{\Gamma}_{\mathrm{eff}}^x =\sqrt{2}|\lambda_{\mathrm{eff}}^{xx}|/\bar{\Gamma}_{\mathrm{eff}}^x >1$. And separability is established for $\eta_\mathrm{DP} <1$.

\begin{figure}[htbp]
    \centering
    \includegraphics[width=\textwidth]{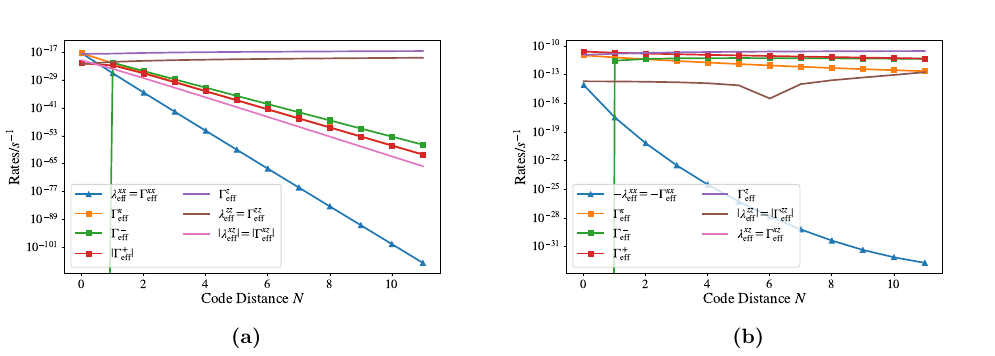}
    \caption{The effective error rates $\Gamma_{\mathrm{eff}}^x,\Gamma_{\mathrm{eff}}^-,\Gamma_{\mathrm{eff}}^+,\Gamma_{\mathrm{eff}}^z,\Gamma_{\mathrm{eff}}^{zz},\Gamma_{\mathrm{eff}}^{xx},\Gamma_{\mathrm{eff}}^{xz}$ and the effective interaction rate $\lambda_{\mathrm{eff}}^{xx}$, $\lambda^{xz}_\mathrm{eff}, \lambda^{zz}_\mathrm{eff}$ of the DP model, as functions of the code distance $N$. (a) For a large smearing radius $\sigma$, $\lambda_{\mathrm{eff}}^{xx}$ remains much smaller than the relevant $\sigma^x$-type error rates for $N\ge1$; for $N=0$ the situation is the opposite, rendering the overall dynamics non-separable. (b) For a small smearing radius $\sigma$, $\lambda_{\mathrm{eff}}^{xx}$ is always much smaller than the relevant $\sigma^x$-type error rates for all $N$, reflecting the separable character in the $d\gg 2\sigma$ regime. }
    \label{fig:DP}
\end{figure}

As an example, we take $x_0=2\times10^{-8}\mathrm{m}$, $m_A=m_B=10^{-21}\mathrm{kg}$, and $d=4\times 10^{-7}\mathrm{m}$ as in the GRW and CSL examples above. For the DP model, we first consider a large smearing radius $\sigma=2\times 10^{-6}\mathrm{m}$. The effective interaction and error rates are shown in Fig.~\ref{fig:DP}a. We find that the $\sigma^x$-like local errors remain larger than the effective interaction rate for all $N\ge1$. Only for $N=0$, where the map reduces to a projection rather than a genuine correction protocol, this effective dynamics is non-separable following Thm.~\ref{thm:classification_app}. Moreover, at $N=0$ the sufficient condition for entangling in Prop.~\ref{prop:entangling}, i.e. ${\sqrt{(\lambda_{\mathrm{eff}}^{xx})^2+(\Gamma_{\mathrm{eff}}^{xx})^2}}/(\Gamma_{\mathrm{eff}}^x + \Gamma^-_{\mathrm{eff}}) >1$ is also satisfied. For any code distance $N\ge1$, the effective interaction rate $\lambda_\mathrm{eff}^{xx}$ and the nonlocal-like rate $\Gamma_\mathrm{eff}^{xx}$ decrease much faster with $N$ than the relevant local error rates. The condition $\eta_\mathrm{DP}\ll 1$ ensures the separability of the effective error-corrected Lindbladian.

As a comparison, we also evaluate a small smearing radius $\sigma=r_0\approx 4.8\times 10^{-9}\mathrm{m}$, corresponding to the mass radius for silica at the same parameters used above. The results are shown in Fig.~\ref{fig:DP}b. In this case, the $\sigma^x$-like local errors remain larger than the effective interaction rate already at $N=0$, rendering the effective dynamics separable.

\begin{figure}
    \centering
    \includegraphics[width=0.5\linewidth]{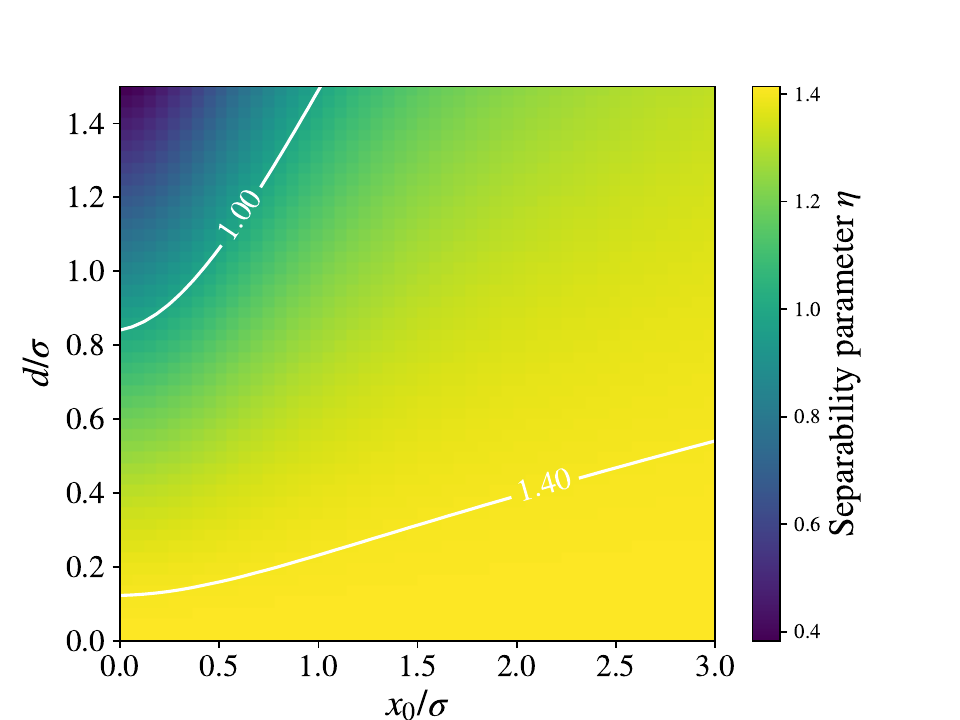}
    \caption{The separability parameter $\eta$ for the DP model, plotted over $(x_0/\sigma,d/\sigma)$. The effective dynamics is non-separable when the $\eta$ exceeds $1$.}
    \label{fig:DP contour}
\end{figure}

We now identify parameters where the DP model becomes non-separable. We restrict to $N=0$ and vary $x_0$, $d$, and $\sigma$. We compute the separability parameter $\eta$, 
which depends only on the dimensionless ratios $x_0/\sigma$ and $d/\sigma$, see Fig.~\ref{fig:DP contour}. The maximal value of this ratio is $\sqrt{2}$. This follows because (i) complete positivity of the effective generator implies $|\Gamma_{\mathrm{eff}}^{xx}|\le\bar{\Gamma}^x_{\mathrm{eff}}$, and (ii) for the DP model one has  $\lambda_{\mathrm{eff}}^{xx}=\Gamma_{\mathrm{eff}}^{xx}$. Hence $|\lambda_{\mathrm{eff}}^{xx}|\le |\bar{\Gamma}^x_{\mathrm{eff}}|$ so without the nonlocal-like term proportional to $\Gamma_{\mathrm{eff}}^{xx}$ the dynamics would remain separable. In this sense, the nonlocal-like measurement structure associated with $\mathcal{D}[x_A-x_B]$ is essential for any non-separable signature in the DP model. In the limit $x_0/\sigma\to 0$, the DP model becomes non-separable when $d/\sigma\lesssim 0.850872$, matching the threshold reported in \cite{PhysRevD.111.L121101}. On the other hand, the non-separability increases as $x_0$ increases, as a consequence of a more delocalized wavefunction. However the data points with big $x_0$ may not be realized physically, considering the overlapping discussions above.

\subsection{Comparison of the effective dynamics and separability classifications}\label{app:Comparison of collapse models}

The preceding sections derived the effective logical interactions and error channels associated with the KTM, GRW, CSL, and DP models. For convenience, we compare here the resulting effective dynamics and separability classifications. This subsection provides a compact overview of the main conclusions of the appendix and highlights the qualitative differences between the various models. Readers primarily interested in the final numerical results may use this subsection as a summary of the effective dynamics obtained after error correction.

\begin{figure}
    \centering
    \includegraphics[width=0.6\linewidth]{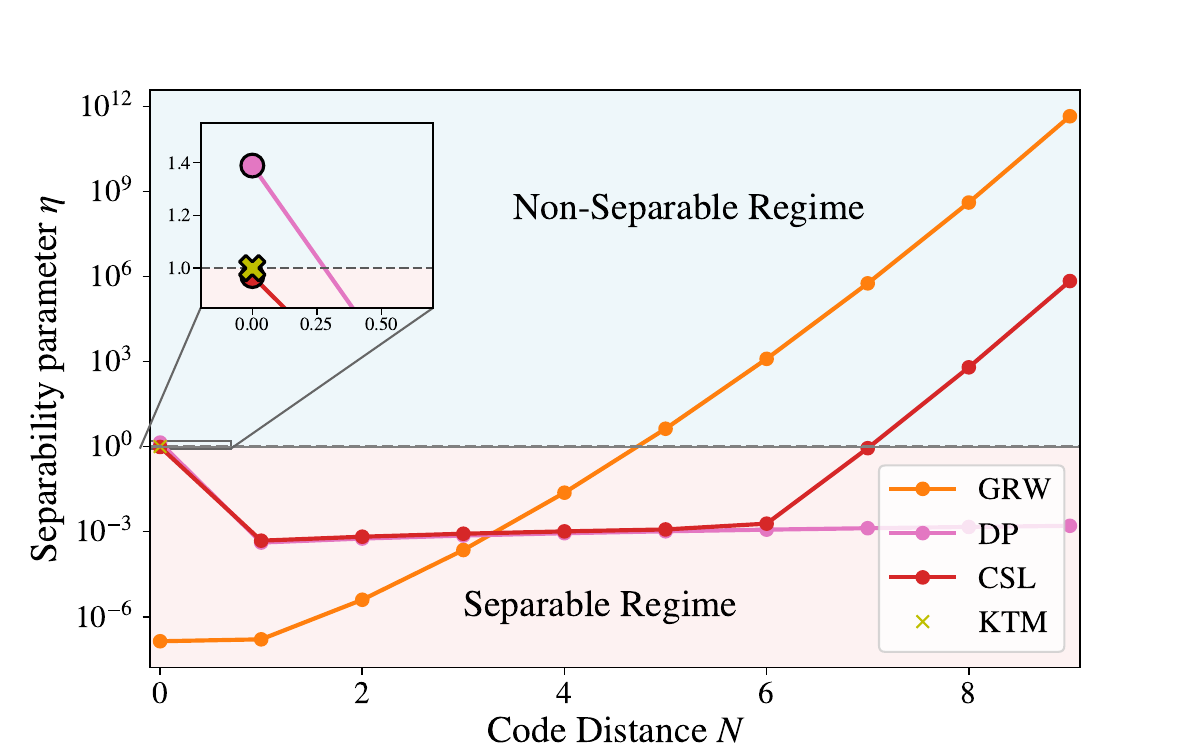}
    \caption{Separability parameter $\eta$ for different models and code distances $N$. The effective error-corrected dynamics is non-separable if and only if $\eta>1$. For sufficiently large localization lengths, Newtonian interaction supplemented with GRW or CSL is non-separable for sufficiently large $N$, whereas the KTM and DP models remain separable except for the $N=0$ DP regime.}
    \label{fig:classification_app}
\end{figure}

The relevant separability parameter $\eta$ governing the separability classification is defined as follows, following from Thm.~\ref{thm:classification_app}:
\begin{equation}
    \eta = \left\|  \begin{pmatrix}
            \frac{\Gamma^{xx}_\mathrm{eff}-i\lambda^{xx}_\mathrm{eff}}{\bar{\Gamma}^x_\mathrm{eff}} & -\frac{\Gamma^{xz}_\mathrm{eff}-i\lambda^{xz}_\mathrm{eff}}{\sqrt{\Gamma^z_\mathrm{eff}\bar{\Gamma}^x_\mathrm{eff}}} \\
            \frac{\Gamma^{xz}_\mathrm{eff}-i\lambda^{xz}_\mathrm{eff}}{\sqrt{\Gamma^z_\mathrm{eff}\bar{\Gamma}^x_\mathrm{eff}}} & \frac{\Gamma^{zz}_\mathrm{eff}-i\lambda^{zz}_\mathrm{eff}}{\Gamma^z_\mathrm{eff}}\\
        \end{pmatrix}\right\|_2,
\end{equation}
such that $\eta>1$ if and only if the corresponding effective error-corrected dynamics is non-separable. Figure~\ref{fig:classification_app} provides a visual comparison of the corresponding behavior as a function of the code distance. More specifically, these models have different additional properties for simplifying the separability parameter, as summarized in Table~\ref{tab:model_comparison}.

\begin{table}[htbp]
\centering
\caption{Comparison of the effective dynamics and separability classifications for the models considered in this work.}
\label{tab:model_comparison}
\renewcommand{\arraystretch}{1.5}
\begin{tabularx}{\textwidth}{
c
@{\hspace{1.2em}}
c
@{\hspace{1.2em}}
X
}
\hline
\textbf{Model} &
\textbf{Additional simplification for $\eta$}  &
\textbf{Effective classification} \\
\hline
KTM & Only $\lambda_{\rm eff},\Gamma_{\rm eff}\neq 0$ contributes to $\eta$, $\Gamma^z_\mathrm{eff}$ does not affect $\eta$ &
Separable by construction (LOCC dynamics), since $\eta=\lambda_{\rm eff}/\Gamma_{\rm eff}\le 1$. \\

GRW & Non-local error rates $\Gamma^{xx}_\mathrm{eff},\Gamma^{xz}_\mathrm{eff},\Gamma^{zz}_\mathrm{eff}=0$.  &
Can become non-separable for sufficiently large code distance $N$ and sufficiently large localization length $r_C$. \\

CSL &
No specific simplification &
Can become non-separable for sufficiently large code distance $N$ and sufficiently large localization length $r_C$.\\

DP & $\lambda_{\mathrm{eff}}^{xx}=\Gamma_{\mathrm{eff}}^{xx}, \lambda_{\mathrm{eff}}^{xz}=\Gamma_{\mathrm{eff}}^{xz}, \lambda_{\mathrm{eff}}^{zz}=\Gamma_{\mathrm{eff}}^{zz}$ &
Generally remains separable for non-trivial code distances $N>0$. Non-separable behavior can occur only in a restricted parameter regime for $N=0$ given our encoding here. \\
\hline
\end{tabularx}
\end{table}

A central distinction emerges between the GRW/CSL models and the DP model. In the GRW and CSL cases, the dominant correctable error channels and the effective interaction originate from different physical kernels. As a result, increasing the code distance suppresses the effective logical error rates more rapidly than the effective interaction rate, allowing the effective dynamics to become non-separable. In contrast, for the DP model the effective interaction and the dominant error channels originate from the same kernel and are therefore suppressed simultaneously by the QEC procedure. Consequently, increasing the code distance does not improve the effective separability classification for the encoding considered here. The KTM model remains separable by construction, providing a consistency check of the framework.

The comparison presented in Table~\ref{tab:model_comparison} and Fig.~\ref{fig:classification_app} shows that error correction affects the different collapse models in qualitatively distinct ways. While the GRW and CSL models admit parameter regimes in which the effective dynamics becomes non-separable after encoding, the DP model does not exhibit the same enhancement because the effective interaction and dominant error channels remain intrinsically linked. This distinction ultimately determines whether QEC can improve the visibility of non-separable dynamics in the corresponding effective description.

\subsection{Analytical bounds on local recoverability}\label{app:Rigorous proof of the entanglement recoverability}

In this section we establish analytically the qualitative distinction between the collapse models observed numerically above. For GRW and CSL noise, we show that the separability parameter can be made arbitrarily large by increasing the code distance, while simultaneously keeping the probability of particle contact exponentially small. By contrast, for the Di\'osi--Penrose model we derive a finite upper bound on the separability parameter under local encoding and recovery, showing that the unbounded enhancement available for GRW and CSL is not possible. We also clarify why increasing the code distance is essential: at fixed finite code distance, varying only the geometric parameters does not in general produce the same asymptotic enhancement.

Establishing whether entanglement generation can be retained while keeping the contact probability of the two particles below a prescribed threshold, is physically important because configurations with center-to-center separation smaller approaching $2r_0$ would require accounting for additional short-range interactions not included in our model. Moreover, for numerical purposes, in our analysis, we regularize the singular Newtonian potential at short distances, since the numerical evaluation involves integration over the full configuration space. We therefore require that the probability of sampling this short-distance region remain negligible, so that the resulting entanglement generation does not rely on either physically overlapping configurations or on the details of the numerical regularization. In this section, we show that this probability can be exponentially suppressed in the large-$N$ limit while retaining an arbitrarily large effective interaction-to-error ratio.

\subsubsection{Unbounded recoverability for GRW and CSL}

We first study the GRW case; the CSL case follows from an equivalent analysis.

\begin{theorem}
\label{thm:recoverability}
    Let us consider the bipartite Fock state code $\mathcal{C}_\alpha= \{|\bar 0\rangle_\alpha = |0\rangle_\alpha, |\bar{1}\rangle_\alpha = |2N+1\rangle_\alpha\}$ and the recovery protocol $R$ introduced in Eq.~\eqref{eq: app recovery GRW}. The effective error-corrected Lindbladian is then described by Eq.~\eqref{eq:NG lind} and Eq.~\eqref{eq:effLind GRW}. We denote $k=2N+1$. For any given $\lambda_{\mathrm{GRW}}$ and $r_C$, let us choose 
    \begin{equation}
        x_0=x_{0,k}=y_0 k^{-1/2}, \qquad d=d_k = 2r_0+ l k^{\gamma}, \qquad 0<\gamma<\frac{1}{4},
    \end{equation}
    where $y_0, r_0, l$ are independent of $k$. Then the probability of the two particles sitting at a distance shorter than $2r_0$ is $o_k < C_0 \exp -C_1 k^{1+2\gamma}$ decreases exponentially with constants $C_0, C_1>0$, while the effective interaction to error ratio $\log (|\lambda_{\rm eff}^{xx}|/\Gamma^{x\prime}_{\rm eff}) \ge  (1/2-2\gamma) k \log k + C_2 k$ with constant $C_2$  can be made arbitrarily large in the formal limit $k\rightarrow \infty$. Here let us denote $\Gamma^{x\prime}_\mathrm{eff}=\Gamma^x_{\mathrm{eff}}+(\Gamma^+_\mathrm{eff}+\Gamma^-_\mathrm{eff})/2$, such that  $\Gamma^{x\prime}_\mathrm{eff}\ge \bar{\Gamma}^x_{\rm eff}=\Gamma^x_{\mathrm{eff}}+\Gamma^+_\mathrm{eff}\Gamma^-_\mathrm{eff}/(\Gamma^+_\mathrm{eff}+\Gamma^-_\mathrm{eff})$.
\end{theorem}
Note that $x_0$ decrease with $k$, which means the frequency of the trap $\omega=\omega_k= (\hbar/2m y_0^2) k$ increases with $k$. This implies the limit $k\rightarrow\infty$ might violate our implicit non-relativistic rigid-body assumption. So this theorem should be understood only as in the formal limit $k\rightarrow \infty$. At the same time, the standard deviation of the Fock state $\sqrt{\langle k|x^2|k\rangle}=\sqrt{2+1/k} y_0$ remains approximately independent of $k$ in the large-$k$ limit, which means the size of the state $|k\rangle$ remains of order $O(1)$. 

Following Thm.~\ref{thm:classification_app}, $|\lambda_{\rm eff}^{xx}|/\bar{\Gamma}^x_{\rm eff}$ is a sufficient condition for the separability parameter $\eta$ to be larger than 1, hence it is a sufficient condition for non-separability of the overall effective dynamics. Moreover, note that $\Gamma^{x\prime}_\mathrm{eff} \ge \min(\Gamma^x_\mathrm{eff}+\Gamma^+_\mathrm{eff},\Gamma^x_\mathrm{eff}+\Gamma^-_\mathrm{eff})$, the entangling sufficient condition in Prop.~\ref{prop:entangling} is also satisfied. Therefore for sufficiently large $N$, the effective dynamics is not just non-separable but also entangling.

\begin{proof}
    The probability of the two particles getting closer than $2r_0$, hereinafter called {\it contact probability}, is given by:
    \begin{equation}
        o_k= \sup_{\rho \in \mathcal{L}(\mathcal{C}_A^k \otimes \mathcal{C}_B^k)} \operatorname{Tr} \rho \mathbbm{1}_{|d+x_A-x_B|\le 2 r_0}
    \end{equation}
    see also Eq.~\eqref{eq:app overlap}, and we have assumed $\rho\ge 0 $ and $\operatorname{Tr}\rho =1$. Note that if $|x_A|,|x_B|< l k^\gamma/2$, then one must have $|d+x_A-x_B|> 2 r_0$. Therefore, the contact probability has an upper bound 
    $$
    o_k \le 2 \sup_{|\psi\rangle_\alpha \in \mathcal{C}_\alpha^k} {}_\alpha\langle \psi| \mathbbm{1}_{|x_\alpha|\ge lk^\gamma/2}|\psi\rangle_\alpha = 2 \operatorname{Pr}_{|k\rangle_\alpha}(|x_\alpha|\ge lk^\gamma/2)
    $$
    Note that $(lk^\gamma/2)/x_{0,k} = l/(2y_0) k^{1/2+\gamma}$. At the same time, the peak position of the dimensionless Hermite polynomial, that is furthest from the origin, grows only at the order $O(\sqrt{k})$. This implies for sufficiently large $k$, $|x_\alpha|>lk^\gamma/2$ is only possible at the exponential decay tail of the wavefunction. With the asymptotic analysis of the Hermite polynomials, we then obtain
    \begin{equation}
        o_k \le 2 \operatorname{Pr}_{|k\rangle_\alpha}(|x_\alpha|\ge lk^\gamma/2) \le C_0 \exp \left[-C_1 k^{1+2\gamma}\right],
    \end{equation}
    where $C_0$ and $C_1$ are some constants. Therefore, we show that the contact probability decreases exponentially as $k$ increases.

    Then let us estimate the scaling of the effective interaction rate. We define the normal modes as $b = (a_A+a_B)/\sqrt{2}$ and $c= (a_A-a_B)/\sqrt{2}$, then the effective interaction rate can be computed as 
    \begin{equation}
        {}_A\langle \bar{0}|{}_B\langle \bar {0}| f_{\mathrm{NG}}(|d+x_A -x_B|) |\bar 1\rangle_A |\bar 1\rangle_B = \langle \mathrm{vac}| f_{\mathrm{NG}} \frac{(b^{\dagger 2}-c^{\dagger 2})^k}{2^k k!} |\mathrm{vac}\rangle_B ={}_c\langle 0|f_{\mathrm{NG}}(|d+x_A -x_B|) (-1)^k \frac{\sqrt{(2k)!}}{2^k k!}|2k\rangle_c,
    \end{equation}
    where $|2k\rangle_c$ is the Fock state in $c$ mode. Let us introduce a projection operator $\Pi_k=\mathbbm{1}_{|x_A-x_B|\le l k^\gamma/2}$. In the support of this projection operator, one has $d+x_A-x_B> 2r_0 >0$. Therefore, in this subspace one has 
    $$\Pi_k f_{\mathrm{NG}}(|d+x_A-x_B|)\Pi_k = \Pi_k f_{\mathrm{NG}}(d+x_A-x_B)\Pi_k,$$
    namely we can remove the absolute value. The residual outside this subspace is controlled as follows:
    \begin{align}
        &|{}_A\langle \bar{0}|{}_B\langle \bar {0}|[f_{\mathrm{NG}}(|d+x_A -x_B|)-\Pi_k f_{\mathrm{NG}}(|d+x_A -x_B|) \Pi_k] |\bar 1\rangle_A |\bar 1\rangle_B|\nonumber\\
        =& |{}_A\langle \bar{0}|{}_B\langle \bar {0}|(\mathbbm{1}-\Pi_k)f_{\mathrm{NG}}(|d+x_A -x_B|) (\mathbbm{1}-\Pi_k) |\bar 1\rangle_A |\bar 1\rangle_B|\nonumber\\
        \le &||f_{\mathrm{NG}}||_\infty \sqrt{{}_A\langle \bar{0}|{}_B\langle \bar {0}|(\mathbbm{1}-\Pi_k)|\bar 0\rangle_A |\bar 0\rangle_B {}_A\langle \bar{1}|{}_B\langle \bar {1}|(\mathbbm{1}-\Pi_k)|\bar 1\rangle_A |\bar 1\rangle_B}\nonumber\\
        \le & \frac{G m_A m_B}{2 r_0} \sqrt{ {}_A\langle k|{}_B\langle k|(\mathbbm{1}-\Pi_k)|k\rangle_A |k \rangle_B}\nonumber\\
        \le & C_2 \exp \left( -C_3 k^{1+2\gamma}\right),
    \end{align}
    where for the first inequality we have used $|f_\mathrm{NG}|\le ||f_\mathrm{NG}||_\infty$ and the Cauchy-Schwarz inequality, and for the last inequality we have used the same argument as $o_k$, and $C_2,C_3$ are constants.   
    
At the same time, let us also consider the complementary residual 
    \begin{align}
        |{}_A\langle \bar{0}|{}_B\langle \bar {0}|(\mathbbm{1}-\Pi_k)\frac{1}{d+x_A -x_B }(\mathbbm{1}-\Pi_k) |\bar 1\rangle_A |\bar 1\rangle_B| .
    \end{align}
    In the region that $x_A-x_B >lk^\gamma/2$, $-d_k+\sqrt{2}x_{0,k}<x_A-x_B<-lk^\gamma/2$ or $x_A-x_B<-d-\sqrt{2}x_{0,k}$, $|1/(d+x_A-x_B)|\le 1/\sqrt{2}x_{0,k}$, similar arguments as above also applies. We only need to be cautious around the pole $x_A-x_B=-d$
    \begin{equation}
        \frac{\sqrt{(2k)!}}{2^k k!}\int_{-d/\sqrt{2}x_0-1}^{-d/\sqrt{2}x_0+1} ds\, {}_c\langle 0|s\rangle \frac{1}{d+\sqrt{2}x_0 s } \langle s|2k\rangle_c =  \frac{\sqrt{(2k)!}}{2^k k!}\int_{-1}^1 \dd s\frac{h_k\left(-\frac{d}{\sqrt{2}x_0}+s\right)}{\sqrt{2}x_0 s}   ,  \text{where }  h_k(s)=H_{e_{2k}}\left(s\right)   \frac{e^{-s^2/2}}{ \sqrt{2 (2k)! \pi }}
    \end{equation}
    where $H_{e_m}$ is the probabilists' Hermite polynomial. The Cauchy principal value of this integral is bounded by $\sqrt{2} \sup_{|s|\le 1} |h'(-d/\sqrt{2}x_0+s)|/x_0$, where we have used $\sqrt{(2k)!}\le 2^k k!$. Since $d_k/\sqrt{2} x_{0,k} \propto k^{1/2+\gamma}$, $h'_k(-d/\sqrt{2}x_0+s)$ is  also in the exponentially decaying tail, yielding the upper bound 
    $$\log |{h'_k(-d/\sqrt{2}x_0+s)}|\le -\frac{d_k^2}{4 x_{0,k}^2}+O\left(k\log\frac{d_k}{\sqrt{2}x_{0,k}}\right)\le \log (C_4)  - C_5 k^{1+2\gamma}$$ 
    with constants $C_4,C_5$ and for $|s|\le 1$. The contribution from the region $x_A-x_B >lk^\gamma/2$ or $x_A-x_B<-d-\sqrt{2}x_{0,k}$ can also be absorbed by changing the constants.

In this situation, the effective interaction is bounded as
    \begin{align}
        &|{}_A\langle \bar{0}|{}_B\langle \bar {0}| f_{\mathrm{NG}}(|d+x_A -x_B|) |\bar 1\rangle_A |\bar 1\rangle_B|\nonumber\\
        =& |{}_A\langle \bar{0}|{}_B\langle \bar {0}| \Pi_k f_{\mathrm{NG}}(|d+x_A -x_B|)  \Pi_k |\bar 1\rangle_A  |\bar 1\rangle_B +{}_A\langle \bar{0}|{}_B\langle \bar {0}| (\mathbbm{1}-\Pi_k) f_{\mathrm{NG}}(|d+x_A -x_B|)(\mathbbm{1}-\Pi_k) |\bar 1\rangle_A  |\bar 1\rangle_B|\nonumber\\
        \ge & |{}_A\langle \bar{0}|{}_B\langle \bar {0}| \Pi_k f_{\mathrm{NG}}(d+x_A -x_B) \Pi_k |\bar 1\rangle_A |\bar 1\rangle_B| -C_2 e^{-C_3 k^{1+2\gamma}}\nonumber\\
        \ge & |{}_A\langle \bar{0}|{}_B\langle \bar {0}|  \frac{G m_A m_B}{d+x_A -x_B}  |\bar 1\rangle_A |\bar 1\rangle_B| -C_2 e^{-C_3 k^{1+2\gamma}} - \frac{G m_A m_B}{\sqrt{2} x_{0,k}}C_4 e^{-C_5 k^{1+2\gamma}}
    \end{align}

     Then let us evaluate the matrix element of $1/(d+x_A-x_B)$. We take the Fourier transform
    $$
    \frac{1}{d+\sqrt{2}x_0(c+c^\dagger)}=\frac{1}{\sqrt{2}x_0} \lim_{\epsilon\rightarrow 0+}\int_0^\infty e^{-\epsilon u}\sin u (d/\sqrt{2}x_0+ (c+c^\dagger)) \dd u.
    $$
    Here and below the inverse should be understood in a principal value sense. Since $|0\rangle_c$ and $|2k\rangle_c$ are both even states, the expectation value of $\sin u (c+c^\dagger)$ is 0, then we get only contribution from
    \begin{align}
    {}_c\langle0| \sin u (d/\sqrt{2} x_0+ (c+c^\dagger)) |2k\rangle_c = &{}_c\langle0| \sin u d/\sqrt{2} x_0 \cos u(c+c^\dagger) |2k\rangle_c\nonumber\\ 
    =& \sin u d/\sqrt{2} x_0  {}_c\langle0| e^{i u (c+c^\dagger)} |2k\rangle_c\nonumber\\
    =& \sin \frac{ud}{\sqrt{2}x_0} (-1)^{k}\frac{u^{2k}e^{-u^2/2}}{\sqrt{(2k)!}}.
    \end{align}
    The effective interaction rate then reads as
    \begin{align}
    \label{eq: integral interaction}
        {}_A\langle \bar{0}|{}_B\langle \bar {0}|  \frac{1}{d+x_A -x_B}  |\bar 1\rangle_A |\bar 1\rangle_B=& \frac{1}{2^k  k! }\frac{1}{\sqrt{2}x_{0,k}} \lim_{\epsilon\rightarrow 0^+}\int_0^{\infty} e^{-\epsilon u} u^{2k} e^{-u^2/2} \sin \frac{ud_k}{\sqrt{2}x_{0,k}} \dd u\nonumber\\
        =& \frac{1}{2^k\sqrt{2} k! x_{0,k}} \int_0^{\infty} u^{2k} e^{-u^2/2} \sin \frac{ud_k}{\sqrt{2}x_{0,k}} \dd u,
    \end{align}
where the limit and the integral are interchangeable in the second equality, since the Gaussian integral is finite. Let us denote $v_k = d_k/(\sqrt{2}x_{0,k})$. Since the integrand is an entire function, we can translate the integral path from $[0,\infty)$ to $[0+i v_k,\infty +i v_k)$ and rewrite the integral as
\begin{align}
    H_k(v) =& \int_0^\infty u^{2k} e^{-u^2/2} \sin (v u) \dd u = (-1)^k I_k(v) + J_k(v),\nonumber\\
    &\text{where } I_k(v) = \int_0^v x^{2k} e^{x^2/2-vx} \dd x, \quad J_k(v) = e^{-v^2/2} \operatorname{Im} \int_0^\infty (x+iv)^{2k} e^{-x^2/2}\dd x.
\end{align}
The first integral has a lower bound as follows:
\begin{align}
    I_k(v) = & \int_0^v x^{2k} e^{x^2/2-vx} \dd x\nonumber\\
    =& \frac{1}{v^{2k+1}}\int_0^{v^2} t^{2k} e^{t^2/(2v^2)-t} \dd t \nonumber\\
    \ge & \frac{1}{v^{2k+1}}\int_0^{v^2} t^{2k} e^{-t} \dd t\nonumber\\
    = & \frac{(2k)!}{v^{2k+1}}-\frac{1}{v^{2k+1}}\int_{v^2}^{\infty} t^{2k} e^{-t} \dd t\nonumber\\
    \ge & \frac{(2k)!}{v^{2k+1}}-\frac{e^{-v^2/2}}{v^{2k+1}}\int_{0}^{\infty} t^{2k} e^{-t/2} \dd t\nonumber\\
    =& \frac{(2k)!}{v^{2k+1}} (1-2^{2k+1} e^{-v^2/2}).
\end{align}
Since $v \propto k^{\gamma+1/2}$, for a sufficiently large $k$ we have  $v^2 \ge 4k$, and hence $2^{2k+1} e^{-v^2/2}\le 2 (2/e)^{2k} \ll 1$. On the other hand,
\begin{align}
    |J_k(v)| \le& e^{-v^2/2}  \int_0^\infty (x^2+v^2)^{k} e^{-x^2/2}\dd x\nonumber\\
    = & e^{-v^2/2} v^{2k} \int_0^\infty (x^2/v^2+1)^{k} e^{-x^2/2}\dd x\nonumber\\
    < & e^{-v^2/2} v^{2k}  \int_0^\infty e^{kx^2/v^2} e^{-x^2/2}\dd x\nonumber\\
    \le & \sqrt{\pi} e^{-v^2/2} v^{2k}
\end{align}
Therefore, their ratio
\begin{equation}
    \log \frac{|J_k|}{I_k} \le \log \sqrt{\pi} \frac{v^{4k+1}e^{-v^2/2} }{(2k)! (1-2^{2k+1} e^{-v^2/2})} = -\frac{l^2}{4 y_0^2} k^{1+2\gamma} -\frac{r_0l}{y_0^2} k^{1+\gamma} + 4\gamma k \log( k) + O(k),
\end{equation}
goes to $-\infty$ when $k$ is sufficiently large. In this situation, there exists a constant $C_6$ such that $|H_k(v_k)| \ge C_6 (2k)!/ v_k^{2k+1}$, hence
\begin{align}
        |\lambda_{\mathrm{eff}}|\ge& \frac{G m_A m_B}{\hbar 2^k\sqrt{2} k! x_{0,k}} \times C_6 \frac{(2k)!} {v_k^{2k+1}}-\frac{C_2}{\hbar} e^{-C_3 k^{1+2\gamma}} - \frac{G m_A m_B}{\sqrt{2}\hbar x_{0,k}}C_4 e^{-C_5 k^{1+2\gamma}}\nonumber\\
        \ge & C_6 \frac{G m_A m_B}{\hbar  } 
        \frac{(2k)! x_{0,k}^{2k}}{k!  d_k^{2k+1}}  -\frac{C_2}{\hbar} e^{-C_3 k^{1+2\gamma}} - \frac{G m_A m_B}{\sqrt{2} x_{0,k}\hbar }C_4 e^{-C_5 k^{1+2\gamma}},
    \end{align}

Therefore, we conclude the scaling of the effective interaction rate as 
    \begin{equation}
        \log |\lambda_{\mathrm{eff}}|\ge \log \frac{(2k)!}{k!}+\log \frac{x_{0,k}^{2k}}{d_k^{2k+1}} + O(1) \ge -2 \gamma k \log (k) -C_7 k,
    \end{equation}
    where $C_7$ is a constant.

    Finally, let us study the scaling behavior of the effective error rate. Equivalently, one can rewrite the dimensionless GRW decoherence channel with respect to mode $a$ as 
    \begin{equation}
        \Phi_{\kappa_k}(\rho)=\int_{-\infty}^\infty \frac{\dd \eta}{\sqrt{\pi \kappa}} e^{-\eta^2/\kappa} e^{i \eta (a+a^\dagger)}\rho e^{-i \eta (a+a^\dagger)}, \qquad \text{where } \kappa_k=  \frac{x_{0,k}^2}{r_C^2}=\frac{y_0^2}{r_C^2} \frac{1}{k}.
    \end{equation}
    Then the transition probability $\langle m|\Phi_{\kappa_k}(|l\rangle\langle l|)|m\rangle$ as introduced in Eq.~\eqref{eq: app GRW transition} can be calculated with the generation functional as follows:
    \begin{align}
        F(s,t) =& \sum_{m,l=0}^\infty \langle m|\Phi_{\kappa_k}(|l\rangle\langle l|)|m\rangle s^m t^l\nonumber\\
        =& \int_{-\infty}^\infty \sum_{m,l=0}^\infty\frac{\dd \eta}{\sqrt{\pi \kappa_k}} e^{-\eta^2/\kappa_k} \operatorname{Tr}[s^m|m\rangle\langle m|e^{i \eta (a+a^\dagger)}|l\rangle\langle l| e^{-i \eta (a+a^\dagger)}t^l]  \nonumber\\
        =& \int_{-\infty}^\infty \frac{\dd \eta}{\sqrt{\pi \kappa_k}} e^{-\eta^2/\kappa_k} \operatorname{Tr}[s^{a^\dagger a}e^{i \eta (a+a^\dagger)}t^{a^\dagger a} e^{-i \eta (a+a^\dagger)}]  \nonumber\\
        =& \int_{-\infty}^\infty \frac{\dd \eta}{\sqrt{\pi \kappa_k}} e^{-\eta^2/\kappa_k} \frac{1}{1-st}\exp \left[-\eta^2\frac{(1-s)(1-t)}{1-st}\right]  \nonumber\\
        =&\frac{1}{\sqrt{(1-st)[1-st+\kappa_k(1-s)(1-t)]}}
    \end{align}
    The error probability of logical $|\bar 0\rangle$ flips to $|\bar 1\rangle$ is then proportional to
    \begin{align}
        a_k =&  \sum_{m=N+1}^\infty \langle m|\Phi_{\kappa_k}(|0\rangle\langle 0|)|m\rangle\nonumber\\
        =& \sum_{m=N+1}^\infty \frac{1}{m!}\left(\frac{\dd }{\dd s}\right)^mF(s,t)\Bigg|_{s=t=0}\nonumber\\
        =& \sum_{m=N+1}^\infty \frac{(2m)!}{m!m! 4^m}\frac{\kappa_k^m}{(1+\kappa_k)^{m+1/2}}\nonumber\\
        \le & \sum_{m=N+1}^\infty \frac{(1+1)^{2m}}{4^m}\frac{\kappa_k^m}{(1+\kappa_k)^{m+1/2}}\nonumber\\
        =& \sqrt{1+\kappa_k}\left(\frac{\kappa_k}{1+\kappa_k}\right)^{N+1}\nonumber\\
        \le& C_a \kappa_k^{N+1},
    \end{align}
    where $C_a$ is some constant. Here and in the following we omit the constant $n\lambda_{\mathrm{GRW}}$ for simplicity. Let us also evaluate the error probability of logical $|\bar 1\rangle$ flips to $|\bar 0\rangle$. Consider $0<s<1$, such that $s^m\ge s^N$ for any $m\le N$, it is proportional to
    \begin{align}
        b_k =& \sum_{m=0}^N \langle m|\Phi_{\kappa_k}(|k\rangle\langle k|)|m\rangle\nonumber\\
        \le& s^{-N} \sum_{m=0}^N \langle m|\Phi_{\kappa_k}(|k\rangle\langle k|)|m\rangle s^m\nonumber\\
        \le & s^{-N} \sum_{m=0}^\infty \langle m|\Phi_{\kappa_k}(|k\rangle\langle k|)|m\rangle s^m\nonumber\\
        = &s^{-N} \frac{1}{k!}\left(\frac{\dd }{\dd t}\right)^k F(s,t)\Bigg|_{t=0}\nonumber\\
        \le & s^{-N} t^{-k}F(s,t)
    \end{align}
    where for the inequalities we have used the fact that all the Taylor expansion coefficients are positive. Let us take $s=\kappa_k$ and $t=1/4\kappa_k$, such that for sufficiently large $k$ we have $s<1$. Then $b_k$ is bounded as
    \begin{align}
        b_k \le \kappa_k^{-N} (4 \kappa_k)^k \frac{1}{\sqrt{\frac{3}{4}[\frac{3}{4}+\kappa_k(1-\kappa_k)(1-\frac{1}{4\kappa_k})]}} = 4^{2N+1} \kappa_k^{N+1} \left(\sqrt{\frac{8}{3}}+O(1/k)\right).
    \end{align}
    In other words, the leading contribution from both $a_k$ and $b_k$ is $\kappa_k^{k/2}$ Therefore, the scaling of the effective error rate introduced in Eq.~\eqref{eq: app GRW transition} is
    \begin{equation}
        \log \left(\Gamma^x_{\mathrm{eff}}+\frac{\Gamma_{\mathrm{eff}}^+\Gamma_{\mathrm{eff}}^-}{\Gamma_{\mathrm{eff}}^+ +\Gamma_{\mathrm{eff}}^-}\right)\le\log \left(\Gamma^x_{\mathrm{eff}} +\frac{\Gamma^+_{\mathrm{eff}}+\Gamma^-_{\mathrm{eff}}}{2}\right) = \log \frac{a_k+b_k}{2}+\log n\lambda_\mathrm{GRW} \le -\frac{k}{2} \log k + O(k).
    \end{equation}

    In summary, we have shown that the contact probability $o_k \le C_0 \exp - C_1 k^{1+2\gamma}$ decreases exponentially as $k$ increases. At the same time, the scaling of the ratio between effective interaction rate and effective error rate is
    \begin{equation}
        \log \frac{|\lambda_{\mathrm{eff}}^{xx}|}{\Gamma^x_{\mathrm{eff}}+\frac{\Gamma_{\mathrm{eff}}^+\Gamma_{\mathrm{eff}}^-}{\Gamma_{\mathrm{eff}}^+ +\Gamma_{\mathrm{eff}}^-}} \ge \log \frac{|\lambda_{\mathrm{eff}}^{xx}|}{\Gamma^x_{\mathrm{eff}}+\frac{\Gamma_{\mathrm{eff}}^+ +\Gamma_{\mathrm{eff}}^-}{2}}\ge \left(\frac{1}{2}-2\gamma\right) k\log k +O(k),
    \end{equation}
    where for simplicity we have omitted the constant $G m_a m_B /(n\times \hbar \lambda_{\mathrm{GRW}})$. Therefore, for $\gamma<1/4$, the ratio can be made arbitrarily large by increasing $k$.
\end{proof}

Note that the scaling behaviour of the proof here is completely different from the result given in App.~\ref{app:Gravity-induced entanglement with spatially-superposed states}. There the leading contribution of the effective interaction rate comes from the contribution in the contact limit, i.e., the regime $|d+x_A-x_B|\approx 2r_0$, but here the major contribution of the effective interaction rate is at the long tail of the inverse potential. In other words, if we increase $d$ as here, the entanglement-based test in App.~\ref{app:Gravity-induced entanglement with spatially-superposed states} would fail to detect entanglement.

For the CSL model, a similar derivation can also be implemented to show the same result. The only qualitative difference is in the nonlocal correlated error $\Gamma^{xx}_{\mathrm{eff}}$. Note that $d_k/r_C = O(k^\gamma)$, such that $\Gamma^{xx}_{\mathrm{eff}}$ would decrease much faster than $\lambda_{\mathrm{eff}}^{xx}$.

\subsubsection{Necessity of increasing the code distance}

Here, let us show that this scaling control is not possible if we fix $N$ and control only $d,x_0$. To avoid the divergence of the contact probability when $x_0\rightarrow 0$, we introduce a gap $d-2r_0 \ge \delta d$ that is independent of $x_0$. A similar analysis shows the contact probability $< \exp -O (\delta d^2/x_0^2)$ can be arbitrarily small when $x_0\rightarrow 0$. By taking the Taylor expansion of Newtonian potential and directly evaluating the matrix element, one can show that 
\begin{equation}
    |\lambda^{xx}_\mathrm{eff}|\le C_{0,N} x_0^{4N+2} + C_{1,N} e^{-C_{2,N}\frac{\delta d^2}{x_0^2}},
\end{equation}
where $C_{i,N}$ are some constants that depend on $N$. The first term is the upper bound on the regime $|x_A -x_B|\le \delta d/2$ and the second term is the upper bound on the regime $|x_A-x_B| \ge \delta d/2$. Therefore in the limit $x_0 \rightarrow 0$, the leading contribution is $O(x_0^{4N+2})$. 

On the other hand, the leading contribution of the error rate is
\begin{align}
        a_k =&  \langle N+1|\Phi_\kappa(|0\rangle\langle 0|)|N+1\rangle + O\left( \left(\frac{x_0}{r_C}\right)^{2N+4}\right) =  \frac{(2N+2)!}{[(N+1)!]^2 2^{2N+2}} \left(\frac{x_0}{r_C}\right)^{2N+2} + O\left( \left(\frac{x_0}{r_C}\right)^{2N+4}\right)\nonumber\\
        b_k =& \langle N|\Phi_\kappa(|k\rangle\langle k|)|N\rangle + O\left( \left(\frac{x_0}{r_C}\right)^{2N+4}\right) =  \frac{(2N+2)!(2N+1)!}{[(N+1)!]^3 N! 2^{2N+2}} \left(\frac{x_0}{r_C}\right)^{2N+2} + O\left( \left(\frac{x_0}{r_C}\right)^{2N+4}\right),\nonumber\\
        \Gamma^x_{\mathrm{eff}} = & c_N\operatorname{Re} \langle N+1|\Phi_\kappa(|0\rangle\langle  k|)|N\rangle + O\left( \left(\frac{x_0}{r_C}\right)^{2N+4}\right) = c_N\frac{(2N+2)!}{[(N+1)!]^{5/2} 2^{2N+2}}\sqrt{\frac{(2N+1)!}{N!}} \left(\frac{x_0}{r_C}\right)^{2N+2} + O\left( \left(\frac{x_0}{r_C}\right)^{2N+4}\right),
\end{align}
where $c_N$ is defined in Eq.~\eqref{eq: GRW ck dk}. Note that 
\begin{equation}
    c_N = c_N^{(0)} + O\left(\left(\frac{x_0}{r_C}\right)^2\right) = \frac{1}{\sqrt{1+\frac{(3N+1)!(N+1)!}{[(2N+1)!]^2}}} + O\left(\left(\frac{x_0}{r_C}\right)^2\right) <1.
\end{equation}
As a result, the effective error rate
\begin{equation}
    \Gamma^x_{\mathrm{eff}}+\frac{\Gamma_{\mathrm{eff}}^+\Gamma_{\mathrm{eff}}^-}{\Gamma_{\mathrm{eff}}^+ +\Gamma_{\mathrm{eff}}^-} = \frac{a_k b_k - (\Gamma^x_{\mathrm{eff}})^2}{a_k + b_k - 2 \Gamma^x_{\mathrm{eff}}} = C_{2,N} x_0^{2N+2} + O(x_0^{2N+4}) \ge C_{2,N}' x_0^{2N+2}.
\end{equation}
Here $C_{2,N} \propto [1-(c_N^{(0)})^2] >0$.

Therefore, we conclude that 
\begin{equation}
    \frac{|\lambda^{xx}|}{\Gamma^x_{\mathrm{eff}}+\frac{\Gamma_{\mathrm{eff}}^+\Gamma_{\mathrm{eff}}^-}{\Gamma_{\mathrm{eff}}^+ +\Gamma_{\mathrm{eff}}^-}}\le C_{3,N} x_0^{2N} + C_{4,N} x_0^{-2N-2} e^{-C_{2,N} \delta d^2/x_0^2},
\end{equation}
where $C_{i,N}$ are some constants that depend on $N$. Therefore, if we decrease $x_0\rightarrow 0$ just as above, we would see this ratio also goes to 0 for $N\ge 1$. For $N=0$, this ratio is at most a finite constant. Similarly, one can also find an upper bound for $x_0>r_C$, where the localization is very strong. In other words, the entanglement generation with fixed $N$ might fail if one increases $\lambda_{\mathrm{GRW}}$.

\subsubsection{Finite recoverability bound for the DP model}

On the other hand, the DP model has an upper bound on the separability parameter $\eta$. Without loss of generality, let us rewrite the interaction/decoherence kernel with Gaussian smearing as
\begin{equation}
    \int \frac{\dd^3 \boldsymbol{k}}{(2\pi)^3} \frac{4\pi e^{-\sigma^2 |\boldsymbol{k}|^2}}{|\boldsymbol{k}|^2} e^{i\boldsymbol{\boldsymbol{k}\cdot(\boldsymbol{x}-\boldsymbol{y})}} =  \sum_{\nu}\int \dd^3 \boldsymbol{k} f_\nu (\boldsymbol{x})f_\nu(\boldsymbol{y}),
\end{equation}
where the existence of this kernel decomposition is guaranteed by the positivity of the kernel in the momentum space, and $\nu= 0,1,$ is a summation over cosine mode and sine mode to ensure $f_\nu$ is real. And we rewrite the unsmeared mass density operator as $M(\boldsymbol{x})_\alpha = m_\alpha \delta(\boldsymbol{x}-\boldsymbol{x}_\alpha)$. This is equivalent to the original DP model. We define operators
\begin{align}
    \alpha_\nu(\boldsymbol{k}) = \sqrt{\frac{G}{\hbar}}\int \dd^3 \boldsymbol{x} f_\nu (\boldsymbol{\boldsymbol{x})}M_\alpha(\boldsymbol{x}).
\end{align}
Then the DP Lindbladian can be rewritten as
\begin{equation}
    \mathcal{L}_{\mathrm{DP}}(\rho)=\sum_{\nu}\int \dd^3 \boldsymbol{k} \frac{i}{2}[(A_\nu \otimes I_B+I_A\otimes B_\nu)^2,\rho] +\mathcal{D}[A_\nu \otimes I_B+I_A\otimes B_\nu](\rho).
\end{equation}

For a general local code subspace defined by the projector $\Pi_{AB}=\Pi_A\otimes \Pi_B$, let us assume the following two conditions: (i) for any $\Pi_{AB}\rho\Pi_{AB}=\rho$, the code state is invariant under the \textit{local} recovery channel $R(\rho)=\rho$; (ii) the recovery ensures there is no leakage error, i.e. $\Pi_{AB}R(\mathcal{L}_{\mathcal{DP}}(\rho))\Pi_{AB}=R(\mathcal{L}_{\mathcal{DP}}(\rho))$. According to Prop.~\ref{prop:KL_and_trivial_unitary}, we can rewrite the Kraus operator of $R(\cdot) = \sum_{ij} R^A_i R^B_j \cdot R^{A\dagger}_i R^{B\dagger}_j $, such that $R^A_i\Pi_A=\delta_{i0} \Pi_A$, $R^B_j\Pi_B=\delta_{j0} \Pi_B$; otherwise we can rewrite the Kraus operators with linear combinations of themselves to ensure this form\footnote{Without loss of generality, assume we have $R^\alpha_i \Pi_\alpha = r_i^\alpha\Pi_\alpha$. The trace preservation implies $\sum_i |r_i^\alpha|^2=1$, i.e. $r_i$ is a unit vector. Then one can rewrite the Kraus operators as $R^\alpha_i \rightarrow \sum_j U_{ij}^\alpha R^\alpha_j$, where $U_{ij}^\alpha$ is a unitary. We choose $U^\alpha$ such that $\sum_j U_{ij}^\alpha r^\alpha_j=\delta_{i0}$. This transformation does not change the recovery channel itself.}. For the local error part, these two conditions then imply for every pure code state $\rho = |\psi\rangle \langle\psi|$,
\begin{align}
    0=&(I_{AB}-\Pi_{AB})R(\mathcal{L}_{\mathcal{DP}}(\rho))(I_{AB}-\Pi_{AB})\nonumber\\
    =& (I_{AB}-\Pi_{AB})\sum_{ij}R^A_i R^B_j\left(\sum_{\nu}\int \dd^3 \boldsymbol{k} \mathcal{D}[A_\nu + B_\nu](\rho)\right)R^{A\dagger}_i R^{B\dagger}_j(I_{AB}-\Pi_{AB})  \nonumber\\
    =&\sum_{\nu}\int \dd^3 \boldsymbol{k}(I_{AB}-\Pi_{AB})\sum_{ij}R^A_i R^B_j\left( A_\nu + B_\nu) |\psi\rangle \langle \psi| ( A_\nu^\dagger + B_\nu^\dagger\right)R^{A\dagger}_i R^{B\dagger}_j(I_{AB}-\Pi_{AB})
\end{align}
Therefore every branch should be zero, i.e.
\begin{equation}
    (I_{AB}-\Pi_{AB})R^A_i R^B_j( A_\nu + B_\nu) |\psi\rangle=0\qquad \forall i,j, \nu,\boldsymbol{k}.
\end{equation}
Since it holds for every pure state, then we take $i=0$ or $j=0$, then it immediately implies the orthogonality
\begin{equation}
    (I_A - \Pi_A)R^A_i A_\nu(\boldsymbol{k}) \Pi_A = (I_B - \Pi_B)R^B_j B_\nu(\boldsymbol{k}) \Pi_B=0, \forall i,j, \nu, \boldsymbol{k}.
\end{equation}
In other words, the recovery map turns every local error into a logical error operator 
\begin{equation}
    E^\alpha_{i,\nu,\boldsymbol{k}} = R^\alpha_i \alpha_\nu(\boldsymbol{k}) \Pi_\alpha, \qquad \text{such that}\quad E^\alpha_{i,\nu,\boldsymbol{k}} = \Pi_\alpha E^\alpha_{i,\nu,\boldsymbol{k}} \Pi_\alpha.
\end{equation}
For simplicity let us assume $\operatorname{Tr} E^\alpha_{i,\nu,\boldsymbol{k}}=0$, otherwise we subtract those norms with these constants times $\Pi_\alpha$, obtaining an equivalent Lindbladian.

In this case, the local error part of error-corrected channel can be rewritten as
\begin{equation}
    \sum_{\nu}\int \dd^3 \boldsymbol{k} R(\mathcal{D}[\alpha_\nu](\rho)) = \sum_{i,\nu}\int \dd^3\boldsymbol{k} {D}[E^\alpha_{i,\nu,\boldsymbol{k}}](\rho).
\end{equation}
Specifically, note that $\Pi_\alpha =  \sum_i R^{\alpha\dagger}_i R^\alpha _i \Pi_\alpha =  R^{\alpha\dagger}_0 R^\alpha_0 \Pi_\alpha = R^{\alpha\dagger}_0 \Pi_\alpha$ and hence $E^\alpha_{0,\nu,\boldsymbol{k}}=\Pi_\alpha R^\alpha_0 \alpha_\nu \Pi_\alpha = \Pi_\alpha \alpha_\nu \Pi_\alpha$. 

Let us now investigate the nonlocal effective interaction part, which can be rewritten as 
\begin{equation}
    \sum_{\nu}\int \dd^3 \boldsymbol{k} \text{ } iR([A_\nu\otimes B_\nu,\rho ]) = \sum_{\nu}\int \dd^3 \boldsymbol{k} \text{ } i R^A_0 R^B_0([A_\nu\otimes B_\nu,\rho ])R^{A\dagger}_0 R^{B\dagger}_0=\sum_{\nu}\int \dd^3 \boldsymbol{k} \text{ }  i[\Pi_A A_\nu \Pi_A \otimes \Pi_B B_\nu \Pi_B, \rho].
\end{equation}
Let us denote $\Pi_A A_\nu(\boldsymbol{k})\Pi_A = a_{\nu,\boldsymbol{k}}^0\Pi_A + \sum_{i=1} a^i_{\nu,\boldsymbol{k}} O^i_A$, where $\{O^i_A\}$ is an orthonormal basis of non-trivial Hermitian traceless logical operators defined in $\Pi_A$, i.e. $\operatorname{Tr} O^{i\dagger}_A O^j_A = \delta_{ij}$, and write the vector $\boldsymbol{a}_{\nu,\boldsymbol{k}}=(a^1_{\nu,\boldsymbol{k}},a^2_{\nu,\boldsymbol{k}},\cdots)$. Similarly we define $\{O^i_B\}$ and $\boldsymbol{b}_{\nu,\boldsymbol{k}}$. Then the nonlocal effective Hamiltonian can be further rewritten as
\begin{equation}
    H_{AB} = -\sum_{ij} J_{ij} O^i_A \otimes O^j_B, \quad \text{where}\quad J=\sum_\nu \int \dd^3 \boldsymbol{k} \,\boldsymbol{a}_{\nu, \boldsymbol{k}}\boldsymbol{b}^T_{\nu,\boldsymbol{k}}.
\end{equation}
To account for the largest contribution to the interaction, let us make a singular value decomposition of $J$, and write its largest singular value as $\lambda_1$ with left singular vector $\boldsymbol{u}$ and right singular vector $\boldsymbol{v}$. The singular value can be rewritten as
\begin{equation}
    \lambda_1(\boldsymbol{u},\boldsymbol{v}) = \boldsymbol{u}^T J \boldsymbol{v} =\sum_\nu \int \dd^3 \boldsymbol{k} \,(\boldsymbol{u}\cdot\boldsymbol{a}_{\nu, \boldsymbol{k}})(\boldsymbol{v}\cdot\boldsymbol{b}_{\nu,\boldsymbol{k}})
\end{equation}

Then let us investigate local logical error rates along these two operator axes. Similarly, let us also write down the decomposition of error operators as $E^\alpha_{i,\nu,\boldsymbol{k}}=\boldsymbol{e}^\alpha_{i,\nu,\boldsymbol{k}}\cdot \boldsymbol{O}_\alpha$. Specifically, $\boldsymbol{e}^A_{0,\nu,\boldsymbol{k}} = \boldsymbol{a}^i_{\nu,\boldsymbol{k}}$. Then along the operator axis $\boldsymbol{w}\cdot \boldsymbol{O}_\alpha$, the weight of logical error rate is defined as 
\begin{equation}
    \Gamma_\alpha(\boldsymbol{w}) = \sum_{i,\nu} \int \dd^3 \boldsymbol{k} \big|\operatorname{Tr}(\boldsymbol{w}\cdot \boldsymbol{O}_\alpha)^\dagger E^\alpha_{i,\nu,\boldsymbol{k}}\big|^2 .
\end{equation}
Specifically, along the axis $u$ for $A$ and along the axis $v$ for $B$ we have 
\begin{equation}
   \Gamma_A(\boldsymbol{u}) \ge \sum_{\nu} \int \dd^3 \boldsymbol{k} \left(\boldsymbol{u}\cdot \boldsymbol{a}_{\nu,\boldsymbol{k}}\right)^2, \qquad \Gamma_B(\boldsymbol{v}) \ge \sum_{\nu} \int \dd^3 \boldsymbol{k} \left(\boldsymbol{v}\cdot \boldsymbol{b}_{\nu,\boldsymbol{k}}\right)^2.
\end{equation}
where we have taken only $i=0$.

Given these considerations above, we can observe that the single-axis separability parameter is always bounded as
\begin{equation}
    \eta_{\boldsymbol{u},\boldsymbol{v}} = \frac{\sqrt{\lambda_1^2(\boldsymbol{u},\boldsymbol{v}) + \Gamma^2_{AB}(\boldsymbol{u},\boldsymbol{v})}}{\sqrt{\Gamma_A(\boldsymbol{u})\Gamma_B(\boldsymbol{v})}}\le \sqrt{2},
\end{equation}
where the nonlocal error rate $|\Gamma_{AB}(\boldsymbol{u},\boldsymbol{v})|=|\lambda_1(\boldsymbol{u},\boldsymbol{v})| $ for the decoherence term $\mathcal{D}[A_\nu \otimes I_B+I_A\otimes B_\nu]-\mathcal{D}[A_\nu ]-\mathcal{D}[B_\nu]$, and we have used the Cauchy-Schwarz inequality
\begin{equation}
    \Bigg|\sum_\nu \int \dd^3 \boldsymbol{k} \,(\boldsymbol{u}\cdot\boldsymbol{a}_{\nu, \boldsymbol{k}})(\boldsymbol{v}\cdot\boldsymbol{b}_{\nu,\boldsymbol{k}})\Bigg|^2 \le \left(\sum_{\nu} \int \dd^3 \boldsymbol{k} \left(\boldsymbol{u}\cdot \boldsymbol{a}_{\nu,\boldsymbol{k}}\right)^2\right)\left(\sum_{\nu} \int \dd^3 \boldsymbol{k} \left(\boldsymbol{v}\cdot \boldsymbol{b}_{\nu,\boldsymbol{k}}\right)^2\right).
\end{equation}
And actually this $\sqrt{2}$ upper bound also holds for arbitrary operator axes $\boldsymbol{u}$ and $\boldsymbol{v}$, not limited to the largest one. Therefore, we have shown that a direct generalization of the separability parameter matrix in Theorem~\ref{thm:classification_app} has its all diagonal elements bounded by $\sqrt{2}$, which is finite. This is more general than the result of the 2-qubit encoding we considered above, while the derivation of the exact separability classification parameter is beyond our scope here.

Taken together, these results establish a qualitative distinction between the collapse models. For GRW and CSL, increasing the code distance permits an unbounded enhancement of the effective entangling scale relative to the logical errors while maintaining an exponentially small contact probability. For DP, by contrast, the common kernel underlying the coherent interaction and decoherence imposes a finite upper bound on the separability parameter. Thus the different recoverability behavior observed numerically is rooted in the analytical structure of the corresponding noise kernels.

\newpage
\section{Suppressing $p^2$ decoherence in the Anastopoulos--Hu model} \label{app:Asymptotic entangled state in the AH model}

As discussed in App.~\ref{app:Collapse models}, gravity-induced decoherence may also arise from the interaction of matter with a quantized gravitational field, as described, for example, by the Anastopoulos--Hu (AH) model~\cite{anastopoulos2013master}. From the perspective of the present work, the AH model is distinguished by the fact that the Newtonian interaction and the gravitational decoherence are generated by distinct operators. In contrast to the collapse models discussed previously, where the effective interaction and error channels arise from closely related operators and therefore require a more detailed analysis, the AH model provides a particularly simple setting in which these channels can be more readily distinguished. As a result, the construction of a suitable error-correcting encoding can be illustrated in a comparatively transparent manner. In this appendix, we show that a stabilized Schr\"odinger-cat code can suppress the dominant decoherence channel while preserving a non-trivial interaction, and that the resulting asymptotic state becomes entangled. This provides a direct certification that the model is non-separable.

We first show that the Schr\"odinger-cat code suppresses the $p^2$-induced decoherence while preserving the Newtonian interaction within the encoded subspace. We then introduce a local stabilization mechanism and demonstrate numerically that the resulting asymptotic state becomes entangled. Without loss of generality, we consider two identical masses $m_A=m_B=m$ separated by a distance $d$. We denote their displacement operators from the equilibrium positions by $x_A$ and $x_B$, and their respective momentum operators by $p_A$ and $p_B$. We assume both masses are confined in harmonic traps with frequency $\omega$. The full Lindbladian can be written as
\begin{align}
    \mathcal{L}_{\mathrm{tot}}(\rho)=&\mathcal{L}_{\mathrm{NG}}(\rho)+\mathcal{L}_{\mathrm{AH}}(\rho)\nonumber\\
    =&\frac{i}{\hbar}\Big[\frac{G m_A m_B}{|d+x_A-x_B|},\rho\Big]
    +\frac{8\pi G \Theta}{9 \hbar m^2 c^5}\left( \mathcal{D}[p_A^2]+\mathcal{D}[p_B^2]\right)(\rho)\nonumber\\
    \approx&-\frac{2 i G m_A m_B}{\hbar d^3}[x_A x_B,\rho]
    +\frac{8 \pi G \Theta}{9 \hbar m^2 c^5}\left( \mathcal{D}[p_A^2]+\mathcal{D}[p_B^2]\right)(\rho),
\end{align}
where we ignore the local free Hamiltonian for simplicity. Note that the nonlocal interaction is now based on displacement operators $x_{A,B}$, while the local decoherence effect is based on momentum operators $p_A^2,p_B^2$. Because the interaction acts through $x$ while the dominant error channel acts through $p^2$, one can seek an encoding that approximately trivializes the latter while preserving the former. A simple example is provided by the Schr\"odinger-cat code~\cite{GERRY1993185}, whose code states are defined as follows:
\begin{equation}
    |\bar 0\rangle =\frac{|\alpha\rangle+|-\alpha\rangle}{\sqrt{2(1+e^{-2|\alpha|^2})}}, \quad
    |\bar 1\rangle =\frac{|\alpha\rangle-|-\alpha\rangle}{\sqrt{2(1-e^{-2|\alpha|^2})}}, \quad
    |\alpha\rangle = e^{-\frac{|\alpha|^2}{2}} \sum_{i=0}^\infty \frac{\alpha^i}{\sqrt{i!}}|i\rangle.
\end{equation}
Note that these two coherent states are not orthogonal, i.e. $\langle -\alpha|\alpha\rangle=e^{-2|\alpha|^2}$, so we introduce the two orthonormalized bases $|\bar 0\rangle,|\bar 1\rangle$. Here we choose a real $\alpha=\alpha^*$, and denote $p_0=\sqrt{m\hbar \omega/2}$, $x_0=\sqrt{\hbar/2m \omega}$, the Pauli operators $\sigma^z=|\bar 0\rangle \langle \bar 0|-|\bar 1\rangle \langle \bar 1|$ and $\sigma^x=|\bar 0\rangle \langle \bar 1|+|\bar 1\rangle \langle \bar 0|$, and the projection operator into this code subspace as $\Pi=|\bar 0\rangle\langle \bar 0|+|\bar 1\rangle \langle \bar 1|$. Then in the code subspace, the relevant operators can be rewritten as
\begin{align}
    \Pi p^2 \Pi/p_0^2 =&1+\frac{4|\alpha|^2 e^{-4|\alpha|^2}}{1-e^{-4|\alpha|^2}}-\frac{4|\alpha|^2 e^{-2|\alpha|^2}}{1-e^{-4|\alpha|^2}}\sigma^z,\\
    \Pi p^4\Pi/p_0^4=& 3- \frac{(16 \alpha^4 - 24\alpha^2) e^{-4|\alpha|^2}}{1-e^{-4|\alpha|^2}}+\frac{(16\alpha^4-24\alpha^2)e^{-2|\alpha|^2}}{1-e^{-4|\alpha|^2}}\sigma^z,\\
    \Pi x \Pi/x_0=&\frac{2\alpha}{\sqrt{1-e^{-4|\alpha|^2}}}\sigma^x.
\end{align}
Note that the interaction term is of order $O(\alpha^2)$ in the code space, while the non-trivial part of the error operator (the term proportional to $\sigma^z$) is of order $O(e^{-2|\alpha|^2})$. This implies that in the limit $\alpha\rightarrow\infty$, the error correction condition Eq.~\eqref{eq:qec cond} in the main text holds while the nonlocal interaction remains non-trivial in the code subspace, so entanglement generation is in principle possible.

We now consider an error-avoiding mechanism~\cite{GERRY1993185}. Note that $|\alpha\rangle$ is the unique right null state of the operator $a-\alpha$. In this spirit, the code space spanned by $|\bar 0\rangle,|\bar 1\rangle$, which are linear combinations of $|\alpha\rangle$ and $|-\alpha\rangle$, forms a decoherence-free subspace of the encoding operator $a^2-\alpha^2$. This protects the leakage of quantum information from the code space. As a result, we consider the full Lindbladian consisting of Newtonian interaction, AH gravitational decoherence, and the encoding Lindbladian:
\begin{align}
    \frac{\dd}{\dd t}\rho=&\mathcal{L}_{\mathrm{NG}}(\rho)+\mathcal{L}_{\mathrm{AH}}(\rho)+\mathcal{L}_{\mathrm{encode}}(\rho),\\
    \mathcal{L}_{\mathrm{NG}}(\rho)=&-\frac{2 i G m_A m_B}{\hbar d^3}[x_A x_B,\rho],\nonumber\\
    \mathcal{L}_{\mathrm{AH}}(\rho)=&\frac{8\pi G \Theta}{9 \hbar m^2 c^5}\left( \mathcal{D}[p_A^2]+\mathcal{D}[p_B^2]\right)(\rho),\nonumber\\
    \mathcal{L}_{\mathrm{encode}}(\rho)=&\Gamma_{\mathrm{encode}} \left(\mathcal{D}[a^2_A-\alpha^2]+\mathcal{D}[a^2_B-\alpha^2]\right)(\rho).\nonumber
\end{align}
We consider the strong encoding limit $\Gamma_{\mathrm{encode}}\gg \lambda_{\mathrm{NG}}\equiv 2Gm^2 x_0^2/\hbar d^3$ and $\Gamma_{\mathrm{GD}}\equiv 8 \pi G \Theta p_0^4/9\hbar m^2 c^5$ (gravitational decoherence rate). In this regime, $\mathcal{L}_{\mathrm{encode}}$ strongly stabilizes the system in the code subspace spanned by $\mathcal{C}_{AB}=\operatorname{span}\{|\bar i\rangle_A\otimes |\bar j\rangle_B\}_{i,j=0,1}$, suppressing leakage induced by $\mathcal{L}_{\mathrm{NG}}$ and $\mathcal{L}_{\mathrm{AH}}$.

In the following, we study the asymptotic state of this Lindbladian. For the numerical simulations, we choose $\alpha = 1/2$, which is sufficient to illustrate the mechanism while keeping the Hilbert-space dimension required for convergence manageable. In this case, the matrix elements of the relevant operators read
\begin{align}
    \Pi p^2 \Pi/p_0^2 =&\frac{e}{e-1}-\frac{ \sqrt{e}}{e-1}\sigma^z,\\
    \Pi p^4\Pi/p_0^4=&\frac{3e + 2 }{e-1}-\frac{5\sqrt{e}}{e-1}\sigma^z,\\
    \Pi x \Pi/x_0=&\frac{1}{\sqrt{1-e^{-1}}}\sigma^x.
\end{align}
We initialize the system in $|\bar 0\rangle_A\otimes |\bar 0\rangle_B$. Numerically, we find that after projection onto the code space, the asymptotic state is supported only on the subspace spanned by $\{\ket{\bar 0 \bar 0}, \ket{\bar 1 \bar 1} \}$ due to parity symmetry of the Lindbladian, leading to the structure
\begin{equation}
    \Pi_A\otimes \Pi_B \rho(\infty) \Pi_A\otimes \Pi_B=
    \begin{pmatrix}
        a & 0 & 0 & bi\\
        0 & 0 & 0 & 0\\
        0 & 0 & 0 & 0 \\
        -bi& 0 & 0 & c
    \end{pmatrix},
\end{equation}
where $a,b,c$ are non-negative real numbers that depend on the interaction strength, decoherence strength, and stabilization strength, and we have $b^2\le ac$. This post-selected state can be normalized as
\begin{align}
    \rho^\infty_{\mathrm{post}}
    &\equiv
    \begin{pmatrix}
       \frac{a}{a+c} & 0 & 0 &\frac{ bi}{a+c}\\
        0 & 0 & 0 & 0\\
        0 & 0 & 0 & 0 \\
        -\frac{bi}{a+c}& 0 & 0 & \frac{c}{a+c}
    \end{pmatrix}\nonumber\\
    &=\frac{a}{a+c}|\bar 0\bar 0\rangle\langle \bar 0 \bar 0|
    +\frac{bi}{a+c}\left(|\bar 0 \bar 0\rangle \langle \bar 1\bar 1|-|\bar 1\bar 1\rangle \langle \bar 0 \bar 0|\right)
    + \frac{c}{a+c}|\bar 1\bar 1\rangle \langle \bar 1\bar 1|.
\end{align}

\begin{figure}
    \centering
    \includegraphics[width=0.5\linewidth]{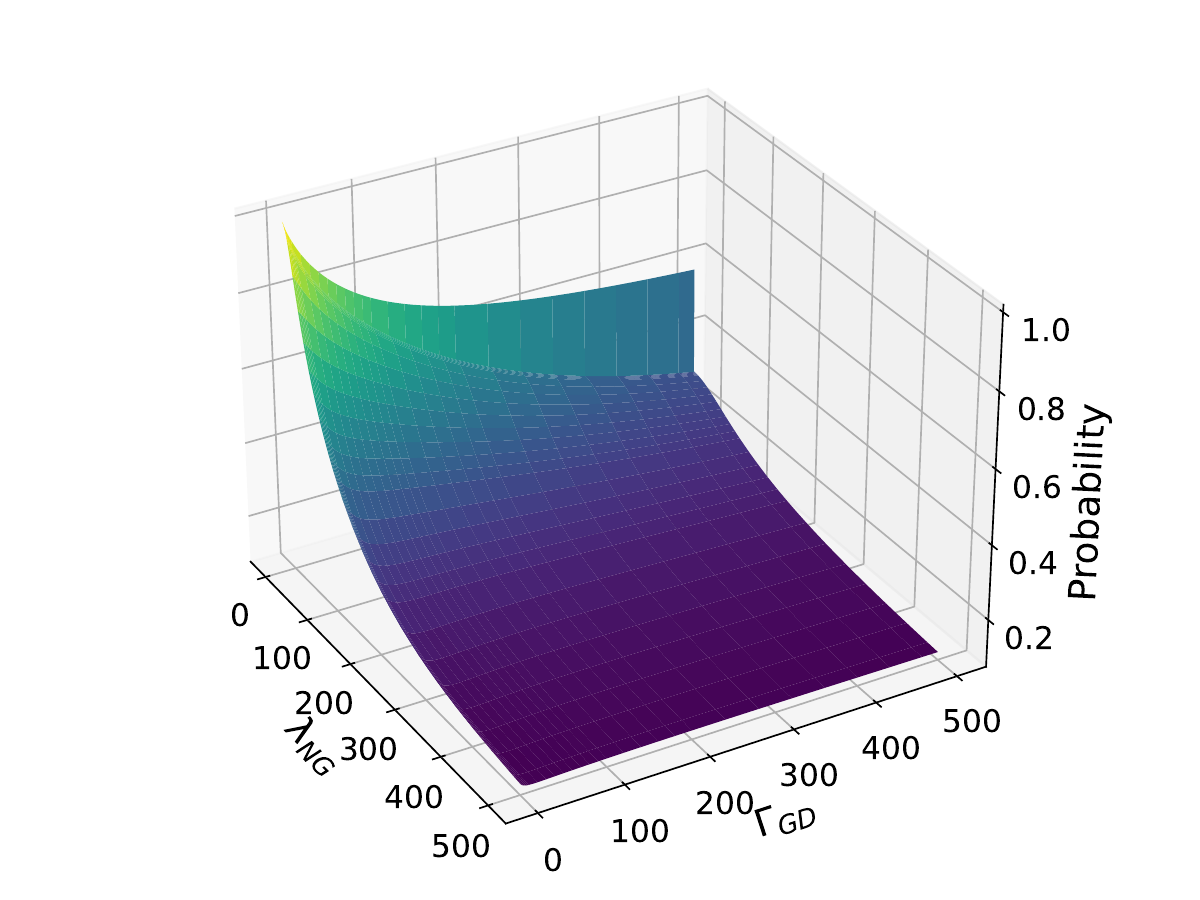}
    \caption{Probability for the post-selection into the code subspace $\{|\bar i\rangle_A\otimes|\bar j\rangle_B\}_{i,j=0,1}$, varied over different interaction strengths $\lambda_{\mathrm{NG}}$ and gravitational decoherence strengths $\Gamma_{\mathrm{GD}}$. Here we choose $\Gamma_{\mathrm{encode}}=100$.}
    \label{fig:post_prob}
\end{figure}

\begin{figure}[htbp]
    \centering
    \includegraphics[width=\textwidth]{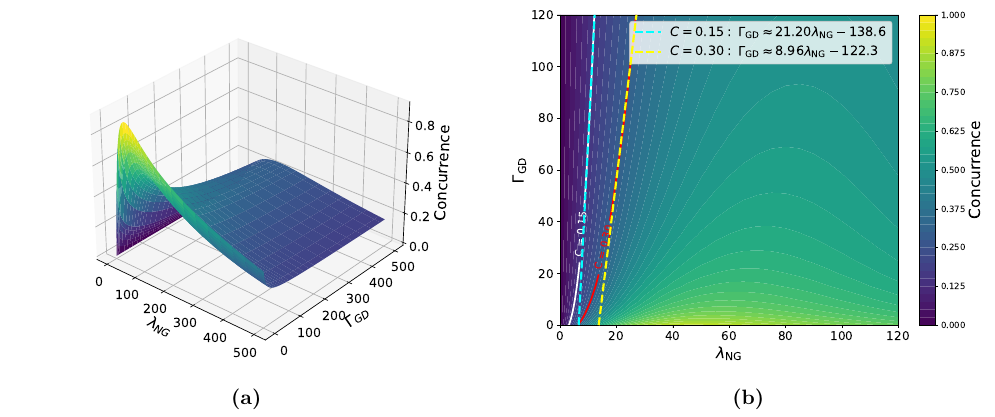}
    \caption{Concurrence $C(\rho^\infty_{\mathrm{post}})$ of the post-selected state projected into the code subspace $\{|\bar i\rangle_A\otimes|\bar j\rangle_B\}_{i,j=0,1}$, varied over different interaction strengths $\lambda_{\mathrm{NG}}$ and gravitational decoherence strengths $\Gamma_{\mathrm{GD}}$. Here we choose $\Gamma_{\mathrm{encode}}=100$. (a) A surface plot over a wide range. (b) A contour plot in the weak interaction and weak gravitational decoherence regime. The contours for $C(\rho^\infty_{\mathrm{post}})=0.15,0.30$ can be well fitted by a linear relation between $\lambda_{\mathrm{NG}}$ and $\Gamma_{\mathrm{GD}}$ for large $\Gamma_{\mathrm{GD}}$.}
    \label{fig:concurrences}
\end{figure}

The amount of entanglement for this family of states can be characterized by the concurrence~\cite{PhysRevLett.80.2245}. The post-selected asymptotic state has non-zero concurrence whenever $b\neq 0$:
\begin{equation}
    C(\rho^\infty_{\mathrm{post}})=\frac{2|b|}{a+c}.
\end{equation}
We now examine how $a,b,c$ depend on the interaction strength $\lambda_{\mathrm{NG}}=2Gm^2 x_0^2/\hbar d^3$, the gravitational decoherence strength $\Gamma_{\mathrm{GD}}=8 \pi G \Theta p_0^4/9\hbar m^2 c^5$, and the stabilization strength $\Gamma_{\mathrm{encode}}$. We fix $\Gamma_{\mathrm{encode}}=100$ and focus on (i) the post-selection probability
\begin{equation}
    \operatorname{Tr}\left(\Pi_A \Pi_B \rho(\infty) \Pi_A \Pi_B\right)=a+c,
\end{equation}
and (ii) the concurrence of the post-selected state $C(\rho^\infty_{\mathrm{post}})=2b/(a+c)$. First, varying $\lambda_{\mathrm{NG}}$ and $\Gamma_{\mathrm{GD}}$, the post-selection probability is shown in Fig.~\ref{fig:post_prob}. It decreases as $\lambda_{\mathrm{NG}}$ and $\Gamma_{\mathrm{GD}}$ increase. For sufficiently large $\Gamma_{\mathrm{encode}}$, the probability remains appreciable; even for $\lambda_{\mathrm{NG}}=\Gamma_{\mathrm{GD}}=\Gamma_{\mathrm{encode}}$, we have $a+c\approx 0.31$ and $C\approx 0.52$. Second, the concurrence $C(\rho^\infty_{\mathrm{post}})$ is shown in Fig.~\ref{fig:concurrences}. The maximum concurrence is achieved around $\lambda_{\mathrm{NG}}/\Gamma_{\mathrm{encode}}\approx 0.46$ and $\Gamma_{\mathrm{GD}}=0$, where $a+c\approx 0.69$ and the post-selected state approaches a Bell pair as $C\approx 0.91$. To resolve the weak interaction behavior, Fig.~\ref{fig:concurrences}b displays the contour plot in the small $\lambda_{\mathrm{NG}}$ regime. When $\lambda_{\mathrm{NG}}$ is strong enough, the concurrence saturates to a constant level even if $\Gamma_{\mathrm{GD}}\gg \Gamma_{\mathrm{encode}}$, as seen in Fig.~\ref{fig:concurrences}a. In the weak interaction limit, the contour can be well approximated by a linear relation; for instance, the fitted relation indicates that when
\begin{equation}
    \Gamma_{\mathrm{GD}}\le 20\lambda_{\mathrm{NG}}- \frac{7}{5}\Gamma_{\mathrm{encode}},
\end{equation}
one can achieve an observable concurrence $C(\rho^\infty_{\mathrm{post}})\ge 0.15$. Equivalently, the interaction strength requirement can be written as
\begin{equation}
    \lambda_{\mathrm{NG}}\ge \frac{\Gamma_{\mathrm{GD}}}{20}+\frac{7\Gamma_{\mathrm{encode}}}{100},
\end{equation}
which implies that entanglement can remain observable even in the regime $\lambda_{\mathrm{NG}}\ll \Gamma_{\mathrm{GD}}$.

In summary, the AH model provides a particularly transparent example in which the dominant $p^2$-type decoherence channel can be suppressed while preserving a non-trivial Newtonian interaction, such that entanglement generation is well established.

\end{document}